\documentclass[useAMS,usenatbib,]{mn2e}
\usepackage{times}
\usepackage{natbib}
\usepackage[dvips]{graphicx}
\usepackage{subfigure}
\newcommand{\iee}{{\it i.e.}}
\newcommand{\eg}{{\it e.g.}}

\newcommand{\msun}{$M_{\odot}$}
\title[]
{A non-ideal MHD Gadget: Simulating massive galaxy clusters}
\author[A. Bonafede, K. Dolag, F. Stasyszyn, G. Murante, S. Borgani]
{A. Bonafede$^{1,2,3}$\thanks{E-mail: a.bonafede@jacobs-university.de},
K. Dolag$^{4,5}$, F. Stasyszyn$^{4,5}$, G. Murante$^6$, S. Borgani$^{7,8,9}$\\
$^1$ Jacobs University Bremen, Campus Ring 1, D-28759 Bremen, Germany  \\
$^2$INAF Istituto di Radioastronomia, via P. Gobetti 101, I-40129 Bologna, Italy\\
$^2$Universit\`a di Bologna, Dip. di Astronomia, via Ranzani 1, I-40126 Bologna, Italy\\
$^4$ Universit\"atssternwarte M\"unchen, M\"unchen, Germany\\
$^5$ Max-Planck-Institut f\"ur Astrophysik, Garching, Germany\\
$^6$ INAF -- Osservatorio astronomico di Torino, Str. Osservatorio
25,  I-10025, Pino Torinese, Torino, Italy\\
$^7$ Universit\'a di Trieste, Dip. di Fisica, Sezione di Astronomia, via Tiepolo 11, I-34131 Trieste, Italy\\
$^8$ INAF -- Osservatorio astronomico di Trieste, via Tiepolo 11, I-34131 Trieste, Italy\\
$^9$ INFN -- Istituto nazionale di Fisica nucleare, Trieste, Italy }
\begin{document}

\date{Accepted ???. Received ???; in original form ???}


\maketitle

\label{firstpage}

\begin{abstract}
 Magnetic fields in the intra-cluster medium of galaxy clusters have
 been studied in the past years through different methods. So far, our
 understanding of the origin of these magnetic fields, as well as
 their role in the process of structure formation and their interplay
 with the other constituents of the intra-cluster medium is still
 limited. In the next years the up-coming generation of radio
 telescopes is going to provide new data that have the potential of
 setting constraints on the properties of magnetic fields in galaxy
 clusters. \\ Here we present zoomed-in simulations for a set of
 massive galaxy clusters ($M_{v} \geq 10^{15} h^{-1 }M_{\odot}$). This
 is an ideal sample to study the evolution of magnetic field during
 the process of structure formation in detail. 
 Turbulent motions of the gas within the ICM will manifest
  themselves in a macroscopic magnetic resistivity $\eta_m$, which has
  to be taken explicitly into account, especially at scales below the
  resolution limit.  We have adapted the MHD {\tt GADGET} code by
  Dolag \& Stasyszyn (2009) to include the treatment of the magnetic
  resistivity and for the first time we have included non-ideal MHD
  equations to better follow the evolution of the magnetic field
  within galaxy clusters. We investigate which value of the magnetic
  resistivity $\eta_m$ is required to match the magnetic field profile
  derived from radio observations. We find that a value of $\eta_m
  \sim 6 \times 10^{27}$ cm$^2$s$^{-1}$ is necessary to recover the
  shape of the magnetic field profile inferred from radio observations
  of the Coma cluster. This value agrees well with the expected level
  of turbulent motions within the ICM at our resolution limit.  The
  magnetic field profiles of the simulated clusters can be fitted by a
  $\beta-$model like profile (Cavaliere \& Fusco-Femiano 1976), with
  small dispersion of the parameters.  We find also that that the
  temperature, density and entropy profiles of the clusters depend on
  the magnetic resistivity constant, having flatter profiles in the
  inner regions when the magnetic resistivity increases.


\end{abstract}



\section{Introduction} \label{sec:intro}
Magnetic fields are an important ingredient to understand the physical
processes taking places in the intra-cluster medium (ICM) of galaxy
clusters. Their presence is demonstrated by radio observations, which,
since the last 30 years, have revealed diffuse and faint radio sources
filling the central Mpc$^3$ of some galaxy clusters \citep[radio
  halos, see \eg][]{2009A&A...507.1257G,2008A&A...484..327V}. These
sources arise because of the interaction of highly relativistic
electrons with the ICM magnetic fields.  About 30 radio halos are
known so far, and all of them are found in clusters with clear
signatures of on-going or recent merger activity
\citep[\eg][]{2001ApJ...553L..15B,2001A&A...369..441G,
  2010ApJ...721L..82C}. The origin of the relativistic particles still
needs to be understood, although several models have been
proposed. Shocks and turbulence associated with merger events are
expected to inject a considerable amount of energy in the ICM, that
could compress and amplify the magnetic field and (re-)accelerate
relativistic electrons, giving thus rise to the observed radio
emission \citep[see][ for recent reviews of the
  subject]{2008SSRv..134...93F,2008SSRv..134..311D}. Understanding the
magnetic field amplification and evolution during the process of
structure formation is then mandatory for modeling the acceleration,
transport and interactions of non-thermal energetic particles and thus
to understand the observed emission. In addition, an accurate modeling
of the magnetic field properties is necessary to understand both the
heat transport and the dissipative processes in the ICM.\\ The
properties of magnetic fields in the ICM have been investigated in the
past through cosmological simulations, performed with different
numerical codes
\citep{1999A&A...348..351D,2002A&A...387..383D,2005JCAP...01..009D,
  2008A&A...482L..13D, 2009MNRAS.398.1678D, 2010ApJS..186..308C} and
also through Faraday Rotation measures analysis
\citep[\eg][]{2004A&A...424..429M, 2006A&A...460..425G,
  2005A&A...434...67V, 2008MNRAS.391..521L, 2010A&A...513A..30B}. The
comparison with observed data is necessary to constrain the main
magnetic field properties, and it is starting now to be feasible
thanks to the progress that has been done in the recent years. One key
aspect is that, so far, large scale radio emission is mainly detected
in very massive clusters. Such massive systems are not easily studied
by numerical simulations, since the size of the density fluctuations
responsible for the formation of massive halos is large, \iee  $\sim$
20 $h^{-1}$Mpc, and the value of the cosmological parameter
$\sigma_{8}$ in the standard $\Lambda CDM$ model requires that
statistically a total volume of $\sim 200 h^{-1} Mpc^3$ needs to be
sampled by simulations in order to produce at least one cluster as
massive as $\sim 10^{15} h^{-1}$\msun.  An important step for studying
non-thermal phenomena is to perform simulations based on extremely
large cosmological volumes, \eg 1 Gpc side-length. Such large volumes
cannot be simulated at the resolution reached by observations, so that
re-simulation techniques have been developed (\eg GRAFIC
\citealt{1995astro.ph..6070B}; ZIC \citealt{1997MNRAS.286..865T};
\citealt{2010MNRAS.403.1859J}). When such high resolution is reached,
the physic of the baryonic component must be followed with particular
care. The magnetic field amplification, in particular, depends on the
small scale motions of the gas. Hence, as the resolution increases
smaller scale motions are revealed, and the magnetic field
amplification increases accordingly \citep[see \eg
][]{2008SSRv..134..311D}. \\ In this paper we present a set of galaxy
clusters extracted by a low resolution DM simulation and re-simulated
at high resolution (the softening length is $\sim$5 kpc $h^{-1}$)
within a cosmological framework in order to resolve scales comparable
to those reached by observations. This work is focused on the 24 most
massive galaxy clusters ($M_{200}>10^{15} h^{-1} Mpc$) of our
sample. Simulations are performed for the first time relaxing the
assumption of ideal MHD, and including a resistivity term in the
induction equation ($\eta_m$). Our sample of simulated galaxy cluster
is publically available for further studies\footnote{contact
  a.bonafede@jacobs-university.de or kdolag@mpa-garching.mpg.de}. In
this paper we present the simulated cluster sample: the MHD
implementation with some test problems (Section. \ref{sec:nonideal}),
the re-simulation technique used (Section. \ref{sec:clusters} and more
detailed in the appendix); the effect of different values for $\eta_m$
are analyzed and discussed in Section \ref{sec:etam}, where the main
properties of the clusters are also presented.  Finally, discussion
and conclusions are reported in Section \ref{sec:disc}.\\ This is a
first paper aimed at presenting the cluster sample, the zoom-in
initial conditions, and the non-ideal MHD implementation in the {\tt
  GADGET} code. This sample has also been used by
\citet{2011arXiv1102.2903F} for a study of the scaling relations of
X--ray mass proxies. In a future paper the authors will investigate in
more detail the cluster properties, and the interplay between thermal
and non-thermal components in the ICM.


\section{Non-ideal MHD simulations}
\label{sec:nonideal}
Within the last decade, cosmological simulations of structure
formation have shown that the observed properties of magnetic fields
in galaxy clusters are direct consequences of turbulent amplification
driven by the the structure formation process
\citep{1999A&A...348..351D, 2002A&A...387..383D, 2005JCAP...01..009D,
  2005ApJ...631L..21B, 2008A&A...482L..13D, 2009MNRAS.398.1678D,
  2010ApJS..186..308C}. Simulations performed with different codes
reach good agreement in predicting that the ratio of the bulk kinetic
energy to the thermal energy has an upper limit of $\sim$10-20\%
\citep[see e.g. the review by][ and references
  therein]{2009arXiv0906.4370B}. Recently, non-cosmological MHD
simulations of merging galaxies
\citep{2009MNRAS.397..733K, 2010ApJ...716.1438K} predict that the
magnetic field is amplified up to a level close to $\sim$10-20\% of
the thermal energy.  The same is expected for the ICM of galaxy
clusters. Although the properties of magnetic fields in galaxy
clusters are still not strongly constrained from the observational
point of view, present data suggest that the magnetic field energy
content is not amplified up to the level of the kinetic energy. In the
Coma cluster, for example, the turbulent energy content is $\sim$10\%
of the thermal one \citep{2004A&A...426..387S}, whereas the magnetic
energy content associated within the observed magnetic field of
$4.7\mu G$ \citep{2010A&A...513A..30B} is only 1.6\% of the thermal
one. Dissipative processes could possibly explain the saturation of
magnetic fields far below the level of the kinetic energy. Such
dissipative processes, driven by the physical properties of the ICM
plasma, are not investigate in numerical simulations so far, but
\citet{2009MNRAS.398.1678D} have shown that dissipative processes
driven by numerical diffusivity may alter the central properties of
the magnetic field profiles obtained by numerical simulations. \\ The
simulations we present in this paper were carried out with {\tt
  GADGET-3} \citep{2005MNRAS.364.1105S}, the current version of the
parallel TreePM+SPH simulation code {\tt GADGET}
\citep{2001NewA....6...79S}. It includes an entropy-conserving
formulation of SPH \citep{2002MNRAS.333..649S}, the implementation of
ideal MHD \citep{2009MNRAS.398.1678D} and an implementation of a
divergence cleaning scheme
\citep{2009MNRAS.398.1678D,2001ApJ...561...82B}. The cosmological
simulations presented here assume an initially homogeneous magnetic
field of $10^{-12}$G co-moving.\\ In previous works it was usually
assumed that the electric conductivity of the gas is infinite, meaning
that the second term of the induction equation (Eq. \ref{eq:Induct1})
vanishes ($\eta_m=0$).
\begin{equation}
\label{eq:Induct1}
\frac{\mathrm{ \partial}\vec{B}}{\mathrm{ \partial}t} = 
\vec{\nabla} \times ( \vec{v} \times \vec{B} ) +
\vec{\nabla} \times ( \eta_m \vec{\nabla} \times \vec{B}).
\end{equation}
This assumption results in a magnetic field frozen into the gas. We
have extended the treatment of the induction equation to cover the
resistive MHD equation. Here we will assume for simplicity a
  spatially constant resistivity term $\eta_m$. In Section
  \ref{sec:Whyeta20} the physical origin of this term is analyzed and
  the assumption will be discussed. Under the constraint
  $\vec{\nabla}\cdot\vec{B}=0$, and $\eta_m$ spatially constant, the
  induction equation for resistive MHD can then be written as:
\begin{equation}
\label{eq:Induct2}
\frac{\mathrm{ \partial}\vec{B}}{\mathrm{ \partial}t} = 
(\vec{B}\cdot\vec{\nabla})\vec{v} -
\vec{B}(\vec{\nabla}\cdot\vec{v}) + \eta_{m} \vec{\nabla}^2 \vec{B}.
\end{equation}
The resistivity dependent terms have been implemented in the code
  following the approach adopted for the artificial dissipation by
  \citet{2004MNRAS.348..123P,2004MNRAS.348..139P,2005MNRAS.364..384P}. In
  particular, we refer to \citet{2009MNRAS.398.1678D} where the
  artificial dissipation term has been implemented in the {\tt GADGET}
  code. More specifically, the resistive term is included in the
  induction equation as
\begin{equation}
\frac{\mathrm{ \partial}\vec{B_i}}{\mathrm{ \partial}t}|_{res} = \frac{\eta_m
  \rho_i}{Ha^2} \sum_{j=1}^{N}\frac{m_j}{\hat{\rho}_{i,j}^2}\left(\vec{B_i}-\vec{B_j} \right) \frac{\vec{r_{i,j}}}{|\vec{r_{i,j}}|} \cdot \vec{\nabla_i}W_i
\end{equation}
Where $i$ and $j$ refer to two generic particles in the simulation,
$\vec{r}_{i,j}$ is the distance between particle $i$ and $j$, $W$ is
the SPH kernel, and the factor $(Ha^2)^{-1}= \frac{dt}{da}$ takes into
account that the internal time variable in {\tt GADGET} is the
expansion parameter $a$. The resistivity term implemented in the
induction equation causes a change in the Entropy $A$ at the rate
\begin{equation}
\frac{d A_i}{dt}|_{res}=-\frac{\gamma -1}{2 \mu_0 \rho_i^{\gamma-1}} \sum_{j=1}^{N}\frac{m_j}{\hat{\rho}_{i,j}^2}\left(\vec{B_i}-\vec{B_j} \right)^2\frac{\vec{r_{i,j}}}{|\vec{r_{i,j}}|} \cdot \vec{\nabla_i} \overline{W}_{i,j},
\end{equation}
where $\overline{W}_{i,j}$ indicates the mean between the two kernels
$W_i$ and $W_j$, $\gamma$ is the adiabatic index of the gas. We refer
to \citet{2009MNRAS.398.1678D} for more details about the numerical
implementation. Since we basically replaced the artificial dissipation
already implemented and tested \citep{2009MNRAS.398.1678D} with a
physically motivated magnetic resistivity, the only tests that are
left to be performed to validate the numerical scheme are those
regarding the ability of the code to reproduce the correct dissipation
timescale. This can be done, in the case of a spatially constant
$\eta_m$, by investigating the magnetic field evolution for simple
test problems.
\begin{figure}
\begin{center}
\includegraphics[width=\columnwidth]{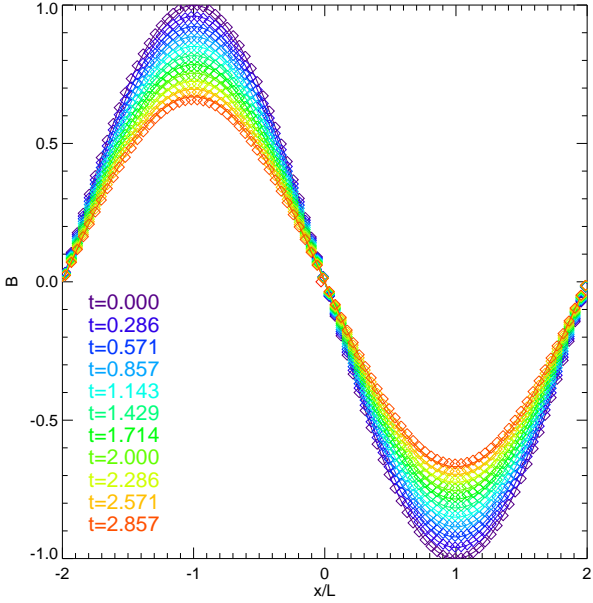}
\end{center}
\caption{ Comparison of the results from the simulations
    (diamonds) to the analytic solution (lines) at different output
    times. The magnetic resistivity $\eta_m$ was set to $1$ in this
    test. For graphical reasons, only one diamond each 8th particle 
    in x-direction was plotted for every time step.}\label{fig:MHDtest1}
\end{figure}
\subsection{Test 1: A one-dimensional slab in a 3D setup}
\label{sec:test1}
We consider first the time evolution of a one-dimensional magnetic
field ($\vec B=B(t)\hat y$) in a one dimensional slab at rest having
side length $4L$. In order to test the code within the configuration
used for cosmological simulations, we performed the test in a 3D setup
 using a glass-like particle distribution and solving a planar
  test problem within this 3D setup \citep{1996clss.conf..349W}. We
  started with $700\times10\times10$ particles, having
a mean inter-particle separation along the x axis of 5.7$\times10^{-3}L$.
Using 64 neighbors within the SPH interpolant
this correspond to roughly 35 resolution elements per length $L$.

The induction equation here reduces to
\begin{equation}
\frac{\mathrm{ \partial}B}{\mathrm{ \partial}t} = 
\eta_m\frac{\mathrm{d}^2B}{\mathrm{d}x^2},
\end{equation}
which has the analytical solution:
\begin{equation}
  B(t) = exp\left(-\eta_m t \left(\frac{2\pi}{4L}\right)^2\right) 
      B_0 \mathrm{sin}\left(\frac{2 x \pi}{4L}\right).
\end{equation}
  We set $B_x=B$ and $B_y=B_z=0$,
and followed the evolution of an initial magnetic field $B = B_0
\mathrm{sin}(2\pi x/(4L))$.\\ In Figure \ref{fig:MHDtest1} the time
evolution of the system is shown. The results obtained from the
numerical simulation (diamonds) are compared with the analytic
solution (lines) for various time steps, showing an excellent
agreement.

\begin{figure}
\begin{center}
\includegraphics[width=0.9\columnwidth]{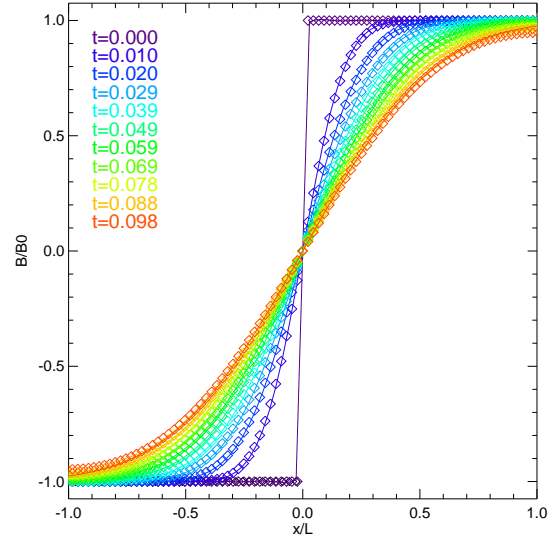}
\end{center}
\caption{ Comparison of the results from the simulations
    (diamonds) to the analytic solution (lines) at different output
    times. The magnetic resistivity $\eta_m$ was set to $1$ in this
    test. For graphical reasons 1 diamonds each 4th particle in x-direction is
    plotted for every time.}\label{fig:MHDtest2}
\end{figure}

\subsection{Test 2: Magnetic diffusion across a step in a 3D setup}
\label{sec:test2}
As a second test, we consider here a one dimensional slab.  The
magnetic field is described by $\vec B= B(x,t)\hat y$, and a step
profile for the magnetic field was included according to:
\begin{equation}
B(x,0)=
\left\{\begin{array}{ll}
 +B_0 & x>0\\
 -B_0 & x<0
 \end{array}\right.
\end{equation}
As in the previous test, the simulation was performed in a full, three
dimensional setup using a glass-like particle distribution and
  solving a planar test problem within this 3D setup. We started with
  700x10x10 particles, having a mean inter-particle separation along the 
  x axis of 5.7$\times10^{-3}L$. Using 64 neighbors within the SPH interpolant
  this correspond to roughly 35 resolution elements per length $L$.
Under the
constrain
\begin{equation}
B(-L,t)=-B(L,t)=B_0,
\end{equation}
meaning that the magnetic field is held fixed at two points
($\pm L$) the solution of the diffusion equation can be written as 
\begin{equation}
B(x,t)=B_0\frac{x}{L}+\frac{2B_0}{\pi} \sum_{n=1}^{\infty}{\frac{1}{n}exp\left(
\frac{-n^2\pi^2\eta_mt}{L^2}\right)sin\left(\frac{n\pi x}{L} \right)}
\end{equation}
\citep[see][]{2005GApFD..99..177W}.  In Figure \ref{fig:MHDtest2} the
results of the numerical simulation (diamonds) are compared to the
analytic solution (lines) at different time steps, as reported in the
figure panel. The magnetic field diffuses rapidly and converges
towards the steady-state solution, $B(x)=B_0(x)/L$.  Since we have not
implemented the necessary boundary conditions to keep $B=B_0$ fixed at
the borders, this simulation was stopped early.

\subsection{Total energy conservation}
\label{sec:totE}
The two tests described in Section. \ref{sec:test1} and
\ref{sec:test2} demonstrate the ability of the code to correctly solve
the diffusion equation. When real physical
problems are considered, the magnetic energy dissipated is 
  explicitely added to the energy equation, i.e. it is transferred to
  the system, so that the total energy is conserved. Hence, within
the cosmological simulations presented in the following Sections, the
energy of the dissipated magnetic field is transferred into heat. This
energy is added explicitly to the internal energy, similarly to what
is done when artificial magnetic resistivity is used as a
regularization scheme to suppress numerical instabilities
(\citealt{2007MNRAS.374.1347P}, \citealt{2009MNRAS.398.1678D}).
We refer to
  \citet{2009MNRAS.398.1678D} for the details of the numerical
  implementation, and in particular, to section 3.1 and Figure 4
  (upper middle panel) of \citet{2009MNRAS.398.1678D} where such
  issues are discussed and analyzed. In the two tests presented in
  Section \ref{sec:test1} and \ref{sec:test2} the conversion of the
  dissipated magnetic energy into heat has been switched off, since
  the solution we compare with do not include such conversion. \\


\begin{figure*} 
\begin{center}
\includegraphics[width=0.8\textwidth]{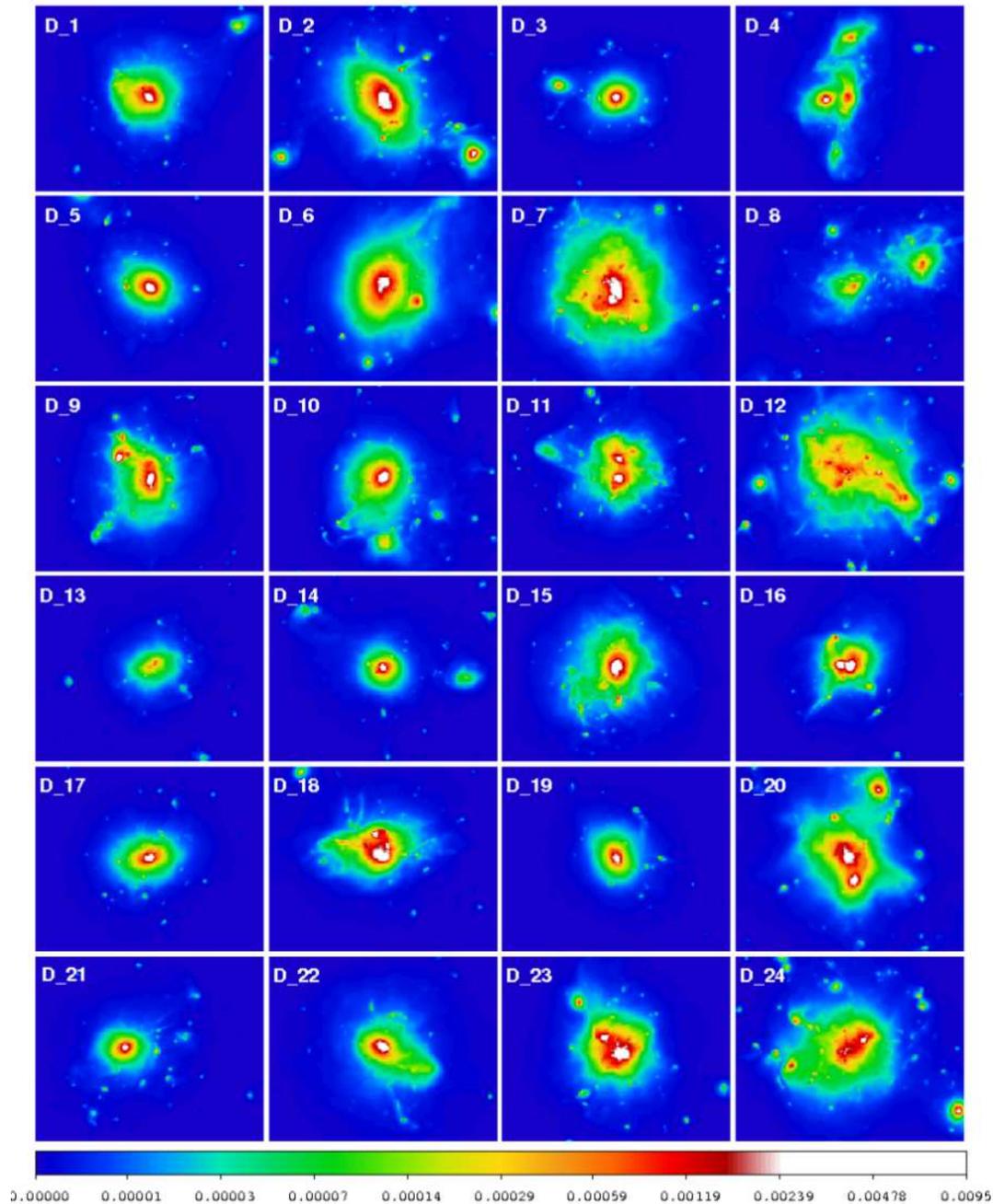}
\caption{Projected X-ray surface brightness of the clusters in our
  sample computed in the range 0.1- 10 keV (square root
  scale). The side of each box corresponds to $\sim$ 1.6 $\times$ 1.4
  $R_\mathrm{vir}$. }
\label{fig:Dianoga_Xmap}
\end{center}
\end{figure*}

\begin{table}
\label{tab:Dianoga_Xray}
\centering
\caption{Properties of the high mass cluster set.}
\begin{tabular}{|c c  c c  c | } 
\hline\hline Cluster & $M_\mathrm{vir}$ & $R_\mathrm{vir}$ & $L_X$ & $T_{MW}$
\\ &\scriptsize{ [$10^{15} h^{-1}$ \msun - $h_{72}^{-1}$\msun ]} & kpc
$h^{-1}$ & [$10^{45}$] erg$s^{-1}$ & [keV] \\

D\_1   &      1.62-2.25  & 2521 & 4.10 &  6.1    \\
D\_2   &      1.51-2.09  & 2442 & 3.60 &  5.0 \\
D\_3   &      1.47-2.04  & 2430 & 6.30 &  6.0 \\
D\_4   &      1.50-2.09  & 2438 & 4.40 &  3.8 \\
D\_5   &      1.53-2.12  & 2455  & 5.34 &  5.6 \\
D\_6   &      1.23-1.70  & 2271 & 1.89 &  5.1 \\
D\_7   &      1.78-2.47  & 2585 & 3.15 &  6.2 \\
D\_8  &      1.85-2.57  & 2707 & 2.58 &  5.6  \\
D\_9  &      1.68-2.33  & 2549 & 5.06 & 6.2 \\
D\_10   &      1.74-2.42  & 2569 & 4.60 & 6.6 \\
D\_11   &      3.09-4.29  & 3133 & 10.5 & 8.7  \\
D\_12  &      1.68-2.32  & 2537 & 2.01 &  4.7  \\
D\_13   &      1.17-1.63  & 2247  & 2.45 &  5.6 \\
D\_14  &      1.56-2.16  & 2484 & 4.87 &  5.5 \\
D\_15   &      1.88-2.61  &  2647 & 4.95 &  5.6 \\
D\_16   &      1.40-1.93  & 2382 & 8.10 &  8.0 \\
D\_17   &      1.81-2.51  & 2626 & 8.95 &  8.4 \\
D\_18  &      1.38-1.92  & 2366 & 6.61 &  6.9 \\
D\_19   &      1.30-1.81  & 2346 & 7.21 &  6.6 \\
D\_20  &      1.07-1.49  & 2165 & 1.86 &  3.5 \\
D\_21  &       1.61-2.23  & 2507 & 4.95 &  5.6  \\
D\_22 &       1.67-2.32  & 2536 & 3.75 &  5.9  \\
D\_23 &       1.90-2.63  & 2648 & 7.43 &  7.9  \\
D\_24  &       1.59-2.21  & 2490 & 2.5  &  4.9  \\
\hline
\multicolumn{5}{l}{\scriptsize Col. 1: Cluster name; Col. 2: Total
  mass inside $R_\mathrm{vir}$; }\\ \multicolumn{5}{l}{ \scriptsize Col. 3:
  Virial radius;}\\ \multicolumn{5}{l} {\scriptsize Col 4: Estimated
  X-ray Luminosity in the band 0.1-10 keV;}\\ \multicolumn{5}{l}{\scriptsize
  Col 5: Mean temperature (mass weighted);}\\ \multicolumn{5}{l}{\scriptsize All quantities are
    computed inside $R_\mathrm{vir}$.}\\
\end{tabular}

\end{table}

\section{Constructing the cluster set}
\label{sec:clusters}
\subsection{The parent simulation}
The clusters were selected from a N-body cosmological simulation
performed according to a flat $\Lambda$CDM cosmological model, with
$\Omega_{m}=$0.24 (the matter density parameter), $\Omega_{bar}=$0.04
(the contribution given by baryons), $h =$0.72, and $\sigma_8=$ 0.8.
The power spectrum for the primordial density fluctuations
$P(k)\propto k^n$ is characterized by $n=0.96$.  This simulation was
carried out with the massively parallel TREE+SPH code {\tt GADGET-3},
the new version of the {\tt GADGET} code
\citep{2001NewA....6...79S,2005MNRAS.364.1105S} and consists of a
periodic box of size 1 Gpc $h^{-1}$. The cluster identification was
performed at $z=0$ using a standard {\it Friend of friends} algorithm
\citep{1985ApJ...292..371D}. The linking length was fixed to 0.17 the
mean inter-particle separation between DM particles, corresponding to
the virial over-density in the adopted cosmological model. This large
simulated cosmological box contains 64 clusters with $M_{FOF}>10^{15}$
$h^{-1}$\msun at $z=0$. Hence, it represents a proper sample to study
the general properties of massive galaxy clusters. Since we want to
analyze the magnetic field properties, and compare our results with
those found from Coma cluster observations, a statistical set of
galaxy clusters with masses similar to the one of Coma is needed. Note
that up to now, no such sample of high resolution re-simulations of
massive galaxy clusters has been constructed.

\subsection{Cluster selection and Initial Conditions}
Clusters were selected from the parent simulation on the basis of
their mass only. We selected the 24 most massive objects among those
with $M_{FOF}>10^{15}$ $h^{-1}$\msun and re-simulated each of these
clusters at higher resolution by using the {\it Zoomed Initial
  Conditions} code \citep[ZIC, ][]{1997MNRAS.286..865T}. In the
appendix the iterative procedure used to obtain the high resolution
initial conditions is described in detail. The setup of initial
conditions was optimized to guarantee a spherical volume around each
cluster with radius of $\sim$ 5-6 virial radii ($R_\mathrm{vir}$)
simulated at high resolution (HR region) and free of contamination by
low resolution, boundary particles.  Two of the cluster initially
selected by the {\it Friend of Friends} algorithm turned out to have a
companion with mass $>10^{15}$ $h^{-1}$\msun.  Other systems are
undergoing a merger event at $z=0$ with less massive companions. In
addition, other clusters with masses between $10^{14}$ $h^{-1}$\msun
and $10^{15}$ $h^{-1}$\msun where found in the HR region of the main
targets.  50 of them are cleaned by low resolution particles inside
their virial radius. Therefore, the final sample consists of 76
clusters with masses larger than $10^{14} h^{-1}$\msun, comprising
both isolated and merging systems. The massive cluster set is shown in
Table \ref{tab:Dianoga_Xray}. In the Appendix (Table
\ref{tab:Dianoga_set}) more details about the cluster surroundings are
given, while in Table \ref{tab:Dianoga_all} the clusters with $M >10^{14}
h^{-1}$\msun, found within 5 $R_\mathrm{vir}$ from the massive
targets, are listed. They are simulated at high resolution
up to 1-5 $R_\mathrm{vir}$, as reported in that Table.\\ The virial
mass of each cluster was defined as the mass contained within a radius
encompassing an average density equal to the virial density,
$\rho_{vir}$, predicted by the top-hat spherical collapse model. For
the assumed cosmology it is $\rho_{vir}= 95 \rho_c$, where $\rho_c$ is
the critical cosmic density \citep{1996MNRAS.282..263E}. In this work
we focus onto the 24 originally selected clusters, as they
represent a statistical well defined, volume and limited sample of
massive clusters. \\ Once the ICs for the DM components have been
obtained, gas particles were added (see appendix for details). The
mass of DM and gas particle is 0.84$\times^{9} h^{-1}$\msun and
0.16$\times10^{9} h^{-1}$\msun respectively. The gravitational
softening length used is 5 kpc $h^{-1}$, which corresponds to the
smallest SPH smoothing length reached in the dense cluster centers.

\subsection{The high mass cluster set}
Simulations of these cluster set including radiative losses and
star-formation are presented in \citet{2011arXiv1102.2903F}. Here, we
focus on non-radiative simulations, since our aim is to study the
effect of the magnetic field, and of non-ideal MHD.\\ From the final
snapshots of these simulations we derived the projected X-ray surface
brightness images, by using a map-making algorithm
\citep{2005xrrc.procE8.10D}.  The predicted emission of every SPH
particle is projected along the line of sight considering an
integration depth of $\pm$ 5 $R_\mathrm{vir}$ around the center of simulated
clusters. The X-ray Luminosity ($L_{X}$) is computed in the range
0.1-10 keV. The X-ray surface brightness images of the clusters are
shown in Figure \ref{fig:Dianoga_Xmap}. The values of $L_{X}$ and of the
gas temperature inside the virial radius are reported in Table
\ref{tab:Dianoga_Xray}. Clusters in different dynamical state belong
to this sample and consequently the X-ray surface brightness images
show quite different morphologies. Several clusters are disturbed in
the very internal part, indicating that a merger event has just
occurred (\eg D\_12), while other clusters have multiple peaks in the
X-ray images, like \eg D\_4.  Some clusters appear to have a regular
shape, and others are going to interact with a smaller halo, that is
visible in the X-ray images (\eg D\_1). In the sample we also found an
ongoing merger event between two massive clusters (D\_8, interacting
with another cluster of $M >10^{15} h^{-1}$\msun).  We note that the
over all range of morphologies found in this mass limited sample
compares qualitatively well with complete, observed samples (like the
REXCESS sample, \citealt{2007A&A...469..363B}), where also such
extremely perturbed systems are found).\\

\begin{figure*} 
\begin{center}
\includegraphics[width=0.7\textwidth]{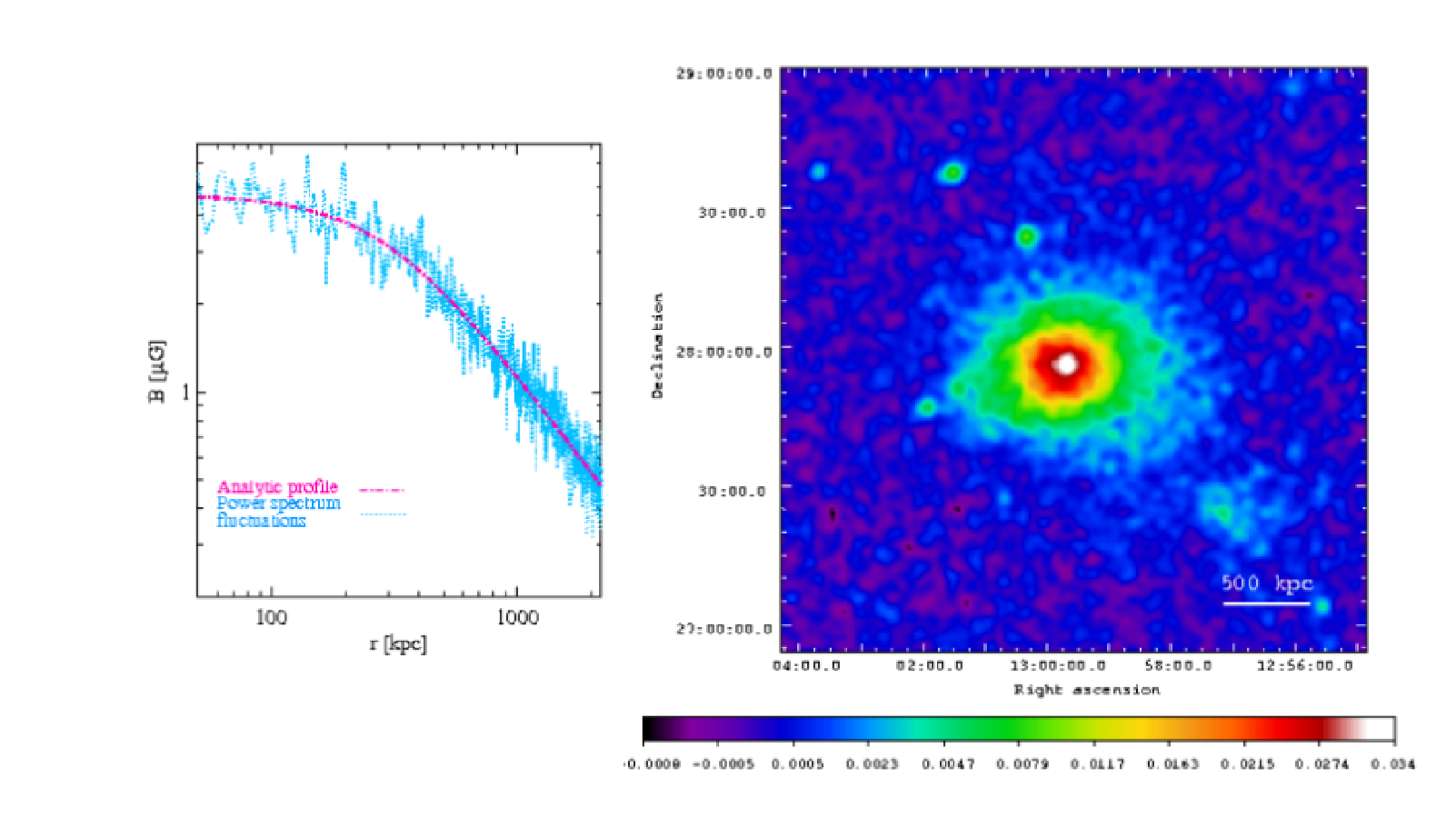}\\

\subfigure{\includegraphics[width=0.47\textwidth]{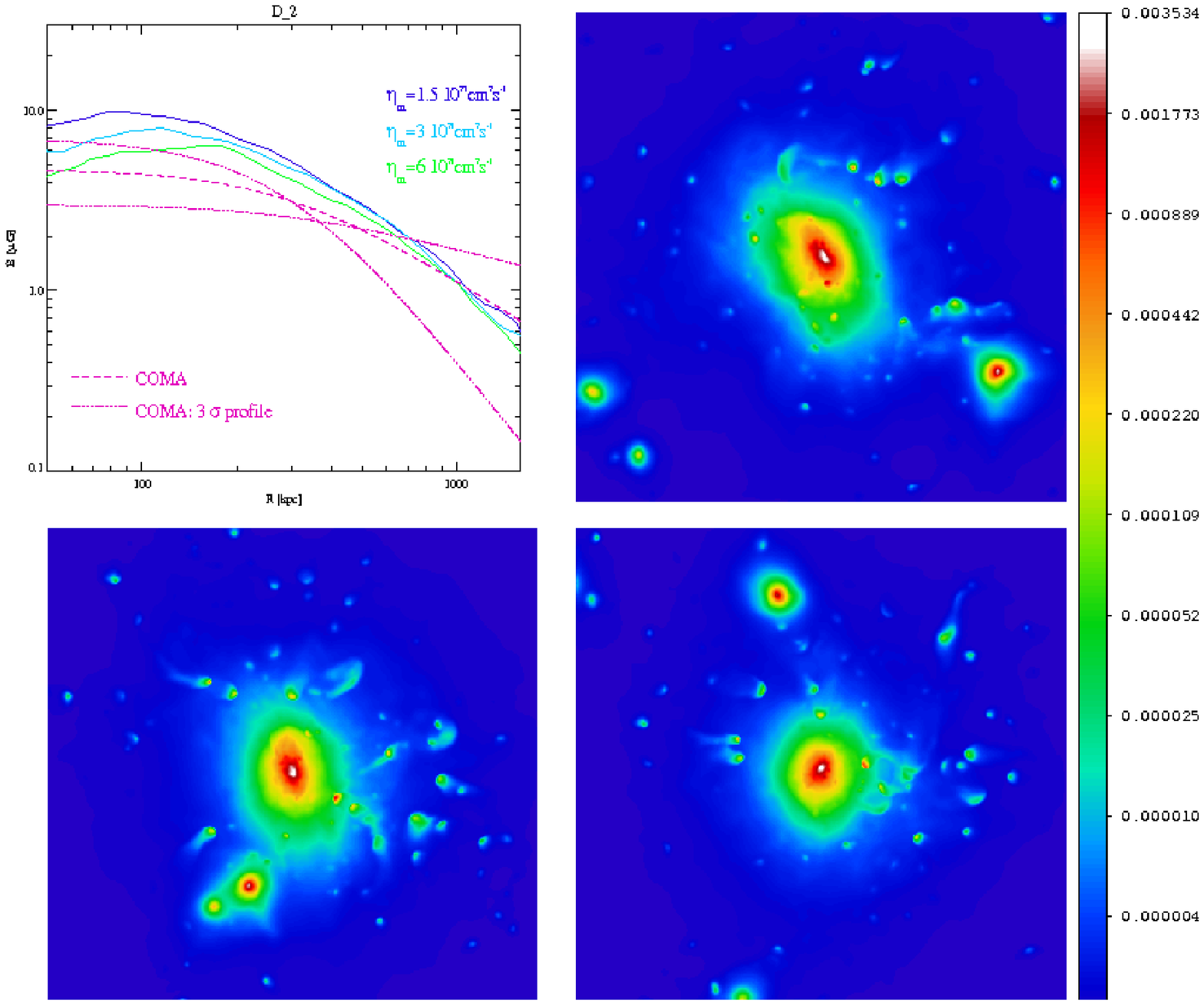}}
\subfigure{\includegraphics[width=0.47\textwidth]{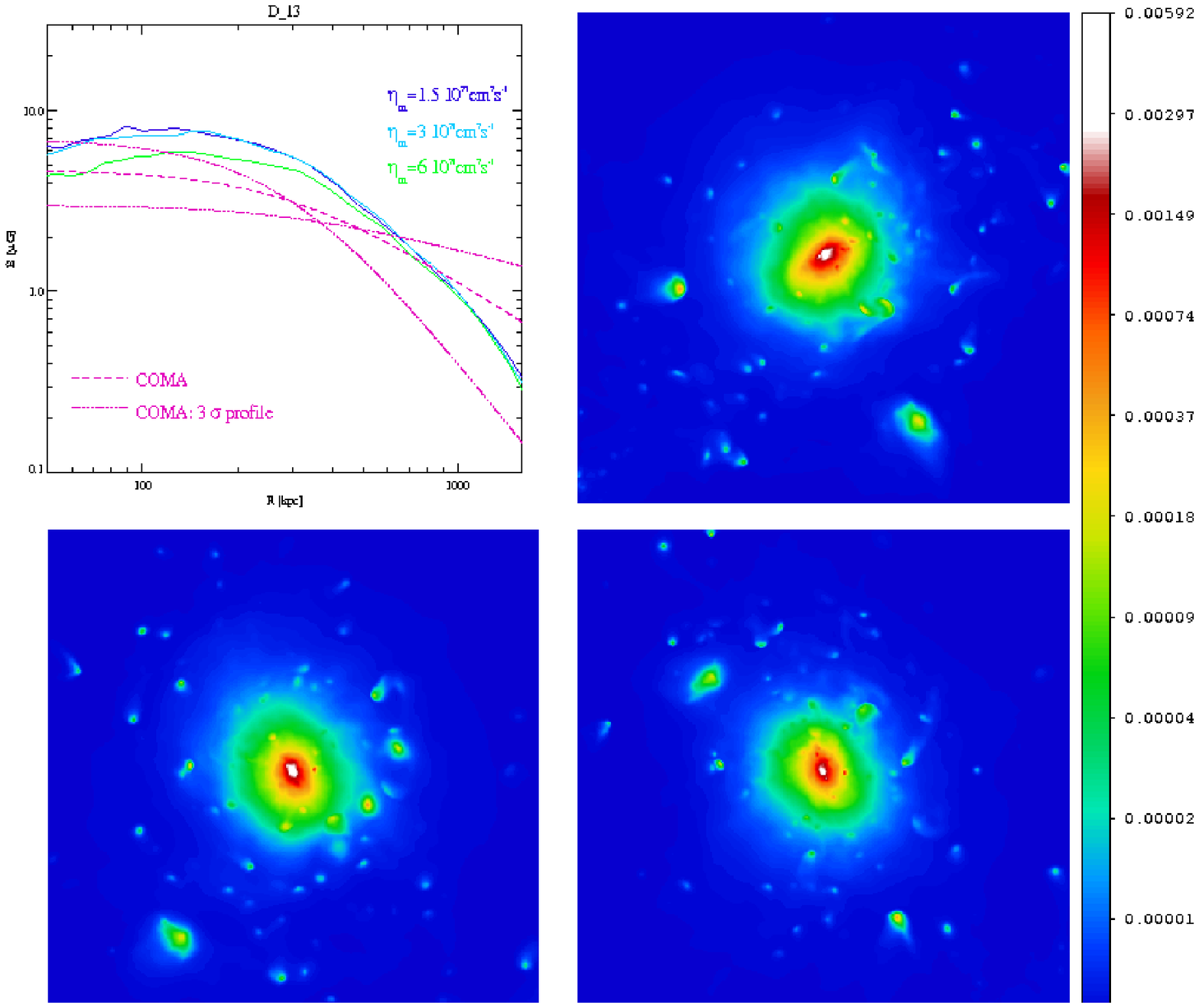}}
\subfigure{\includegraphics[width=0.47\textwidth]{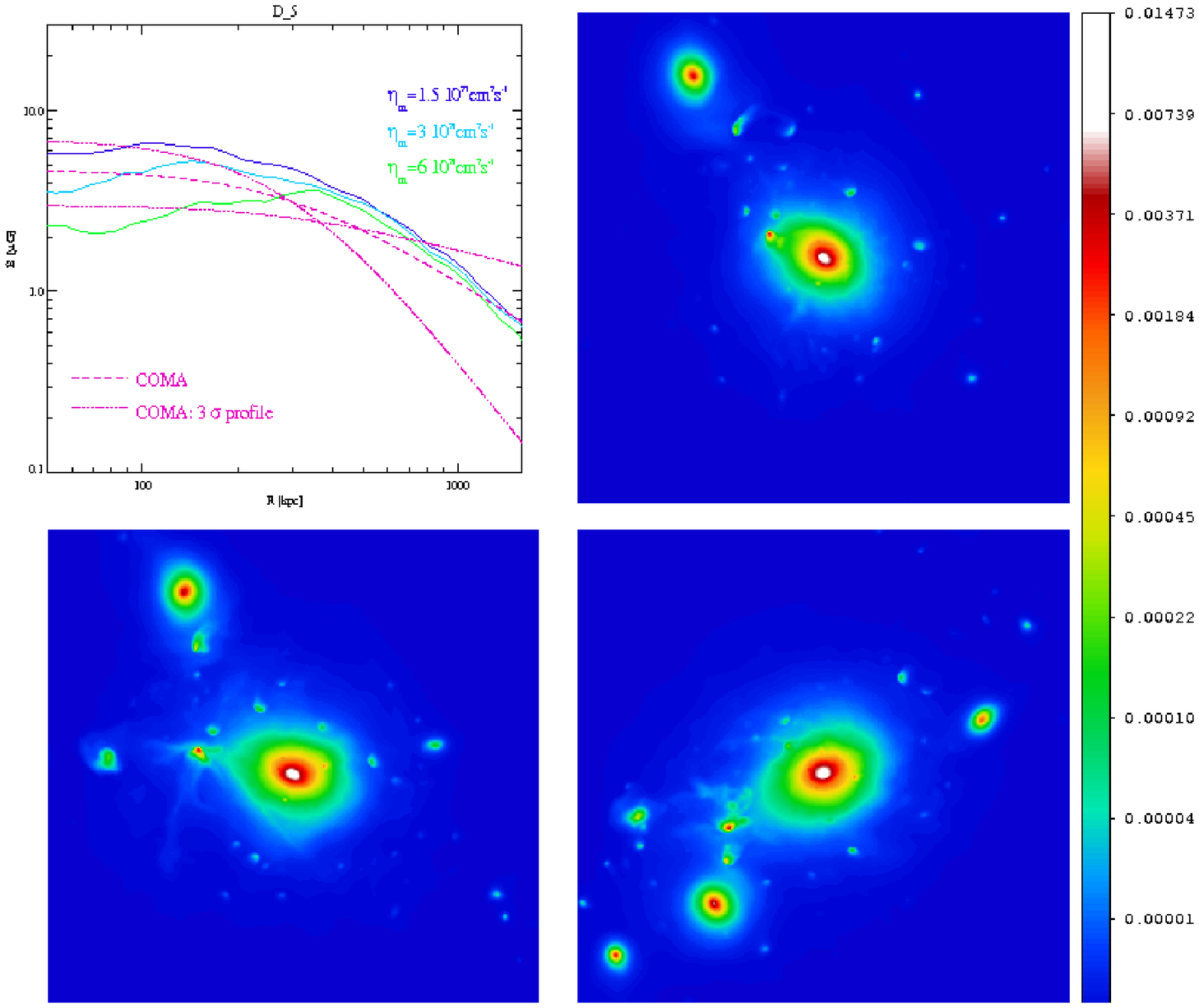}}
\subfigure{\includegraphics[width=0.47\textwidth]{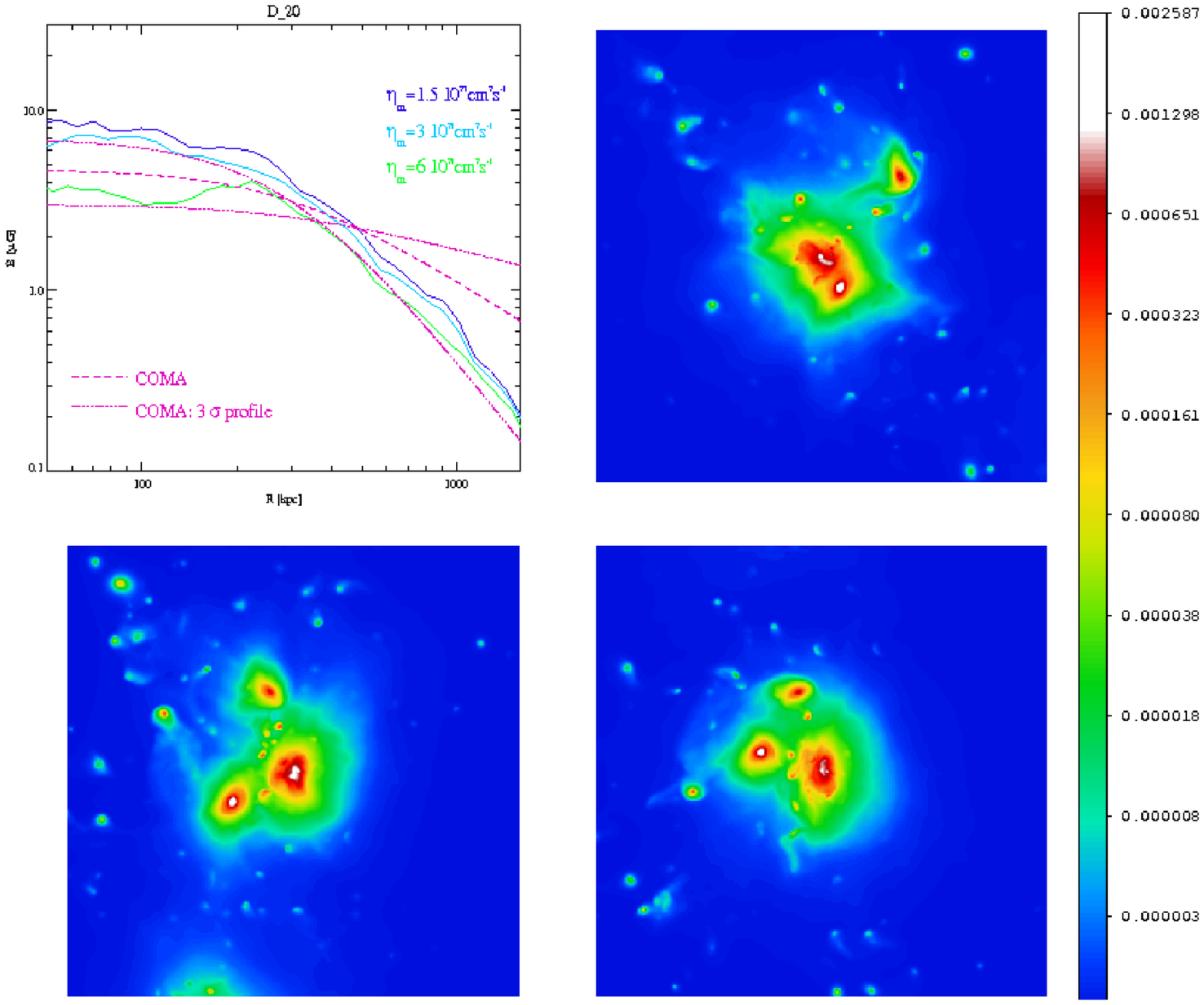}}
\caption{Top: The right panel shows the Coma cluster X-ray surface
  brightness from the ROSAT All Sky Survey, in the energy band 0.1-
  2.4 keV (color coded).  The shown region size is $\sim 3 \times 3$
  Mpc, corresponding to $\sim 1 \: R_\mathrm{vir}\times 1\: R_\mathrm{vir}$. The
  left panel shows the magnetic field profile that gives the best fit
  to Faraday rotation observations (see \citealt{2010A&A...513A..30B}
  for details). Bottom: Projected X-ray surface brightness for the
  clusters used to test the value of $\eta_m$. (Top: D\_2 (left),
  D\_13 (right); Bottom: D\_5 (left), D\_20 (right))x. X Y and Z
  projection are shown in the upper right, lower left and right
  sub-panels of each panel. The size of the projected X-ray images
  corresponds to 2 $R_\mathrm{vir}$. The magnetic field radial profile
  ($\langle B \rangle$) is plotted in the upper left panel for values
  of $\eta_m=$1.3, 3 and 6 $\times 10^{27} \mathrm{cm^2} \:
  \mathrm{s^{-1}}$ (5, 10 and 20 in code units). Magenta lines
  indicate the profile of the Coma cluster as inferred from radio RM
  observations. The shaded line represents the best fit, the
  dotted-shaded lines indicate the flatter and shallower profile that
  are compatible within 3$\sigma$ with radio RM data (see
  \citealt{2010A&A...513A..30B} for further details). }
\label{fig:eta_test}
\end{center}
\end{figure*}

\section{Magnetic fields in massive clusters}
\label{sec:etam}
The properties of the ICM magnetic fields start now to be better
understood, thanks to an increasing effort in analyzing Faraday
Rotation Images of sources located either inside clusters and in their
background (\eg \citealt{2004A&A...424..429M},
\citealt{2004JKAS...37..337C}, \citealt{2004mim..proc...13J},
\citealt{2005A&A...434...67V} \citealt{2006A&A...460..425G},
\citealt{2010A&A...513A..30B}). In general, the magnetic field in
clusters inferred from these observations is found to be consistent
with a magnetic field driven by the turbulence within the ICM and
generally shows a radial decline. Once the density profile $\rho(r)$
has been inferred from X-ray observations, the magnetic field profile
in galaxy clusters is supposed to follow the gas density profile
according to:
\begin{equation}
B(r) = B_0 \rho(r)^\alpha.
\label{eq:Brho}
\end{equation}
The fluctuations within the magnetic field are usually modeled
assuming a power-law power spectrum, described by a slope $\eta$, a
maximum length scale $\Lambda_\mathrm{max}$ (which can be related to
the outer scale of the turbulence in within the ICM) and a minimum
length scale $\Lambda_\mathrm{min}$ (which in case it is resolved,
could be related to dissipative scales, either viscous or
resistive). These model parameter are inferred by comparing the
expected Rotation Measure statistics (mean, dispersion,
auto-correlation function and structure function) and the polarization
properties of the radio galaxies to the observed ones. So far the
magnetic field in the Coma cluster is the one that is best
constrained. It has been inferred from RM observations of seven
radio-sources located at projected distances of 50 to 1500 kpc from
the cluster center. The best fit model results to be the one with
$B_0=4.7^{+0.7}_{-0.8}\mu G$, $\alpha=0.5^{+0.2}_{-0.1}$, and
$\Lambda_\mathrm{min} \sim$ 2 kpc \citep{2010A&A...513A..30B}.
Although previous cosmological MHD simulations of galaxy clusters
produced magnetic field configuration which lead to Rotation Measure
statistics similar to the observed ones \citep{1999A&A...348..351D,
  2002A&A...387..383D,2005JCAP...01..009D}, the magnetic field profile
tended to be steeper, with $\alpha\sim1$ \citep{2001A&A...378..777D}.
In addition, the values of the central magnetic field obtained from
high-resolution simulations resulted to be slightly larger than
observed \citep{2009MNRAS.392.1008D}, but it was noticed that the
magnetic field profiles are significantly altered if the underlying
numerical MHD implementation suffers from the presence of numerical
diffusion \citep{2009MNRAS.398.1678D}.
\begin{table}
\centering
\begin{tabular}{|c c  c c | } 
\hline
$\eta_T [cm^2s^{-1}]$  & $\eta_m$       & $v_{turb}$ [km$s^{-1}$] & $\lambda_{turb}$ [kpc] \\
1.5$\times 10^{27}$  & 5              & 25                 & 2                    \\
3$\times 10^{27}$    & 10             & 50                &  2                    \\
6$\times 10^{27}$    & 20             & 50                &  4                    \\
\hline
\end{tabular}
\caption{Diffusion coefficients used in our simulations (in internal
  and physical units) together example values of turbulent
  length-scales and velocities which would correspond to such values.}
\label{tab:eta1}
\end{table}

\begin{table}
\centering
\begin{tabular}{|c c c | } 
\hline
$\eta_{turb} [cm^2s^{-1}]$   & Process & ref.  \\
3--5 $\times 10^{28}$&CR propagation (value at 1 GeV) & 1\\
3$\times 10^{25}$    & CR driven dynamo in galaxies & 2 \\
1$\times 10^{29}$    & powering the Coma radio halo & 3 \\
2$\times 10^{27}$    & turbulent cascade observed in Coma at 2 kpc & 4 \\
2$\times 10^{28}$    & turbulent cascade from simulations at 7.8 kpc & 5\\
6$\times 10^{27}$--4.5$\times 10^{29}$& iron abundance profile in clusters & 6 \\ 
\hline
\end{tabular}
\caption{Values of the diffusion coefficients commonly used in the
  literature, and observationally inferred. References are: 1:
  \citet{2007ARNPS..57..285S}, 2: \citet{2003A&A...401..809L}, 3:
  \citet{1987A&A...182...21S}, 4: \citet{2004A&A...426..387S}, 5:
  \citet{2009ApJ...707...40M}, 6: \citet{2006MNRAS.372.1840R}. See
  text for details.}
\label{tab:eta2}
\end{table}

\subsection{Testing the effect of the magnetic resistivity}
Having a stable numerical scheme at hand, which does not suffer
  from numerical diffusion outside the SPH smoothing length
(Stasyszyn 2011, in preparation), we can investigate for the first
time the role of a physically motivated resistivity $\eta_m$ in
shaping the ICM magnetic field profile. From our set of massive
clusters, which have all masses comparable to the Coma's one, we
selected four objects that at $z=0$ show X-ray morphologies similar to
the one of Coma. In particular, we avoid selecting clusters with very
spherical morphology as well as clusters with clear multiple X-ray
brightness peaks. Figure \ref{fig:eta_test} shows the X-ray morphology
of the 4 selected clusters for the 3 spatial projection
directions. This sub-set of clusters has been simulated with different
value of $\eta_m$, with the aim of studying the resulting shape and
central value of the magnetic field. Figure \ref{fig:eta_test} shows
the magnetic field profiles of those clusters compared to the best fit
model for the Coma cluster, encompassed by the $\pm \: 3\sigma$
region. Whereas all magnetic field profiles obtained from the
simulations are within the $3\sigma$ region in the outer parts, the
profiles with small magnetic diffusion ($\eta_m=$1.5$\times$10$^{27}
\mathrm{cm}^2 \mathrm{s}^{-1}$) are always above this region towards
the center. For large magnetic diffusion ($\eta_m=6 \times$10$^{27}
\mathrm{cm}^2 \mathrm{s}^{-1}$) half of the simulated profiles are
above the best fit model, the other half below the best fit model in
the central part. From that, we conclude that a value of
$\eta_m=6\times$10$^{27} \mathrm{cm}^2 \mathrm{s}^{-1}$ (20 in the
code internal units) is the one that provides the best match with that
inferred from Coma cluster observations. The numerical diffusion
  inside the SPH smoothing length is of the order of $10^{18} cm^2
  s^{-1}$ that is several orders of magnitudes lower than the one we
  have implemented as magnetic resistivity, and thus does not affect
  our results.

\subsection{Physical origin of the magnetic resistivity}
\label{sec:Whyeta20}
In the previous Sections we have shown that a relatively large value
of $\eta_m$ is required in the induction equation
(Eq. \ref{eq:Induct2}) to match the radial profile and the central
value of the magnetic field inferred from Coma cluster observations.
In order to correctly interpret this result, it must be kept in mind
that the induction equation (Eq. \ref{eq:Induct2}) describes the
evolution of a magnetic field $B$ at our resolution limit (which is of
order of 10 kpc). The turbulent cascade is expected to develop down to
smaller scales, where unresolved turbulent motions would contribute to
the diffusion described by $\eta_m$. Hence, we can define the
diffusion coefficient $\eta_m$ as
\begin{equation} 
\label{eq:etam}
\eta_m = \eta_\mathrm{Coulomb} + \eta_\mathrm{turb},
\end{equation}
with $\eta_\mathrm{Coulomb}$ related to the thermal conductivity $\sigma$ by
\begin{equation}
\eta_\mathrm{Coulomb}=\frac{c^2}{4\pi\sigma}.
\end{equation}
Following \citet{1956pfig.book.....S}, when the mean free path is
determined by Coulomb collisions, the thermal conductivity of the
plasma can be expressed as
\begin{equation}
\frac{1}{\sigma} = \frac{\pi e^2 m_e^{1/2}}{(4\pi\epsilon_0)^2(kT)^{3/2}}\mathrm{ln}(\Lambda),
\end{equation}
$\Lambda$ being the Coulomb logarithm.  For a typical cluster
environment (\eg  densities $n\approx10^{-2} \mathrm{cm}^{-3}$ and temperatures 
$T \approx 10^{8\circ}$K), the diffusion coefficient is:
\begin{equation}
\eta_\mathrm{culomb} \approx 2\times10^{13} T^{-3/2} \approx 20 \; \mathrm{ cm}^2 \mathrm{s}^{-1}.
\end{equation} 
Hence, the diffusion coefficient, arising from the gas thermal
conductivity, does not significantly contribute to the evolution of
$B$ in the induction equation. On the other hand, in a turbulent
plasma the motion of charges will be a random walk characterized by a
length scale $\lambda_\mathrm{turb}$ and by a velocity
$v_\mathrm{turb}$. Following \citet{2005ApJ...622..205D}, the plasma
turbulent diffusion coefficient $\eta_\mathrm{turb}$ can be defined
as:
\begin{equation}
\label{eq:etaturb}
\eta_\mathrm{turb} \sim 0.1 \lambda_\mathrm{turb} \times v_\mathrm{turb}
\end{equation} 
Typical values of $v_{turb}$ at our resolution of several kpc,
corresponding to scales $\lambda_{turb}$ that fall below our
resolution limit, will be several tens of km $s^{-1}$, and will lead
to diffusion coefficients similar to the one that we used in our
simulations (see Table \ref{tab:eta1}).  Estimates of $v_{turb}$ at
these small spatial scales cannot be provided by any observation so
far. However, it is possible to infer such estimates from the values
of $v$ obtained at larger spatial scales, assuming that the turbulent
power spectrum can be described by a single power-law down to the
small scales of interest.  Using X-ray data,
\citet{2004A&A...426..387S} derived pseudo-pressure fluctuations maps
of the gas in the Coma cluster. They revealed the presence of a
scale-invariant pressure fluctuation spectrum, that is consistent with
the Kolmogorov slope, and could estimate the size of the
  turbulent eddies in the range from 40 kpc to 100 kpc. On smaller
scales, the number of photons detected were not sufficient for a
reliable pressure measurement. The energy content associated with
these turbulent motions is estimated to be roughly 10\% of the thermal
one \citep{2004A&A...426..387S}. The sound velocity within the Coma
cluster ($T \sim 10^8$K) is $\sim$ 1500 km$s^{-1}$. Therefore the
turbulent velocities associated with the largest scales ($\approx 100$
kpc) found by \citet{2004A&A...426..387S} would correspond to $\sim$
470 km $s^{-1}$. Assuming a Kolmogorov-like power spectrum, this
translates into a turbulent velocity of $\sim$ 30 km$s^{-1}$ at a
length scale of 2 kpc, that is the minimum scale revealed by Rotation
Measure observations \citep{2010A&A...513A..30B}.  A turbulent
velocity of $\sim$ 30 km$s^{-1}$ at 2 kpc would yield to $\eta_m \sim
2 \times 10^{27} \mathrm{cm^2}s^{-1}$, similar to the value we have
used in the simulations. A sample of clusters for which $v_{turb}$,
$\lambda_{turb}$, and the power spectrum slope are estimated
observationally would of course allow us a better and more reliable
comparison. Although such observations are not available in the
literature so far, another estimate for $\eta_{turb}$ has been derived
by \citet{2006MNRAS.372.1840R}. The authors have analyzed the effect
of turbulent diffusion on the iron abundance profiles in the ICM for a
sample of clusters, finding $\eta_{turb}$ in the range $\mathrm{6}
\times \mathrm{10}^{27}$ - $\mathrm{4.5} \times
\mathrm{10}^{29}$.\\ Estimates of $v_{turb}$ and $\lambda_{turb}$ can
also be derived by cosmological simulation. Different numerical
schemes \citep[\eg][]{2005MNRAS.364..753D, 2006MNRAS.369L..14V,
  2008MNRAS.388.1089I, 2009A&A...504...33V} have been optimized to
follow the evolution of turbulent flows within the ICM of simulated
galaxy clusters \citep[see also][]{2011arXiv1101.4648Z}. These works
indicate that the energy in turbulent motions is $\sim 10-20\%$ of the
thermal one at $z=0$ within the virial radius. In particular,
\citet{2009ApJ...707...40M} have measured the spectral properties of
the gas velocity field, finding a good agreement with the Kolmogorov
power spectrum slope over scales ranging from 300 kpc down to the
scale correspondent to the Nyquist frequency. The velocity of the
turbulent eddies at scales of 10 kpc is estimated to be $\sim$ 50-100
km $s^{-1}$, resulting is $\eta_m \sim \times 10^{28}$, in good
agreement with the values adopted in the simulations presented
here. In addition, \citet{2009ApJ...707...40M} have found
$\eta_\mathrm{turb}\sim \mathrm{2} \times \mathrm{10}^{28} \:
\mathrm{cm}^2 \mathrm{s}^{-1}$, using $ \lambda_\mathrm{turb}=7.8$ kpc
h$^{-1}$ and $ v_\mathrm{turb}=60 \: \mathrm{km}\mathrm{s}^{-1}$. 
 It is also worth mentioning that the value of the diffusion
  coefficient $\eta_m$ is within the range $\eta_m \approx
  3\times10^{25} cm^2s^{-1}$ -- $\eta_m \approx 3\times10^{29}cm^2
  s^{-1}$ which are the values needed to operate a cosmic ray driven
  dynamo within a galaxy \citep[see][]{2010A&A...510A..97S} and to
  power the Coma radio halo by an in-situ acceleration model
  respectively\footnote{We note however that the re-acceleration
  model proposed in that work is due to Alfen modes, while more recent
  works indicate that the re-acceleration is due to MHD modes at very
  small scales, see \eg \citet{2011MNRAS.410..127B}.}
\citep[see][]{1987A&A...182...21S}.\\ All these different values for
$\eta_m$ are reported in Table \ref{tab:eta2}. In summary we can conclude
that our inferred value of $\eta_m \approx 6\times10^{27} cm^2 s^{-1}$
at our unresolved scales of $\sim$ 10 kpc is well in range with what
would be expected from turbulent motions within the ICM.

\subsubsection{About the use of a constant $\eta_m$} 

 The Equations \ref{eq:etam} and \ref{eq:etaturb} clarify the
  physical origin of the resistivity term that we have implemented in
  the induction equation.  In this work, the value of $\eta_m$ has
  been kept constant throughout the whole re-simulations. It is clear
  that having a $\eta_m$ that changes as a function of $v_{turb}$ and
  $\lambda_{turb}$ locally would allow one to follow the evolution of
  the magnetic field more properly. Identifying the turbulent motions
  to compute the most correct value of $\eta_m$ at every step during
  the simulation is however not feasible. Different algorithms have
  been developed to identify and analyze the turbulent motions
  \citep[see
    e.g.][]{2005MNRAS.364..753D,2009ApJ...707...40M,2011A&A...529A..17V}. These
  algorithms need to subtract large-scale laminar motions before
  revealing the turbulent patterns, and can then be applied in the
  post processing once the simulation is run.  We can verify the
  validity of the assumed constant value of $\eta_m$ by checking which
  values of $\lambda_{turb}$ and $v_{turb}$ are obtained at different
  distances from the cluster center by the above mentioned works.
  \citep{2009ApJ...707...40M} have computed the profile of the
  turbulent velocity for a simulated galaxy cluster. The velocity
  profile, once scaled at the length scale of 7.8 kpc - the highest
  resolved region of the simulation- shows a rather flat profile, with
  values ranging from $\sim$50 to $\sim$90 Km$s^{-1}$ within the
  cluster virial radius. This implies that the assumption of a
  constant $\eta_m$, although not optimal, is well justified in our
  case. The simulations presented here represent a good starting point
  to investigate for the first time the effects related to such
  resistivity term. 

\begin{figure*}
\begin{center}
\subfigure{\includegraphics[width=\columnwidth]{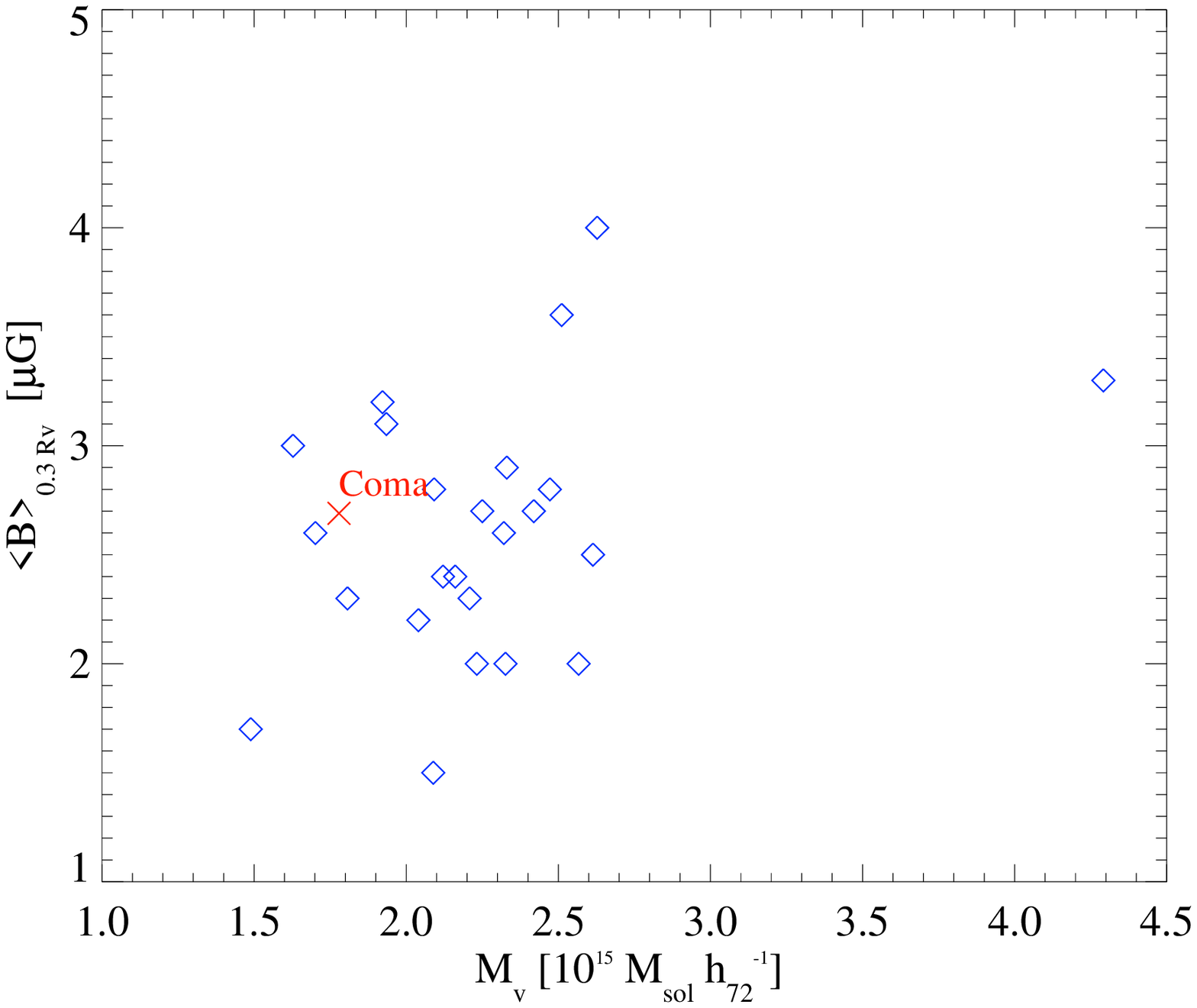}}
\subfigure{\includegraphics[width=\columnwidth]{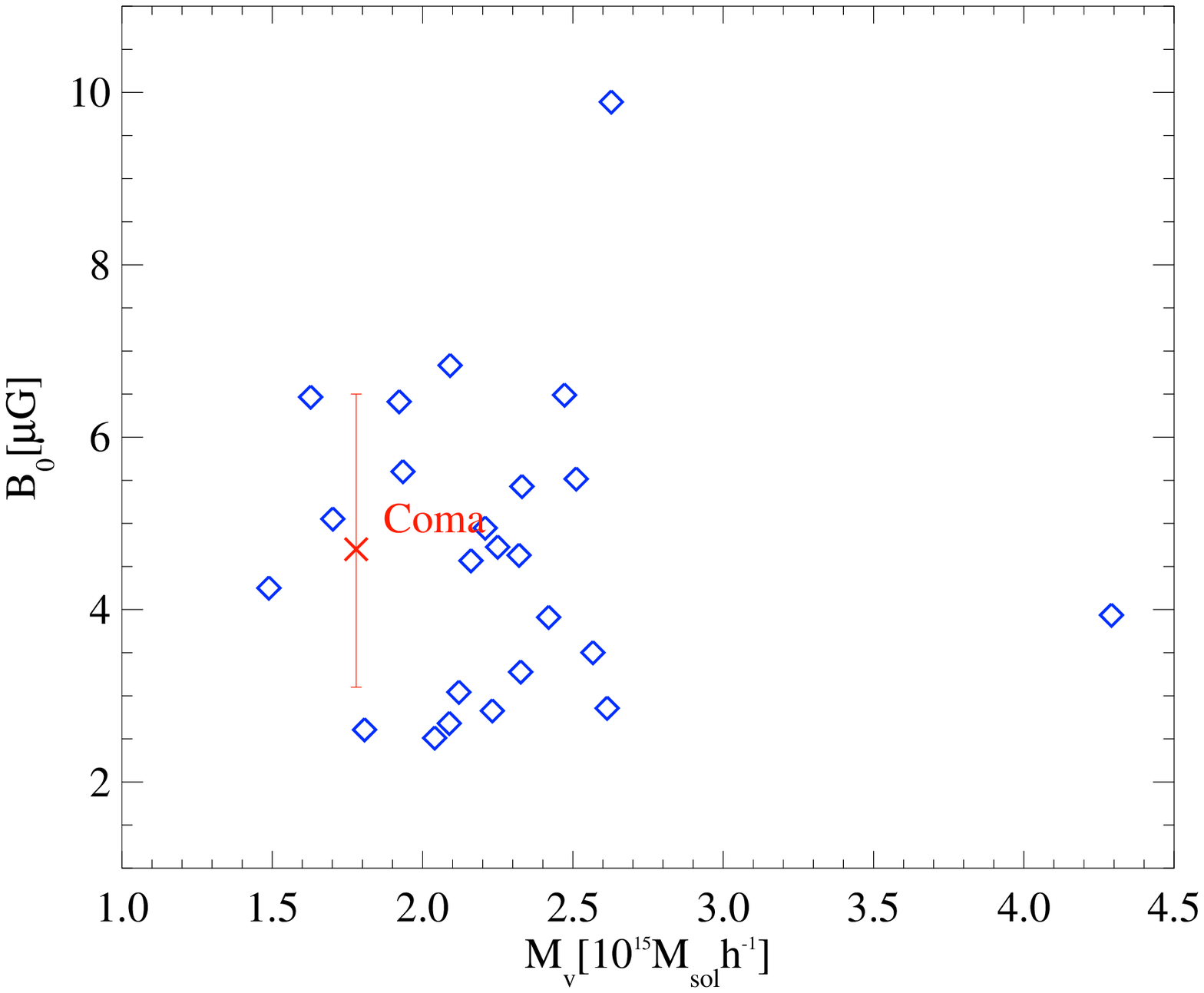}}
\end{center}
\caption{Left: Magnetic field averaged over the central 0.3 $R_\mathrm{vir}$
  versus virial mass of the our cluster set (Blue diamonds). The red cross
  refers to the mean magnetic field for the Coma cluster. The
  error-bar refers to the 3$\sigma$ of the $chi^2$ given by
  \citet{2010A&A...513A..30B}. Right: Magnetic field in the cluster
  center, as results from the fit of a $\beta$-model profile, versus
  the cluster virial mass.}
\label{fig:Bmean_M}
\end{figure*}

\begin{figure*}
\begin{center}
\subfigure{\includegraphics[width=\columnwidth]{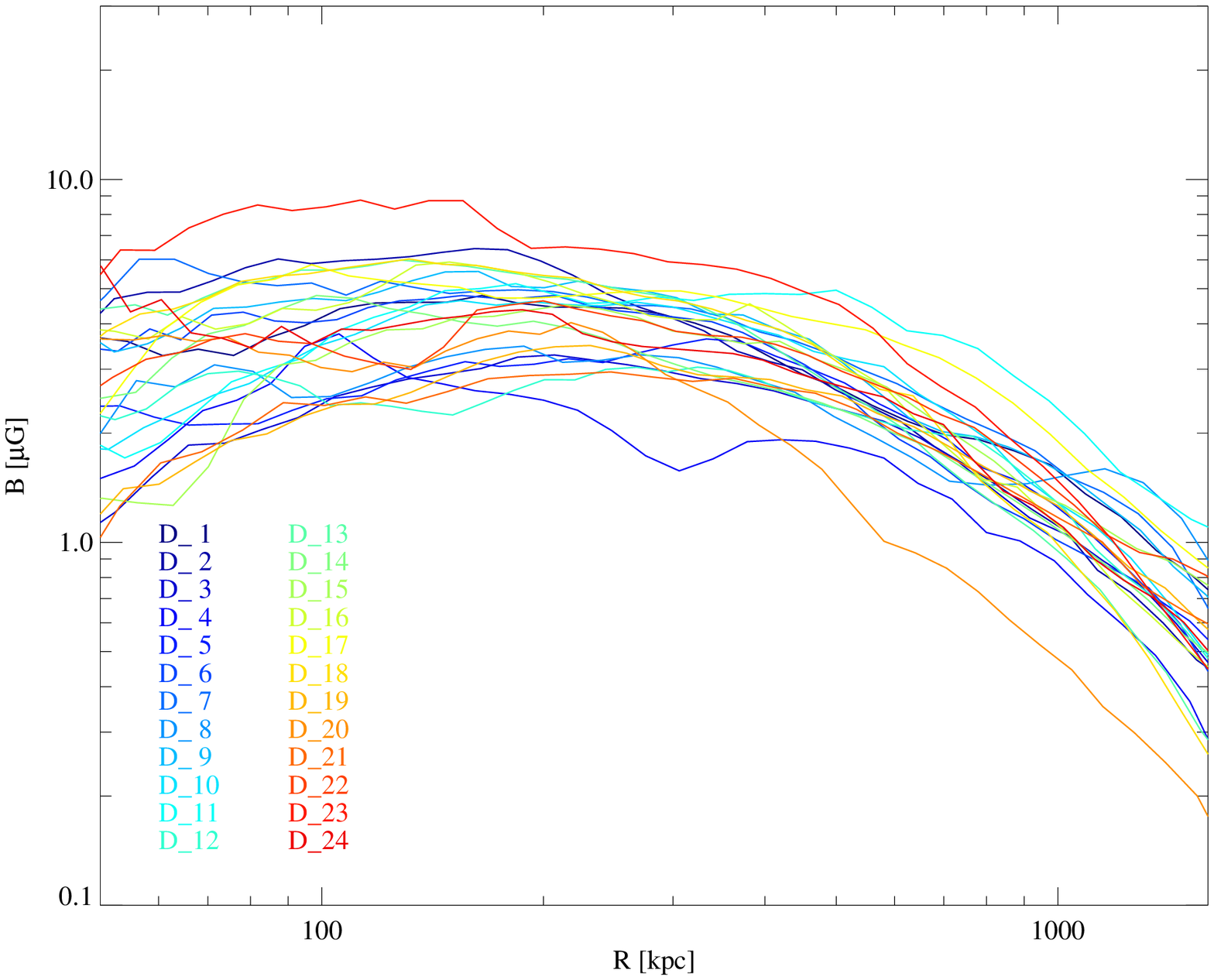}}
\subfigure{\includegraphics[width=\columnwidth]{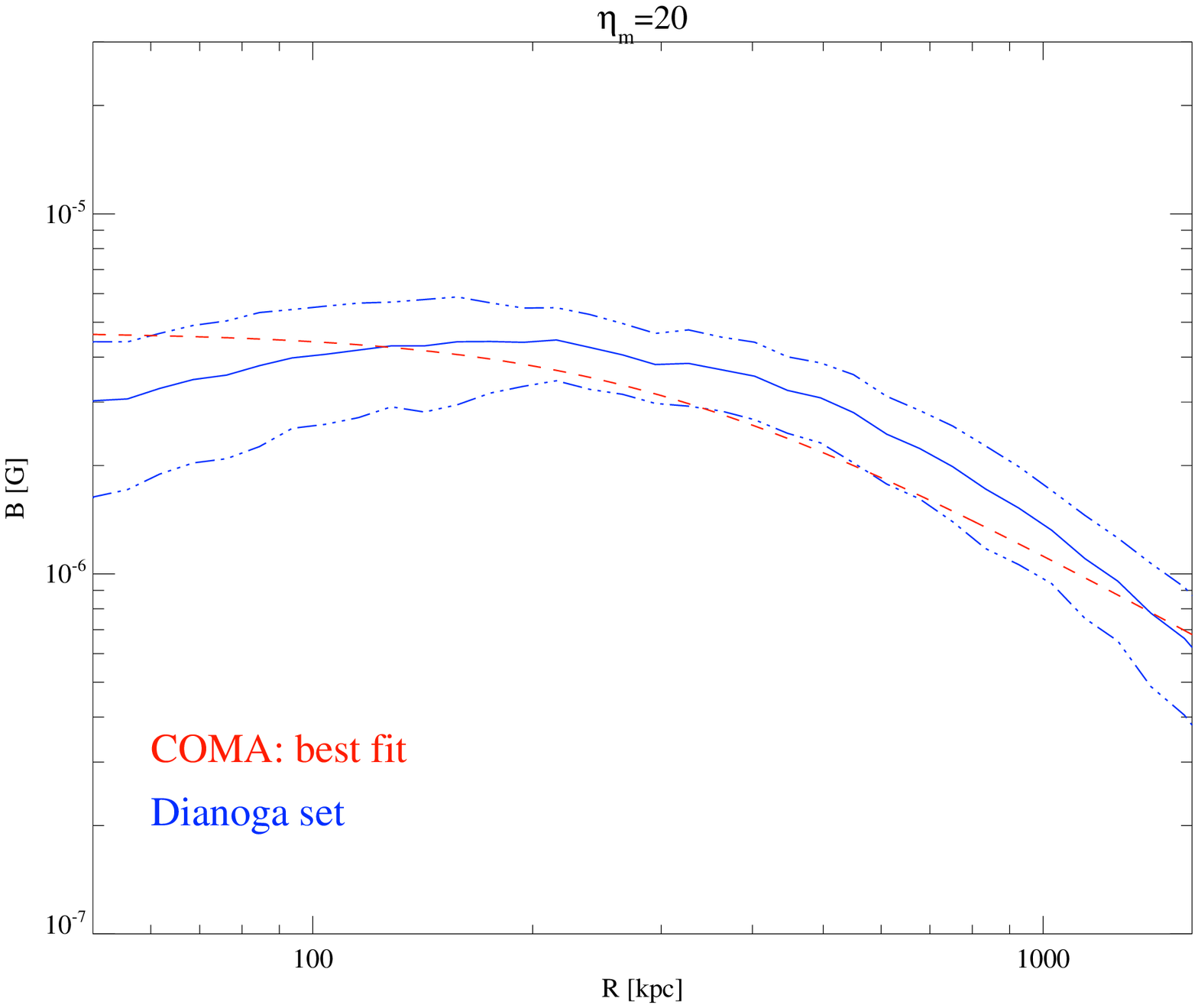}}
\end{center}
\caption{Left: magnetic field strength profile for the whole cluster
  set. Right: in blue mean (continuous line) and dispersion (dot-dashed)
  of the magnetic field strength profile. Red dashed line refers to
  the best fit obtained from RM observations for the Coma cluster
  \citep{2010A&A...513A..30B}.}
\label{fig:cluster_all}
\end{figure*}

\begin{table}
\centering
\begin{tabular}{|c c c c c| } 
\hline
Cluster name & $B_0$ & $r_c$ & $\mu$ & $\chi^2$\\ 
             & $\mu$G & kpc &        &       \\
D\_1   &    4.7  &  339  &    0.40 &  0.9\\
D\_2  &    6.8  &  295  &    0.52 &  0.7\\ 
D\_3   &    2.5  &  361  &    0.33 &  1.8\\
D\_4 &    2.7  &  285  &    0.43 &  1.2\\ 
D\_5   &    3.0  &  346  &    0.35 &  1.7\\
D\_6   &    5.1  &  414  &    0.57 &  0.9\\
D\_7   &    6.5  &  362  &    0.58 &  1.1\\
D\_8   &    3.5  &  342  &    0.34 &  1.0\\
D\_9  &    5.4  &  404  &    0.49 &  0.9\\
D\_10  &    3.9  &  319  &    0.38 &  1.9\\
D\_11 &    3.9  &  293  &    0.31 &  2.0\\ 
D\_12   &    3.3  &  415  &    0.51 &  1.4\\
D\_13   &    6.5  &  332  &    0.59 &  1.0\\
D\_14   &    4.6  &  329  &    0.44 &  0.7\\
D\_15   &    2.9  &  344  &    0.30 &  1.9\\
D\_16   &    5.6  &  354  &    0.54 &  0.9\\
D\_17   &    5.5  &  431  &    0.43 &  1.1\\
D\_18   &    6.4  &  327  &    0.63 &  1.1\\
D\_19  &    2.6  &  341  &    0.29 &  1.7\\
D\_20  &    4.3  &  323  &    0.63 &  0.7\\ 
D\_21   &    2.8  &  311  &    0.43 &  1.9\\
D\_22   &    4.6  &  423  &    0.46 &  0.9\\
D\_23  &    9.9  &  241  &    0.53 &  1.2\\ 
D\_24   &    4.9  &  375  &    0.55 &  1.0\\
&&&&\\
&&&&\\
Mean values &   4.7$\pm$1.7 & 346$\pm$47 & 0.46$\pm$0.11& --\\
Coma Cluster &  4.7$^{+0.7}_{-0.8}$ & 291 $\pm$17 & 0.38$^{+0.17}_{-0.09}$&--\\
&&&&\\
&&&&\\
\hline
\end{tabular}
\caption{Results of the fit of the magnetic field profiles to
  Eq. \ref{Eq:betamodel}}
\label{tab:betamodel}
\end{table}

\subsection{Magnetic properties of massive clusters}
\label{sec:Dianogaeta20}
We finally simulated the whole cluster sample, fixing the magnetic
resistivity to our inferred value of $\eta_m= 6 \times 10^{27}
\mathrm{cm}^2 \mathrm{s}^{-1}$. Therefore, we can study, for the first
time, the scatter of the magnetic field properties in massive clusters
using a volume limited sample. Figure \ref{fig:Bmean_M} shows the mean
magnetic field within $0.3 \times R_\mathrm{vir}$, corresponding to
roughly 1 Mpc for our set of massive clusters. The simulations scatter
mildly around the value inferred from observations of Coma. Our mean
value is 2.6$\mu$G with and rms of 0.6$\mu$G. We point out here again
that the sample of massive clusters comprises objects that have very
different dynamical state at $z=0$. It is interesting to note that the
mean magnetic field, averaged over the central Mpc$^3$, does not
depend on the present dynamical state of the cluster at $z=0$ (see
also Section \ref{sec:disc}.)
\subsubsection{Magnetic field profiles of massive galaxy clusters}
In Figure \ref{fig:cluster_all} the magnetic field profiles are shown
for all the clusters in the sample. In the right panel of that Figure
the mean and the dispersion of the magnetic field profiles are
compared with the best fit for the Coma cluster. It is worth stressing
that the exact shape of the profile inferred from Coma observations is
given {\it ad hoc} as a parametric model to fit the data. Hence, it is
not clear how significant the differences between simulations and
observations in the exact shape are. Nonetheless, the fit to the
observations lies completely within the scatter of the profiles
predicted by our simulations. This is a non-trivial result, confirming
previous findings that the magnetic field in galaxy clusters is shaped
by the (turbulent) motions within the ICM and therefore reflects a
natural prediction of the structure formation process.\\ Although the
mean magnetic field profile shows a good agreement with the one
inferred from Coma observations, there are differences in the shape of
the individual profiles, likely reflecting the dynamical state and the
different morphologies of the individual objects. Magnetic field
profiles are usually compared to the gas density profiles, in order to
derive a scaling with the radial distance from the cluster
center. Here we adopt another approach and fit the magnetic field
profiles directly to a ``$\beta$ model-like'' profile
\citep{1976A&A....49..137C}, that for magnetic fields is usually
written as:
\begin{equation}
B(r)=B_0 \left(1+ \frac{r^2}{r_c'^2}\right)^{-\frac{3}{2}\mu}
\label{Eq:betamodel}
\end{equation}
where $B_0$, $r_c$, and $\mu$ are free parameters. The fits have been
performed in the range 50-2000 kpc, to properly compare with the
results obtained from Coma observations. The results of the fit are
shown in Table \ref{tab:betamodel}. The mean values of $B_0$ and $\mu$
are reported is the last row of the Table and compared to those of the
Coma cluster. While previously numerical simulations indicated a
steeper profile of the magnetic field with the gas density (\iee
$\alpha \sim 1$ in Eq. \ref{eq:Brho}), and thus with the radial
distance from the cluster center, now the effect of the magnetic
resistivity is that of flattening the profiles, reaching a better
agreement with observations. In the case of the Coma cluster, a value
of $\alpha=0.5^{+0.2}_{-0.1}$ gives the best fit with the
observations, corresponding to $\mu \sim 0.38^{+0.17}_{-0.09}$, in
very good agreement with the mean of the best fit for our simulated
clusters, that is $\mu \sim 0.46 \pm 0.11$.  Hence, not only the mean
value of $B$ over the central Mpc$^3$ has a small dispersion in this
high-mass cluster set, but also the central value of $B_0$, and its
slope with the gas density, as derived from the beta-model fit, are
quite similar.



\begin{figure*}
\begin{center}

\includegraphics[width=0.3\textwidth]{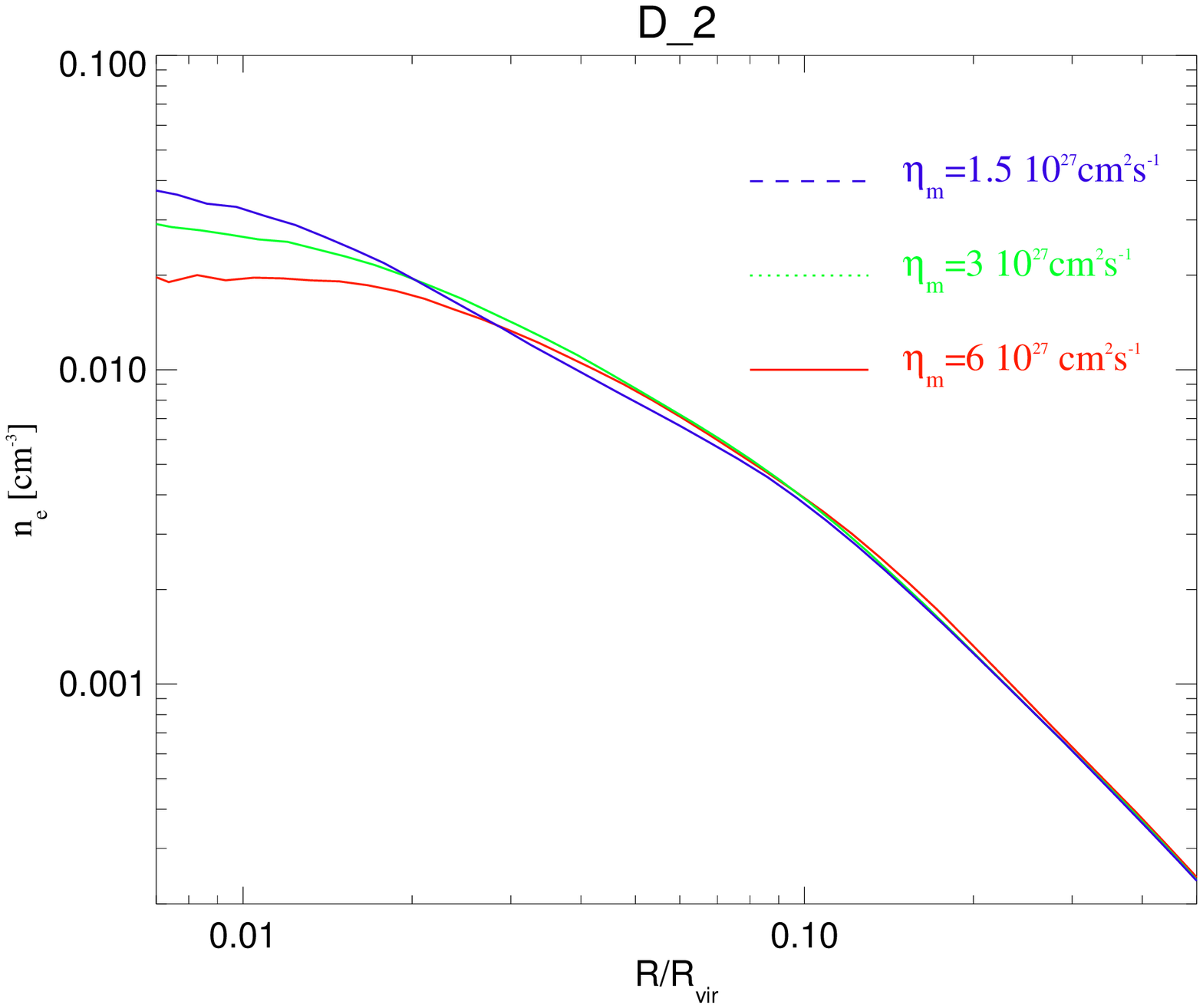}
\includegraphics[width=0.3\textwidth]{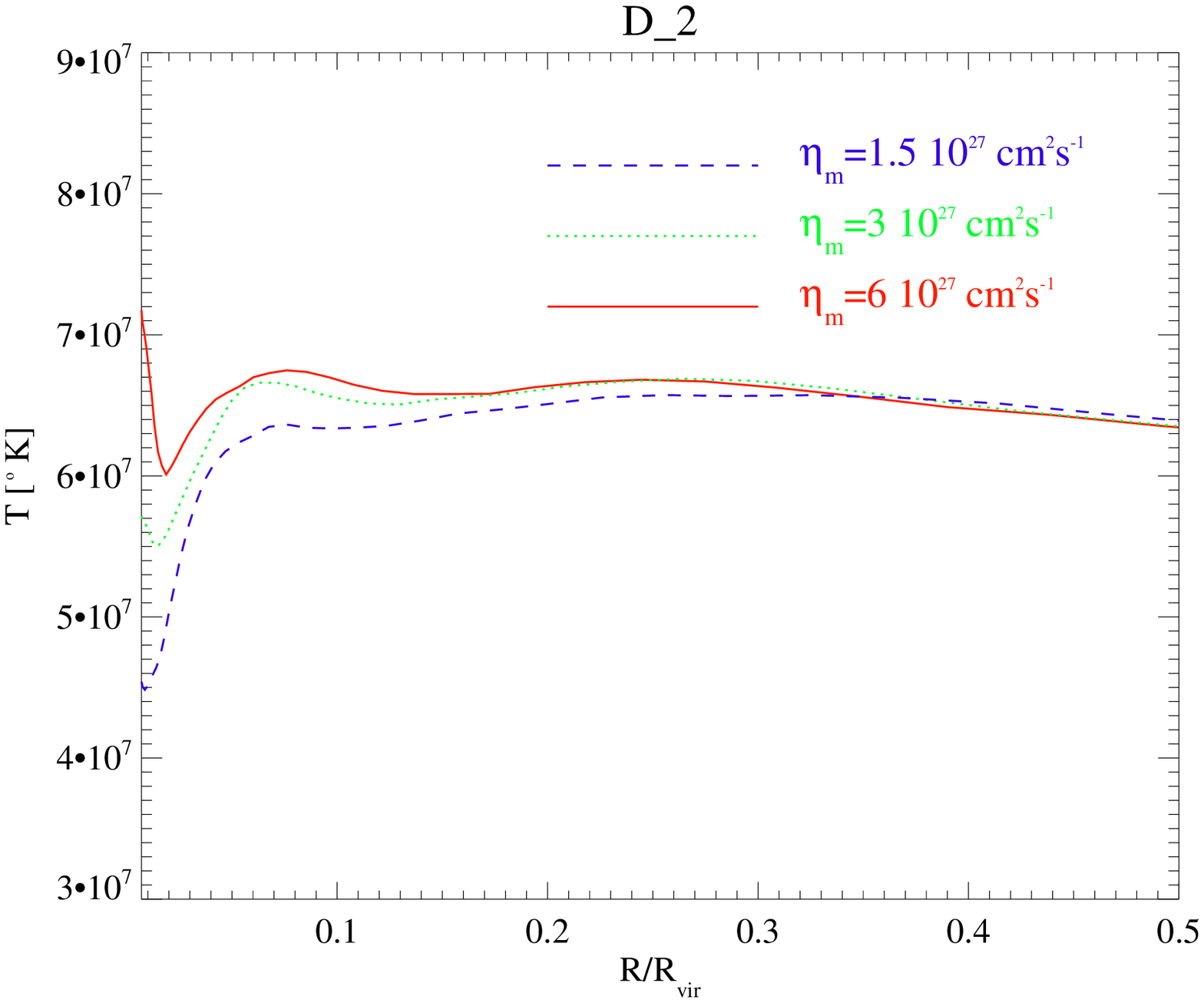}
\includegraphics[width=0.3\textwidth]{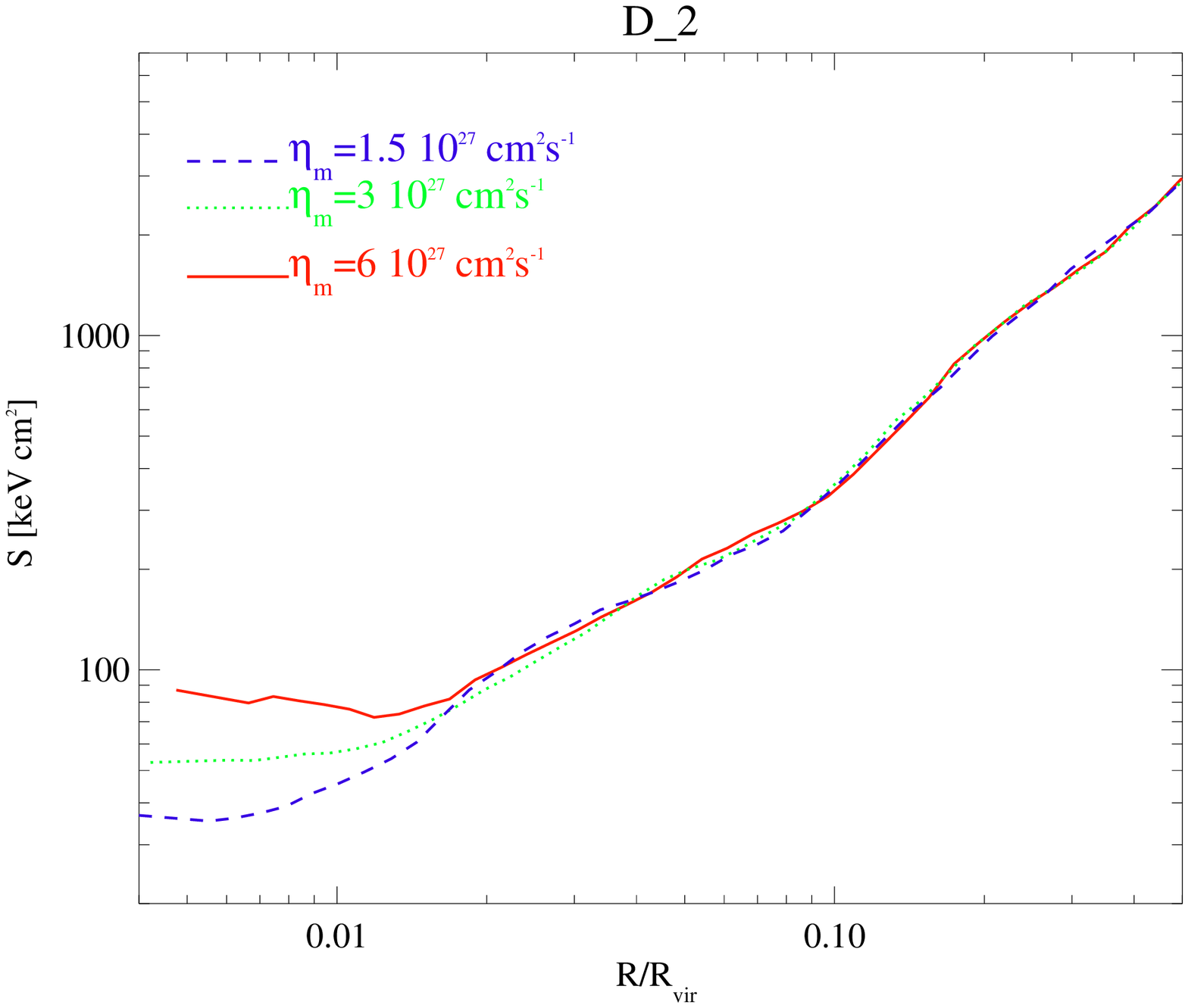}\\

\includegraphics[width=0.3\textwidth]{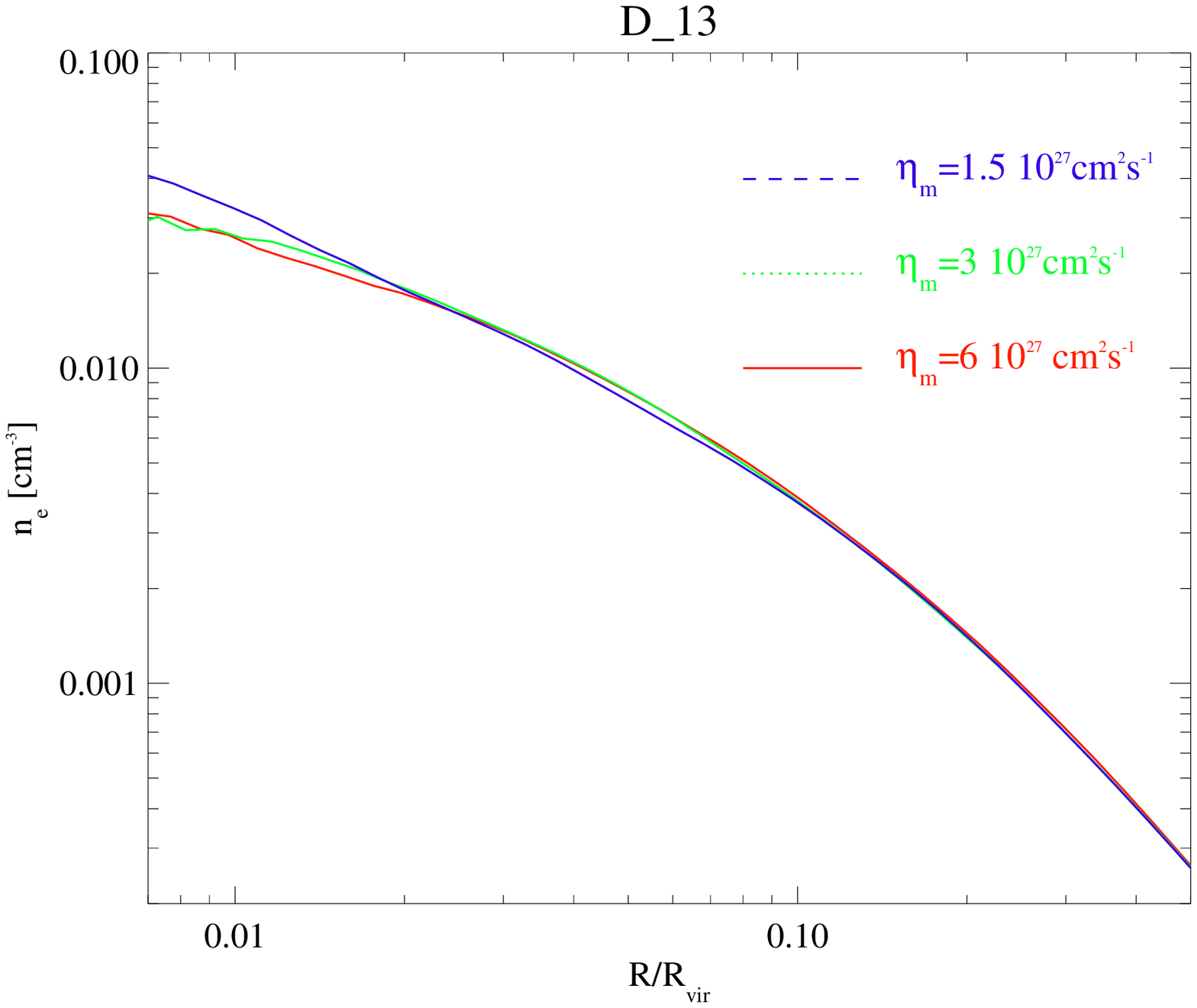}
\includegraphics[width=0.3\textwidth]{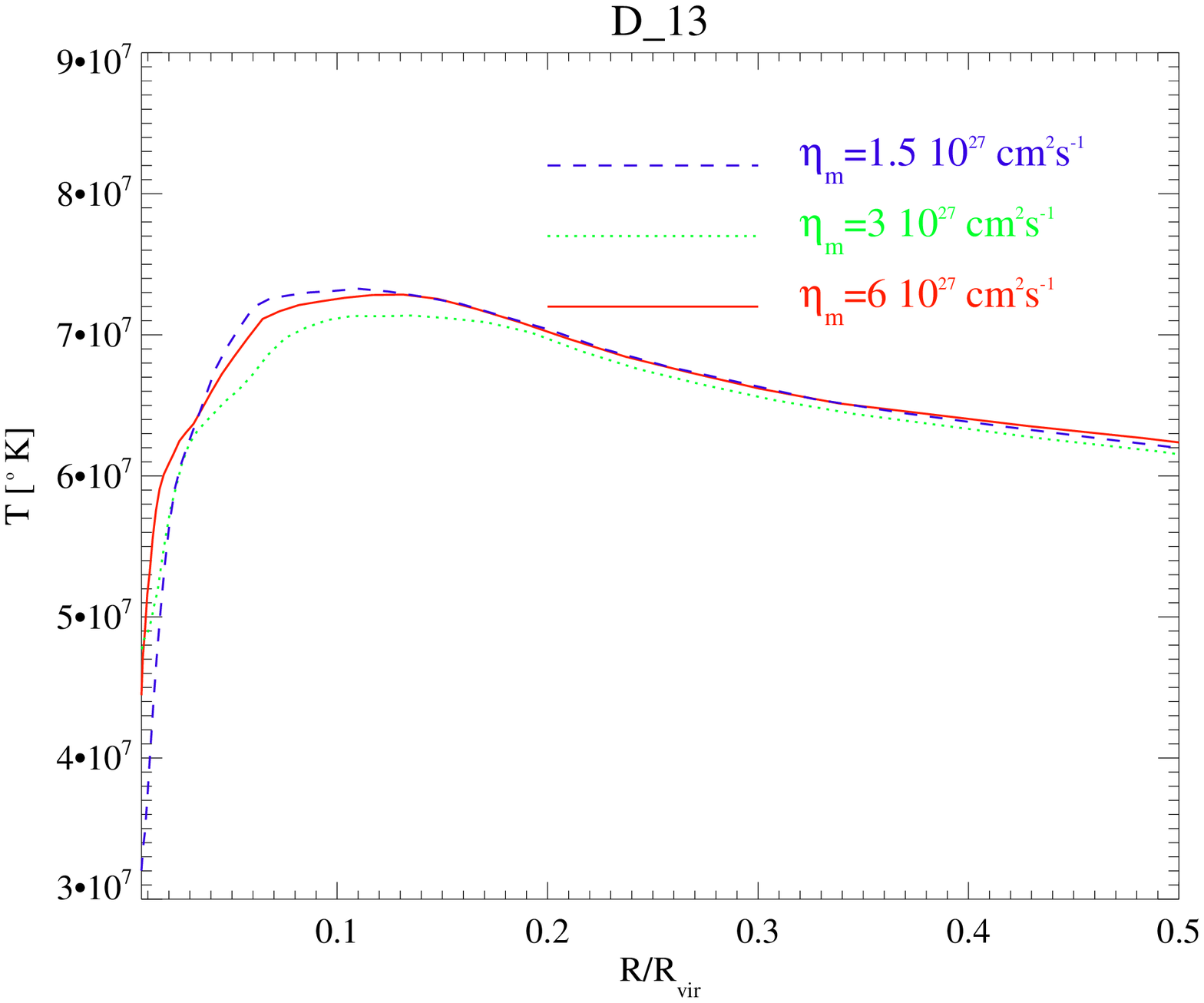}
\includegraphics[width=0.3\textwidth]{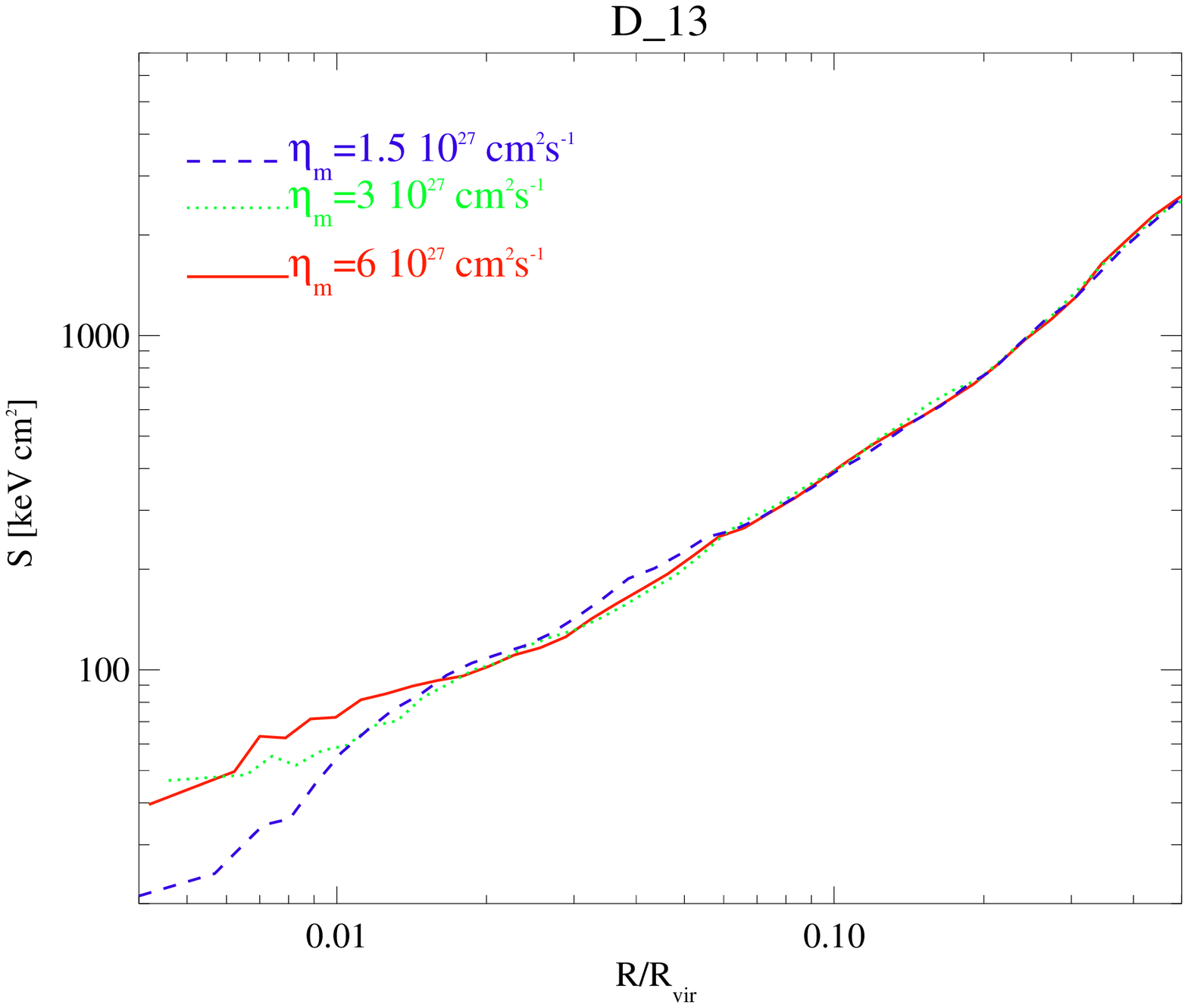}\\

\includegraphics[width=0.3\textwidth]{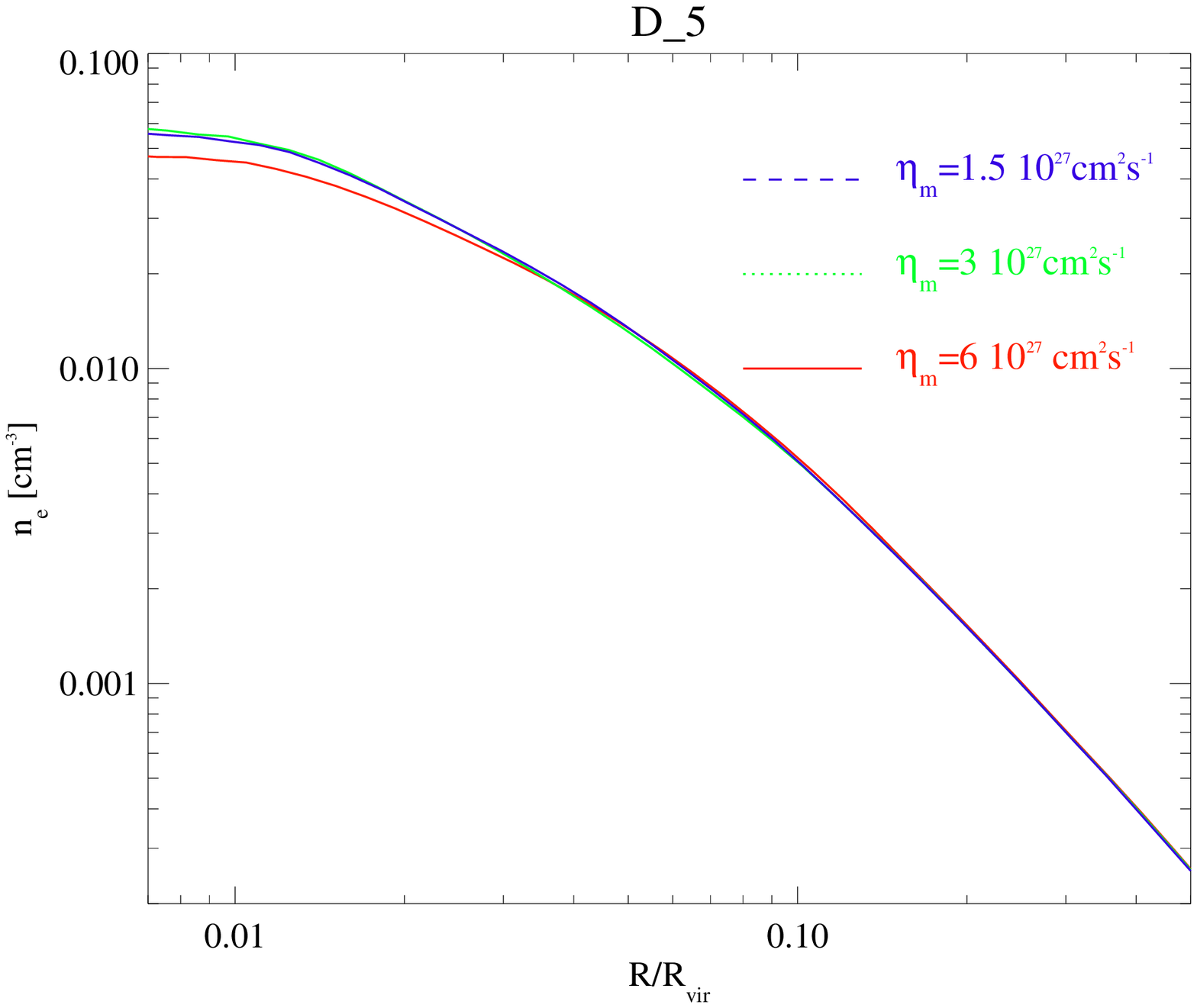}
\includegraphics[width=0.3\textwidth]{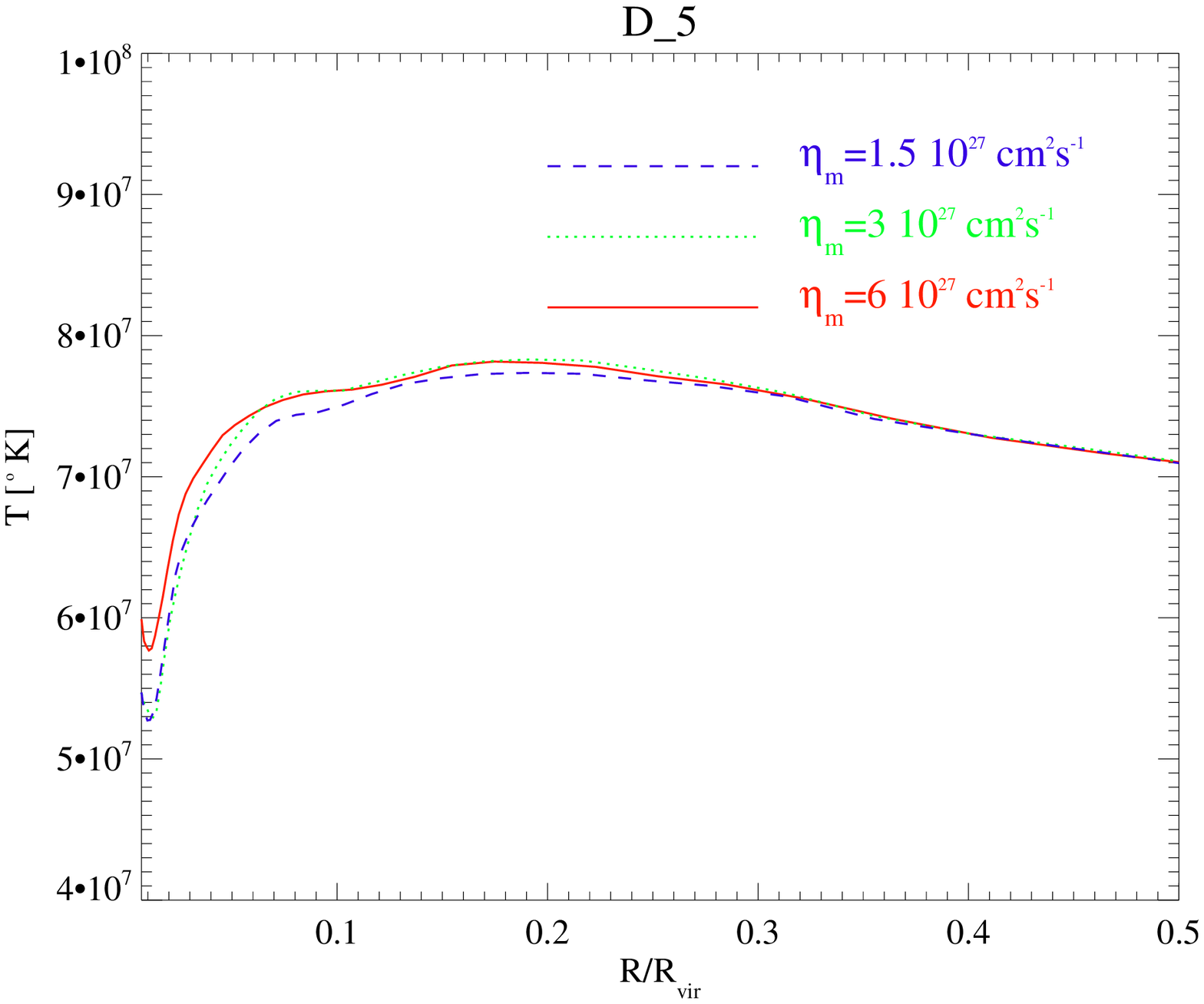}
\includegraphics[width=0.3\textwidth]{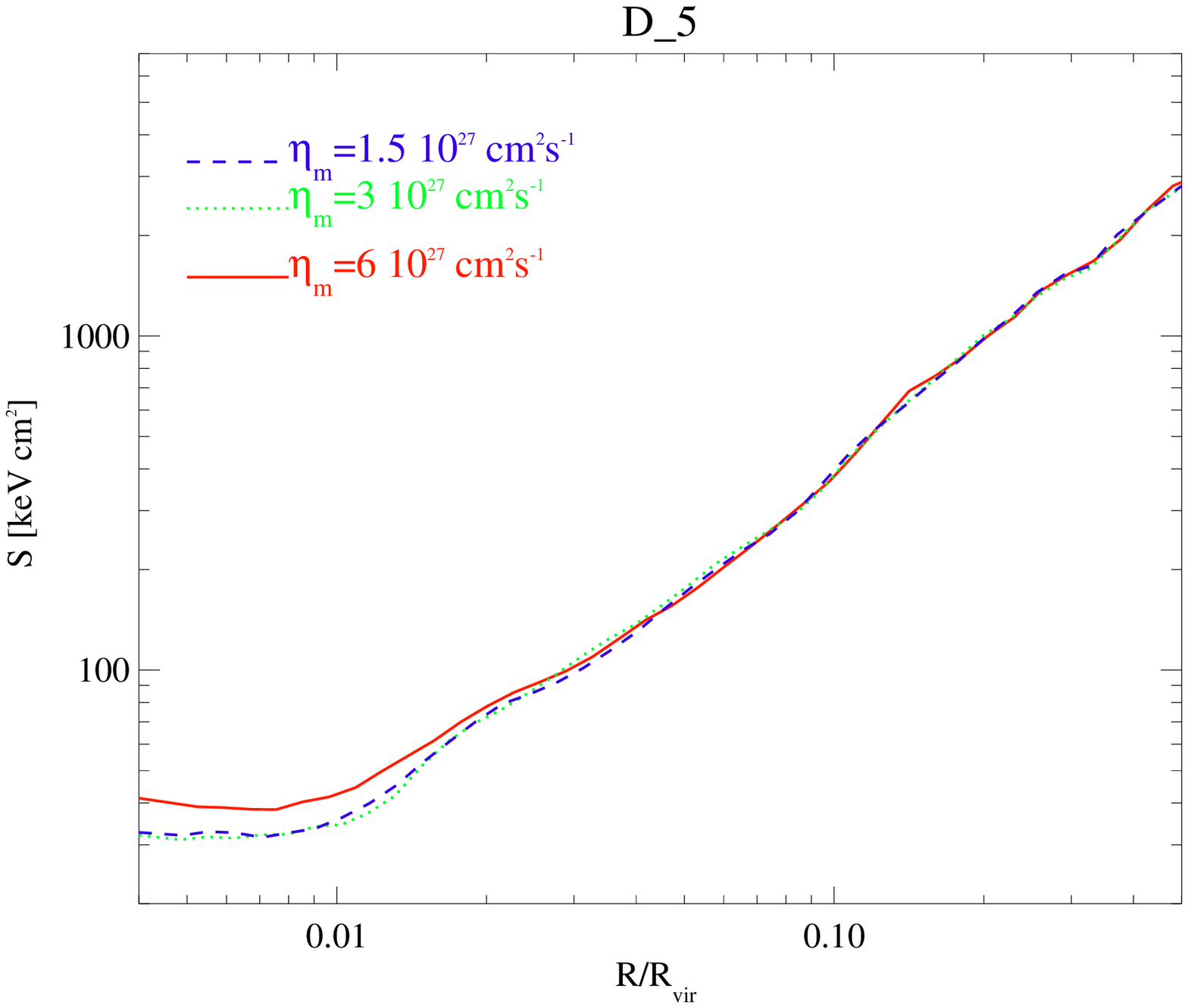}\\

\includegraphics[width=0.3\textwidth]{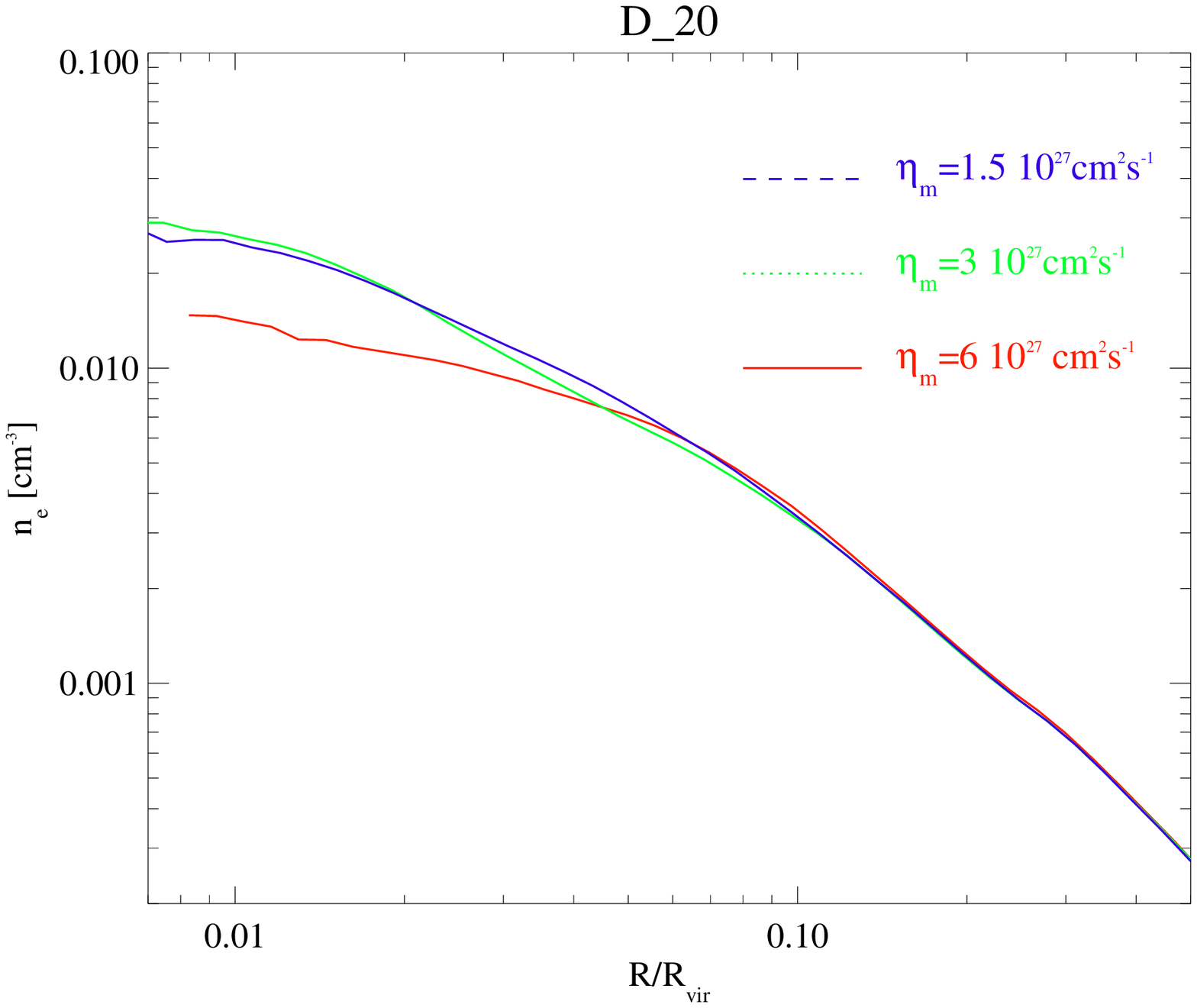}
\includegraphics[width=0.3\textwidth]{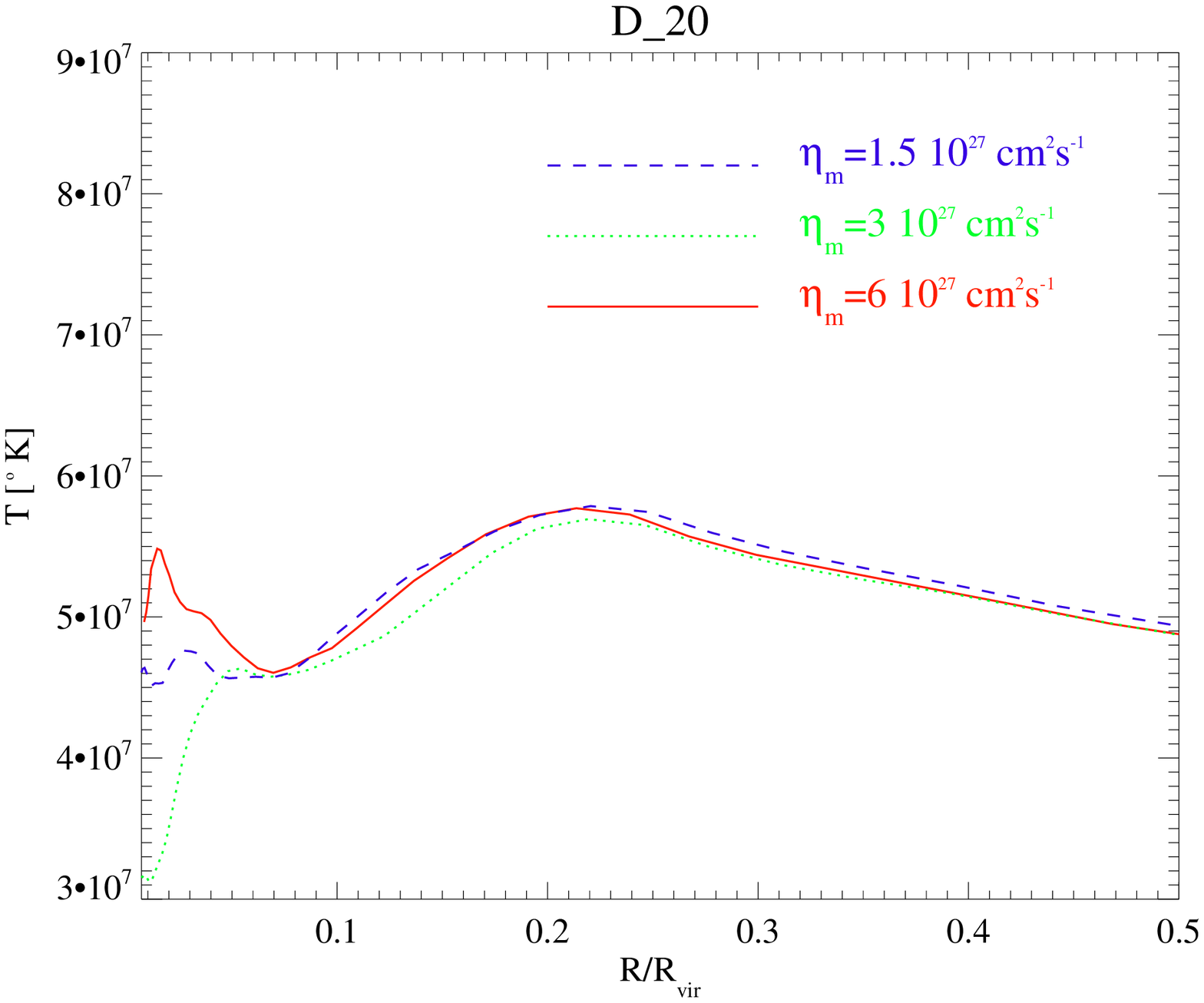}
\includegraphics[width=0.3\textwidth]{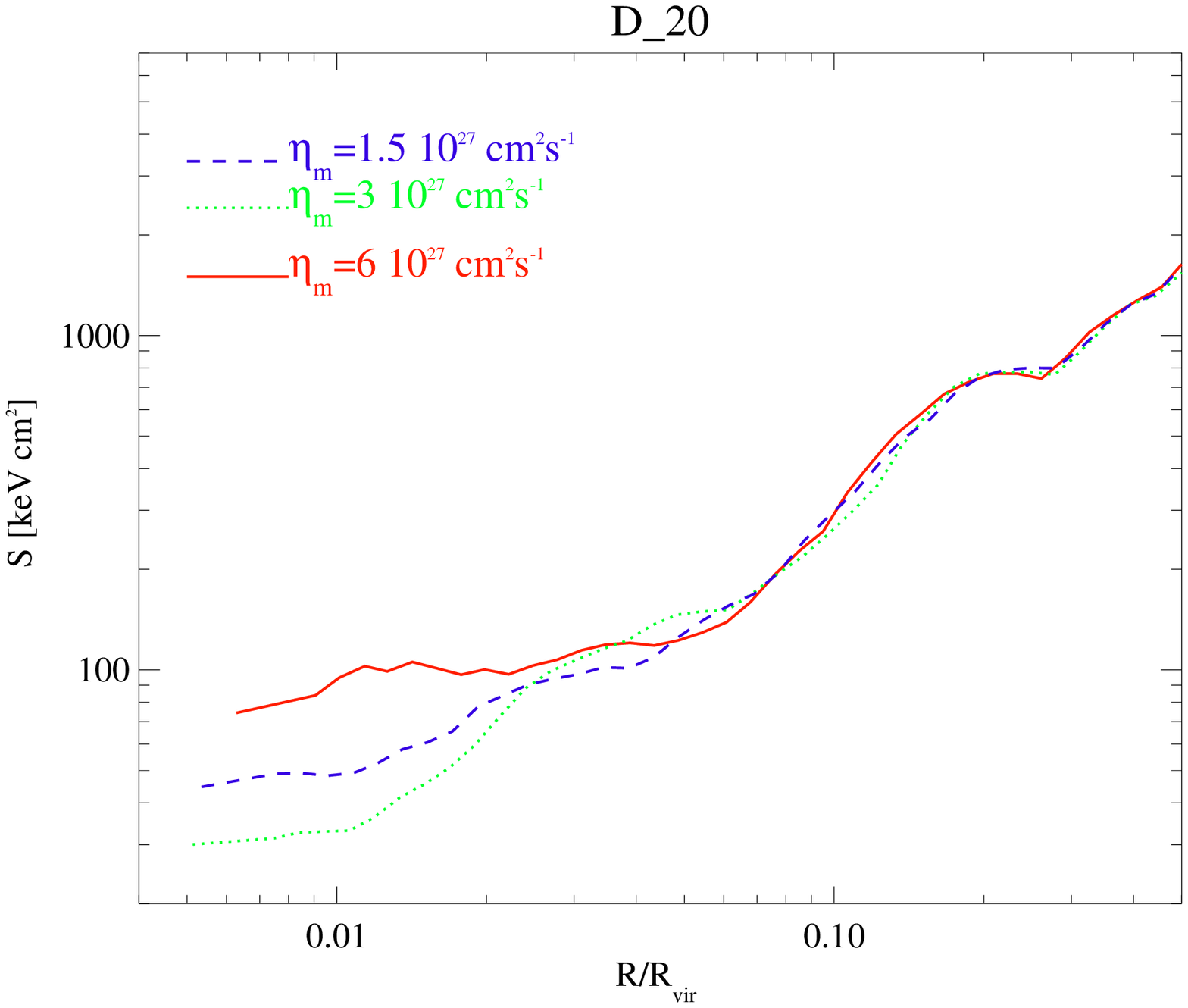}\\

\end{center}
\caption{ Density (left column), temperature (middle column) and entropy
  (right column) profiles for the clusters D\_2, D\_5, D\_13 and D\_20 from
  top to bottom respectively. Different colors refer to
  different values of the magnetic resistivity constant $\eta_m$, as
  indicated in the panels. }
\label{fig:entropy}
\end{figure*}

\begin{figure*}
\begin{center}
\subfigure{\includegraphics[width=\columnwidth]{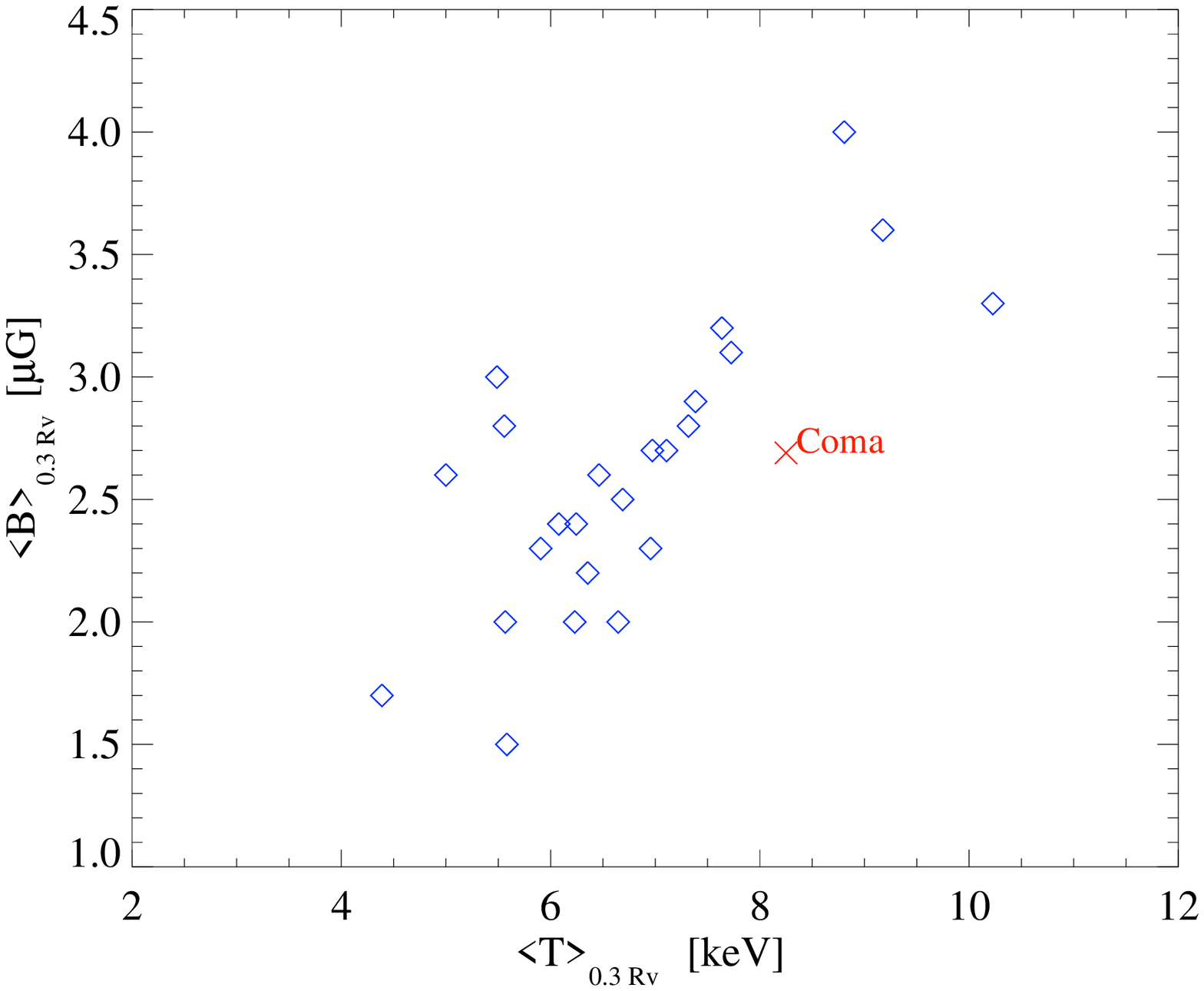}}
\subfigure{\includegraphics[width=\columnwidth]{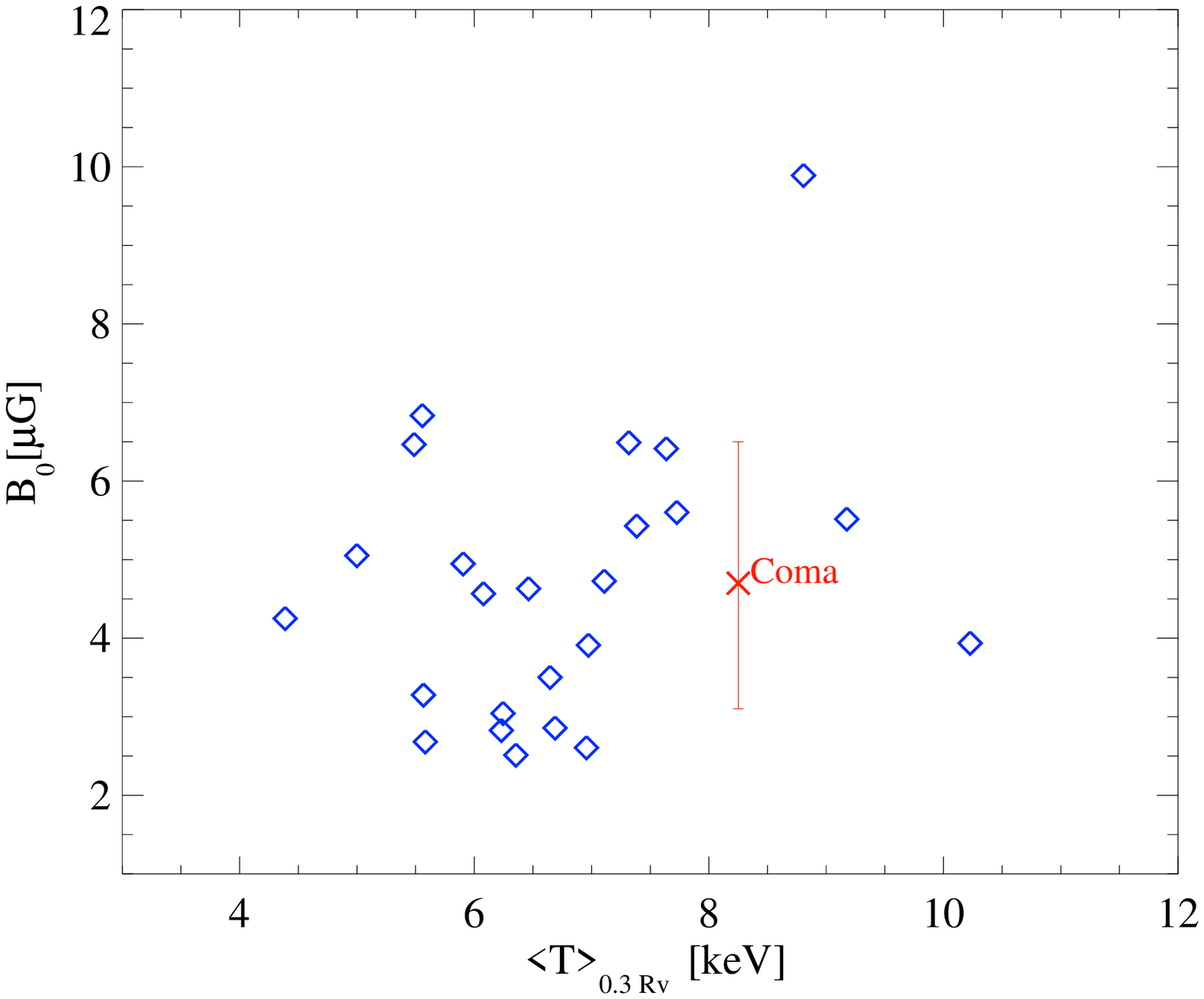}}
\end{center}
\caption{Left: Magnetic field averaged over the central 0.3 $R_\mathrm{vir}$
  versus the mean temperature of the cluster set (Blue diamonds). The
  red cross refers to the mean magnetic field for the Coma
  cluster. The error-bar refers to the 3$\sigma$ of the $chi^2$ given
  by \citet{2010A&A...513A..30B}. Right: Magnetic field in the cluster
  center, as results from the fit of a $\beta$-model profile versus
  the cluster mean temperature inside 0.3 $R_\mathrm{vir}$. The temperature for
  the Coma cluster is the one given by \citet{2001A&A...365L..67A},
  computed inside 0.25 of the Coma virial radius.  All the quantities
  are computed from the dissipative simulation runs with
  $\eta_{m}=6\times 10^{27} \mathrm{cm}^2 s^{-1}$ .}
\label{fig:BT}
\end{figure*}

\subsection{Magnetic field and thermal properties}
We present in this Section a first overview
of the thermal properties of the ICM in the presence of a diffusive
magnetic field. A more detailed analysis will be performed in a second
paper, where the whole sample will be analyzed. Here we
investigate if and how the presence of resistive magnetic field may
affect the ICM properties of the four clusters we have simulated with
different values of $\eta_m$. In Figure \ref{fig:entropy} the density,
temperature, and entropy profiles of these clusters are shown for
different values of $\eta_m$. The profiles converge at distances
larger than few \% of the virial radius, while differ in the very inner
region of the clusters. As mentioned in Section \ref{sec:totE} the
magnetic field energy that is dissipated during the cluster formation
is transformed into heat. Although the dynamical effect of a magnetic
field of the order of $\sim$1-10 $\mu$G in the cluster cores is
negligible, the overall effect of the magnetic force and pressure
integrated over a Hubble time results in a change of the density and
temperature profile. As the resistivity constant $\eta_m$ increases,
the amount of magnetic field energy, that is dissipated and hence
converted into heat, increases accordingly. An additional source of
heating is then present in the cluster central region, that has the
effect of flattening the temperature profile. The higher pressure that
would result from a higher temperature is then balanced by reducing
the gas density in the cluster central region, up to a factor 2. The
temperature and density profiles do not change adiabatically, as
demonstrated by the entropy profiles of the clusters. The entropy,
computed as $S=kT/n^{2/3}$, flattens in the inner region of the
clusters, indicating that the transport of low entropy gas in
inhibited.\\ In Figure \ref{fig:BT}, the magnetic field within $0.3
\times R_\mathrm{vir}$, is plotted versus the cluster mean temperature
computed over the same region. 
Although the sample is small, and the value of the magnetic fields
within $0.3 \times R_\mathrm{vir}$ varies of a factor $\sim$ 2, a
trend is suggested. Magnetic field in higher temperature clusters seem
to be higher. The correlation should be better investigated with a
higher sample of simulated galaxy clusters, since observational data
do not suggest a trend of the $RM$ in clusters depending on the
temperature \citep{2010A&A...522A.105G}. We note also that such a
trend is much less visible when the value of $B_0$, resulting from the
$\beta$-model fit is compared with the cluster mean temperature
(Figure \ref{fig:BT}, right panel).

\section{Discussion and conclusions}
\label{sec:disc}
We have presented a set of simulated galaxy clusters. It consists of
24 massive objects ($M_\mathrm{vir}> 10^{15} h^{-1}$\msun)
re-simulated at high resolution up to 5-6 virial radii, plus 50 more
clusters with $M_\mathrm{vir}> 10^{14} h^{-1}$\msun  that fell within
this high resolution region.  This large set permits to study the
cluster properties in a wide range of masses and at high resolution
\citep[see \eg][]{2011arXiv1102.2903F}. The evolution of the clusters
has been followed using the MHD implementation within the {\tt
  GADGET-3} code \citep{2009MNRAS.398.1678D}, that has been here
modified in order to include the magnetic resistivity term in the
induction equations. It is the first time that this term is analyzed
in the context of cluster formation and evolution.  In this first
paper we have presented the zoomed initial conditions, the non-ideal
MHD implementation, and the average properties of the more massive
clusters when a magnetic resistivity term is included in the MHD
equations.  Further analysis will be performed in a future paper
(Paper II, Bonafede et al. in prep) where the physical implications
will be discussed in more detail.\\ Our main results can be summarized
as follows:
\begin{itemize}
\item{Non-ideal MHD equations have been implemented within the {\tt
    GADGET} code. The tests performed on two different problems show
  that the numerical implementation is accurate and can be used to
  study the effect of the magnetic resitivity.}
\item{The magnetic field profiles obtained with non-ideal MHD can
  reproduce the profile inferred from Faraday Rotation Measures of the
  Coma cluster. Four clusters having X-ray morphologies similar to the
  one of the Coma cluster have been selected to test the effect of
  changing the constant $\eta_m$ used in the induction equation. The
  best agreement with the limits given by observations is achieved
  with $\eta_m= 6\times 10^{27} cm^2 s^{-1}$.}
\item{ The whole sample has been simulated using $\eta_m=6\times
  10^{27} cm^2 s^{-1}$, and the derived magnetic field profiles are
  consistent with the Coma profile. The best-fit found for the Coma profile
lies in fact between the rms of the simulated profiles.}
\item{We have fitted the magnetic field profile with a $\beta$-like
  model, finding that the magnetic field profile of the
  simulated galaxy clusters can be well reproduced by values $B_0= 4.7
  \pm 1.7 $, and $\mu = 0.46 \pm 0.11$ (see Eq. \ref{Eq:betamodel}),
  in good agreement with the value found for the Coma cluster. The
  value of $\mu$ would correspond to a value of $\alpha \sim 0.6$
  (Eq. \ref{eq:Brho}) for a Coma-like gas density profile. }
\item{We have investigated possible correlations of the magnetic field
  strength with the cluster mass. The magnetic field strength,
  averaged over a central spherical volume of 0.3$R_\mathrm{vir}
  h^{-1}$ in radius, is similar for all the clusters in the sample, in
  agreement with what has been recently found by
  \citet{2011A&A...530A..24B}. This indicates that the presence of
  radio halo emission, found in a fraction of massive galaxy
  clusters, cannot be attributed to a difference in the magnetic field
  strength. A mild dependence of the magnetic field strength with
  cluster temperature is indicated by these simulations. }
\item{The density, temperature and entropy profiles of the simulated
  galaxy clusters have been derived for different values of
  $\eta_m$. We find that the effect of a magnetic diffusive constant
  is visible in such profiles, leading to flatter temperature and
  entropy profiles in the inner region of the cluster ($R \leq 0.1
  R_\mathrm{vir}$ at maximum). }
\end{itemize}
The cluster sample and the new MHD-implementation we have presented is
suitable to investigate other issues that are not discussed here, and
that will be studied in a future paper, such as the interplay of the
magnetic field with the thermal gas of the ICM (\eg how is the thermal
conduction modified, the role of the magnetic pressure in suppressing
the cooling in the inner regions). In the next years, the LOw
Frequency ARray (LOFAR) and the Expanded Very Large Array (EVLA) will
allow us to improve our knowledge of the non-thermal component of the
ICM, and a larger sample of data will be soon available for a more
complete comparison.

\section*{acknowledgments}
We thank R. Brunino and C. Gheller for their support at CINECA
Supercomputing Center, G. Tormen for providing the ZIC code, and
M. Br\"uggen and F. Vazza for the useful discussions and comments. We
thank the referee for his/her comments that helped is clarifying the
paper in several parts. A.B. thanks the MPA in Garching for the
hospitality and the support by the Marco Polo exchange program of the
Bologna University. We acknowledge the use of computational resources
under the CINECA-INAF 2008-2010 agreement. AB and FS acknowledge
support by the DFG Research Unit 1254 ``Magnetization of interstellar
and intergalactic media: the prospect of low frequency radio
observations''.  K.D. acknowledges the support by the DFG Priority
Program 1177 and additional support by the DFG Cluster of Excellence
``Origin and Structure of the Universe''.

\bibliographystyle{mn2e}
\bibliography{master}

\begin{appendix}

\section{Generating the zoomed initial conditions}
\subsection{Dark matter re-simulations}
\begin{figure*}
\begin{center}
\includegraphics[width=0.99\textwidth]{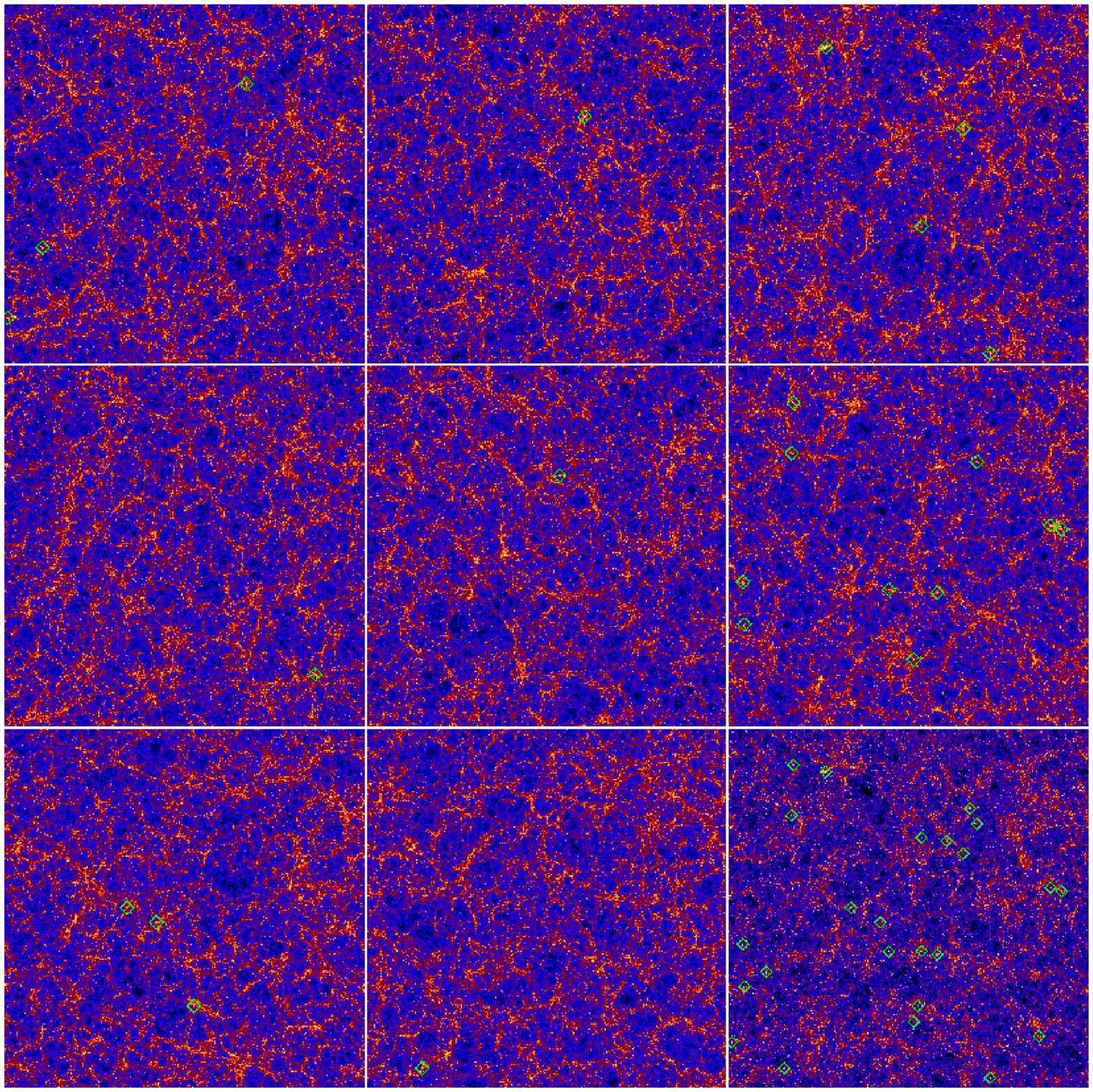}
\end{center}
\caption{125 Mpc$h^{-1}$ thick slices through the parent DM simulation at $z=0$ 
  showing the projected density. The Diamonds indicate the positions of the 
  24 most massive clusters extracted for high resolution re-simulations. 
  The bottom right panel shows the projected density through the whole box.}
\label{fig:app_slice}
\end{figure*}

Creating zoomed initial conditions is essential to extend the
dynamical range accessible through cosmological simulations, which is
needed to study the detailed structure of objects, like \eg galaxy
clusters, with appropriate resolution. Since hydrodynamic simulations
are sensitive to boundary conditions, regions around galaxy clusters
have to be re-simulated with high resolution as well. In the last
years the peripheral regions around galaxy clusters are also
attracting more and more interest, given the increased sensitivity of
modern instruments. Here we have optimized our initial conditions to
study a statistical sample of massive clusters with reasonable
computational resources. Our procedure is based on the {\it ZIC} code
(\citealt{1997MNRAS.286..865T}) and we describe here the iterative
procedure that we have used to obtain such highly optimized, zoomed
initial conditions for our cluster sample.\\ We started from a large,
cosmological, dark-matter only simulations, performed according to the
`concordance' $\Lambda$CDM cosmological model ($\Omega_{\Lambda}=$
0.76, $\Omega_{0}=$0.24, $h =$0.72 and $\sigma_8=$ 0.8). The spectral
index of power spectrum for the primordial density fluctuations
($P(k)\propto k^{n}$) is $n=0.96$. This simulation, that we refer to
as `parent simulation', was carried out with the massively parallel
TREE+SPH code {\tt GADGET-2} \citep{2005MNRAS.364.1105S} and consists
of a periodic box of 1 $h^{-1}$ Gpc size. The cluster identification
was performed at $z=0$ using a standard {\it Friend of Friends}
algorithm. The linking length was fixed to 0.17 of the mean
inter-particle separation between DM particles, to reflect the virial
over-density for the adopted cosmology. Given the large volume this
cosmological box contains a large sample of 64 clusters with
$M_{FOF}>10^{15}$$h^{-1}$ \msun at $z=0$. We selected the 24 most
massive clusters for high resolution re-simulations. Figure
\ref{fig:app_slice} shows the projected density within 125 $h^{-1}$
Mpc slices of the parent simulation at z=0. The positions of the 24
most massive clusters used in this work are marked by diamonds. From
the final output of the DM only run, all of the particles out to a
distance of $\approx 5-7 R_\mathrm{vir}$ around the cluster center
were selected and then traced back to their initial positions. The
corresponding Lagrangian region was enclosed in a box of side
$L_\mathrm{HR} \sim 62.5$ Mpc, the high resolution (HR) region. Since
the volume occupied by the HR particles, $V_\mathrm{HR}$, is usually
only a fraction of the volume of the box ($L_\mathrm{HR}^3$), we
sampled the box with $64^3$ cells, and we marked cells which were
actually occupied by the particles. In order to obtain a volume with a
concave shape and no holes in it, some more cells were marked
around/within $V_\mathrm{HR}$. The particles that occupy the marked
cells were then traced back to the initial redshift of the
simulation. The right panel of Figure \ref{fig:app_gas} shows a cut
through the $L_\mathrm{HR}$ volume. The blue cells trace the
$V_\mathrm{HR}$ region, while the additional cells marked to obtain a
concave volume are marked in red and green. This volume (defined as
``occupied volume'') was re-sampled with a higher number of particles
in order to obtain a higher mass resolution (in this case of 1 $\times
10^{9}$$h^{-1}$\msun for DM particles). Particles were displayed
according to a glass-like particle distribution
\citep{1996clss.conf..349W}.  The HR particles were perturbed using
the same power spectrum of the parent simulation, keeping the same
amplitudes and phases. New fluctuations at smaller spatial scales were
added, since smaller frequencies are now sampled by the higher
resolution particles.  The amplitude of the fluctuations are given by
the theoretical power spectrum $P(k)$ of the parent simulation, but
extended to higher $k$.  To minimize any changes in the tidal forces
acting onto the high resolution region, we created a buffer region
around the HR region, and sampled it with the same mass resolution as
the parent cosmological simulation. The remaining volume of the
simulation was re-sampled at lower resolution according to the
following procedure: the density and velocity fields of the LR
particles were re-sampled onto a spherical grid having constant angular
resolution $d\theta$. The size of each cell $dr=r d\theta$ was chosen
to obtain approximately cubic cells through the sphere. The
interpolation onto a spherical grid reduces the number of LR particles
to the minimum necessary to preserve the large-scale tidal field of
the original simulation. We used $d\theta=1.5^{\circ}$, resulting in
$\sim 2 \times 10^6$ low resolution particles, that guarantees an
accurate sampling of the tidal field (see
\citealt{1997MNRAS.286..865T}). By construction, as the distance from
the HR region increases, $dr$ increases too, and the mass of the LR
particles increases accordingly. The overall volume simulated for each
cluster is the same as the parent simulation, ensuring that the
forming structures correspond to the same that formed within the
original cosmological simulation. The new initial conditions were
finally traced back to a higher redshift (\eg $z=70$) to ensure that
the {\it rms} of the particle displacement in the HR region is still
small enough to guarantee the validity of the Zeldovich
approximation. After generating the new IC at higher resolution, we
re-run further dark matter-only re-simulation to verify that the
volume of the HR region around each cluster was free
from contamination of LR particles. Several iterations
(typically 5-7) of the whole procedure were required for each cluster
to obtain a clean, high resolution spherical volume with radius of 5
$R_\mathrm{vir}$, while keeping the total number of high resolution
particles as low as possible. In several cases, additional clusters
close to our target had to be included in the high resolution region.
Hence, all the initial conditions have at least a spherical volume of
radius $5-6 R_\mathrm{vir}$ ``clean'' of low resolution particles, and
centered on the target cluster (see Table \ref{tab:Dianoga_set}). The
total number of high resolution particles needed is typically only
$2-3$ times larger than the number of high resolution particles within
this regions of interest. Two of the selected clusters turned out to
have a close-by companion with a mass larger than
$10^{15}$$h^{-1}$\msun. Whereas the 24 targeted clusters represent a
fair volume-limited sample of galaxy clusters, the whole simulation
sample encompasses in total 26 clusters with masses above
$10^{15}$$h^{-1}$\msun. In addition, many other clusters with masses
between $10^{14}$$h^{-1}$\msun and $10^{15}$$h^{-1}$\msun were found
close to our massive targets. 50 of them are free from low-resolution
particles up to at least 1 $R_\mathrm{vir}$. We also extracted initial
conditions of 5 more isolated cluster, having masses of $\approx
5,7,4,1,1\times10^{14}$$h^{-1}$\msun. Such additional clusters are
of interest when studying scaling relations
\citep{2011arXiv1102.2903F}.
\begin{table*}
\caption{}
\label{tab:Dianoga_set}
\centering
\begin{tabular}{|c c  c c  | } 
\hline\hline 
Cluster &   $R$ cleaned  & $M_{DM}$ & N of nearby clusters\\ 
        &  [$R_\mathrm{vir}$]     &  with M$>10^{14}$]$h^{-1}$\msun \\
\hline
&&&\\
D\_1  & 5.2 &1.618       & 1 \\
D\_2   & 5.4 &1.518       & 3 \\
D\_3   & 5.3 &1.49        & 2 \\
D\_4  & 5.4 &1.482       &  \\  
D\_5  & 5.0 &1.537       & 4 \\ 
D\_6   & 5.0 &1.165       & 4 \\
D\_7   & 5.4 &1.776       & 1 \\
D\_8  & 5.3 &1.993,1.170 & 4 \\ 
D\_9   & 5.2 &1.657       &  \\ 
D\_10   & 5.1 &1.705       & 6\\
D\_11  & 5.3 &3.163       & 1 \\
D\_12  & 5.5 &1.678       & 2 \\
D\_13   & 5.6 &1.171       & 4\\
D\_14   & 6.0 &1.557       & 3\\
D\_15 & 5.5 &1.840       & 1 \\ 
D\_16  & 5.2 &1.385       &  \\ 
D\_17   & 5.5 &1.813       &  \\
D\_18   & 5.1 &1.356       & 1\\
D\_19   & 5.1 &1.316       & 1\\
D\_20  & 5.2 &1.067       & 2 \\
D\_21   & 5.9 &1.623,1.011 & 4\\
D\_22   & 5.2 &1.674       &  \\
D\_23  & 5.1 &1.880       & 3 \\
D\_24   & 5.0 &1.507       & 3\\

&&&\\
\hline
\hline

\multicolumn{4}{l}{\scriptsize Col. 1: Cluster name; Col. 2: Number of
  virial radii cleaned by LR particles; }\\ \multicolumn{4}{l}{
  \scriptsize Col. 3: Mass of the DM component inside the virial
  radius;}\\ \multicolumn{4}{l}{ \scriptsize Col 4: Number of nearby
  clusters within 5 $R_\mathrm{vir}$ with
  $M_{DM}>10^14$$h^{-1}$\msun.}
\end{tabular}
\end{table*}

\begin{figure*}
\begin{center}
\includegraphics[width=0.195\textwidth]{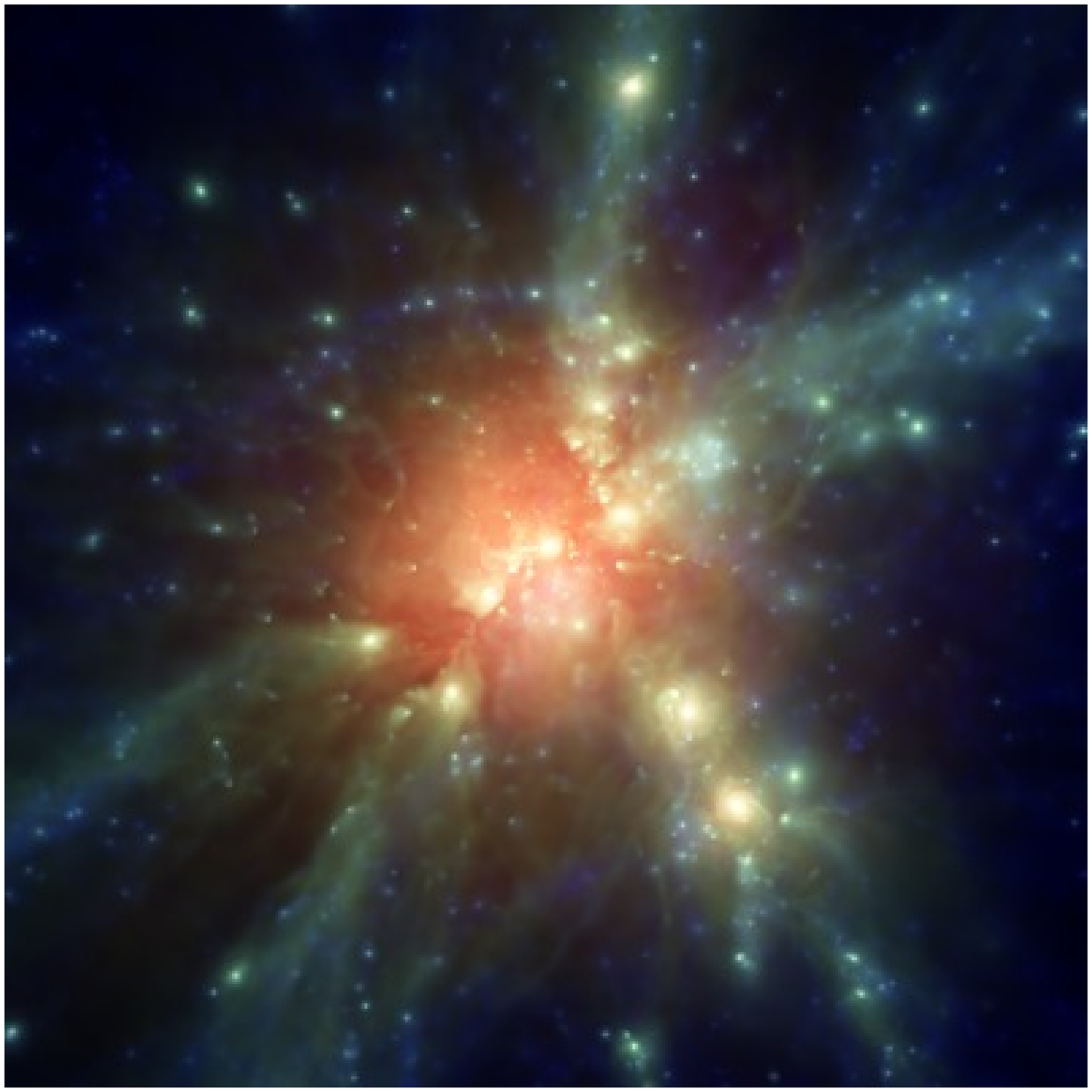}
\includegraphics[width=0.195\textwidth]{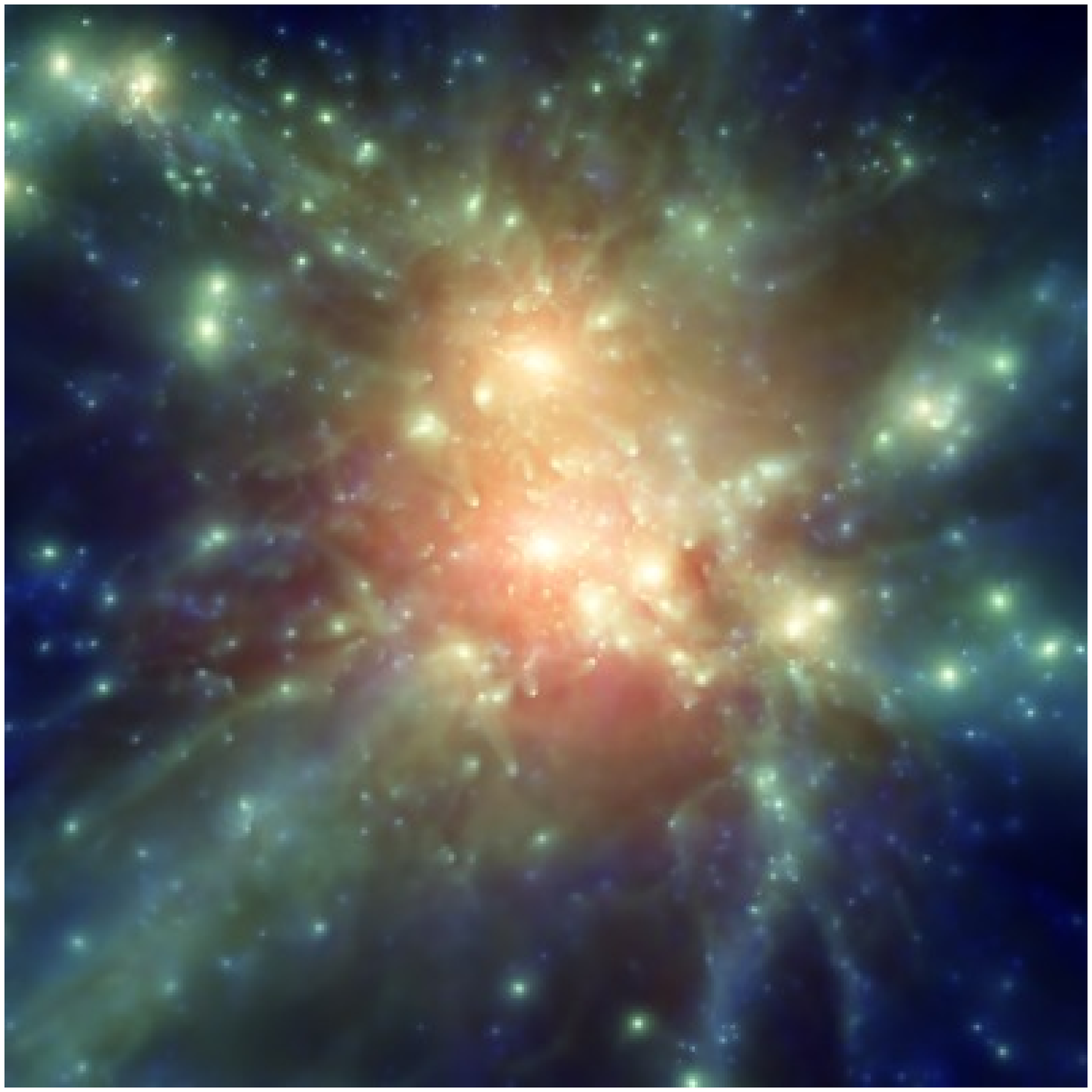}
\includegraphics[width=0.195\textwidth]{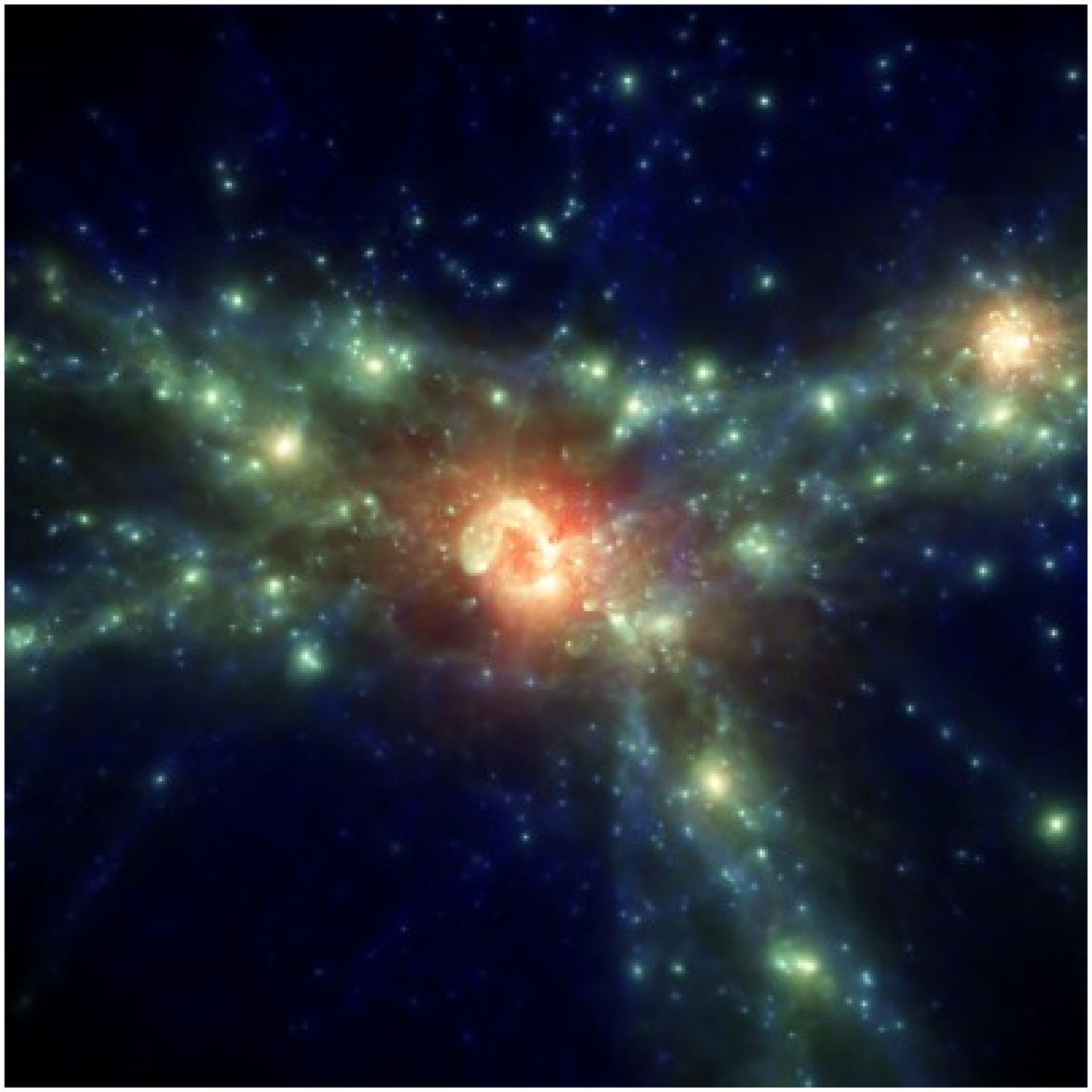}
\includegraphics[width=0.195\textwidth]{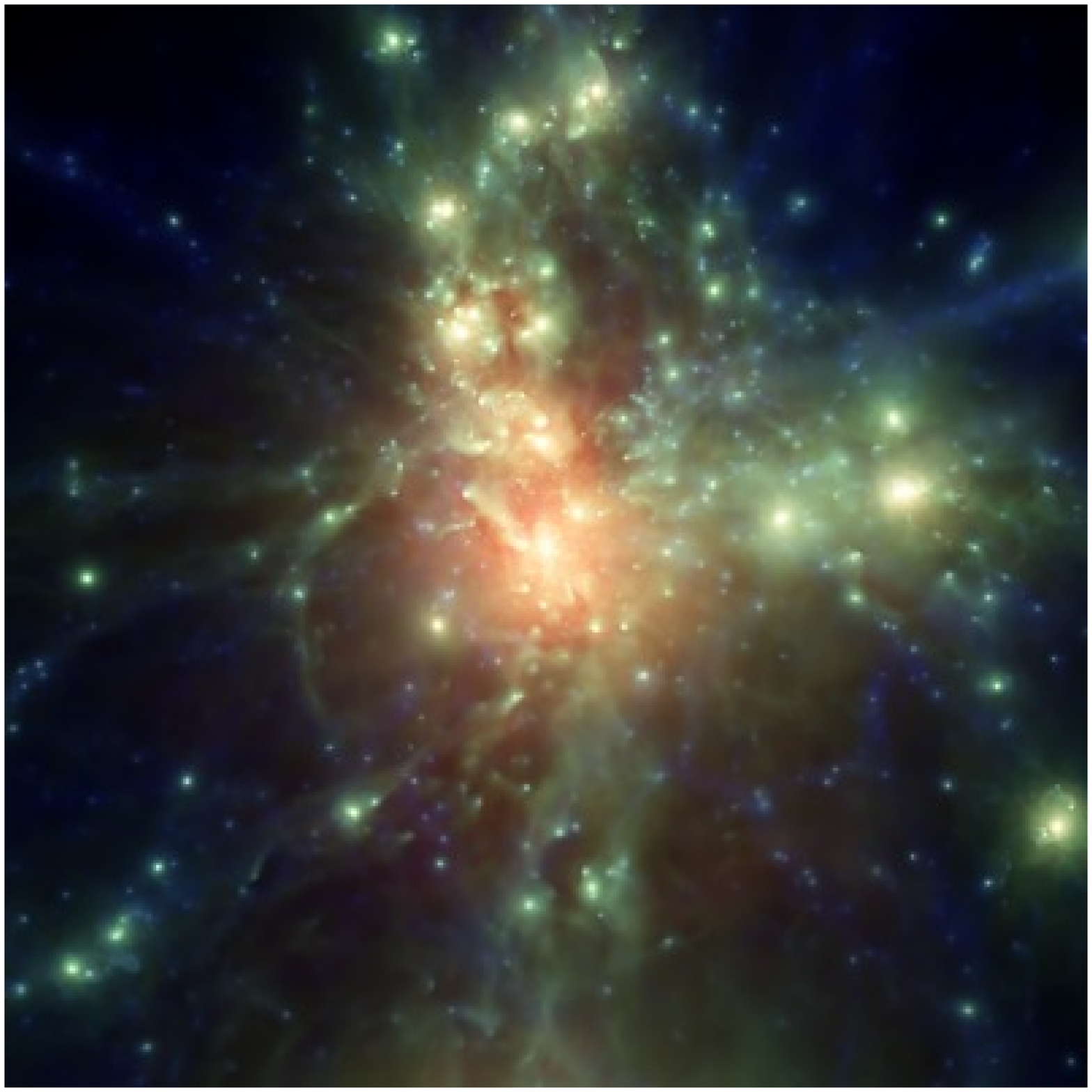} \\
\includegraphics[width=0.195\textwidth]{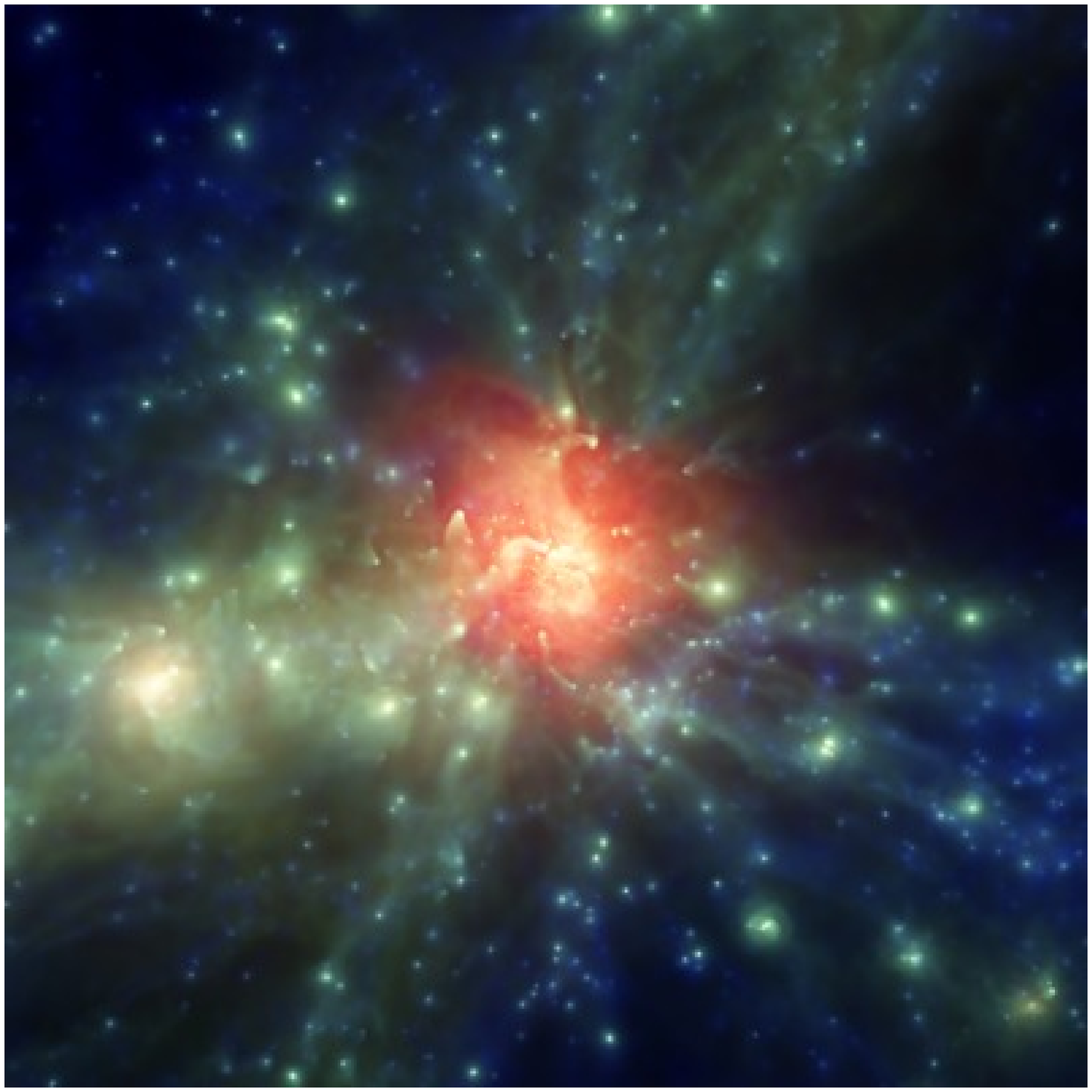} 
\includegraphics[width=0.195\textwidth]{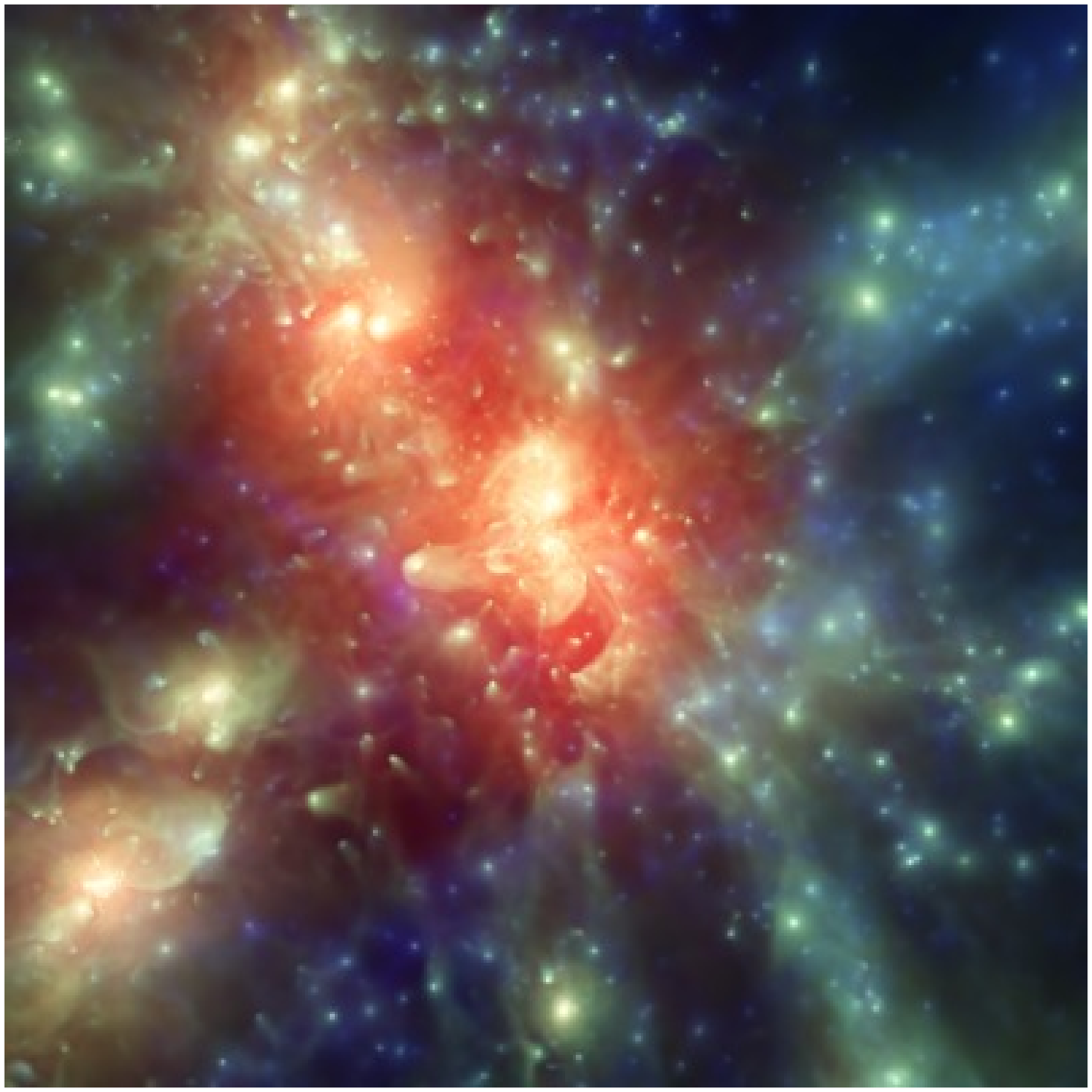}
\includegraphics[width=0.195\textwidth]{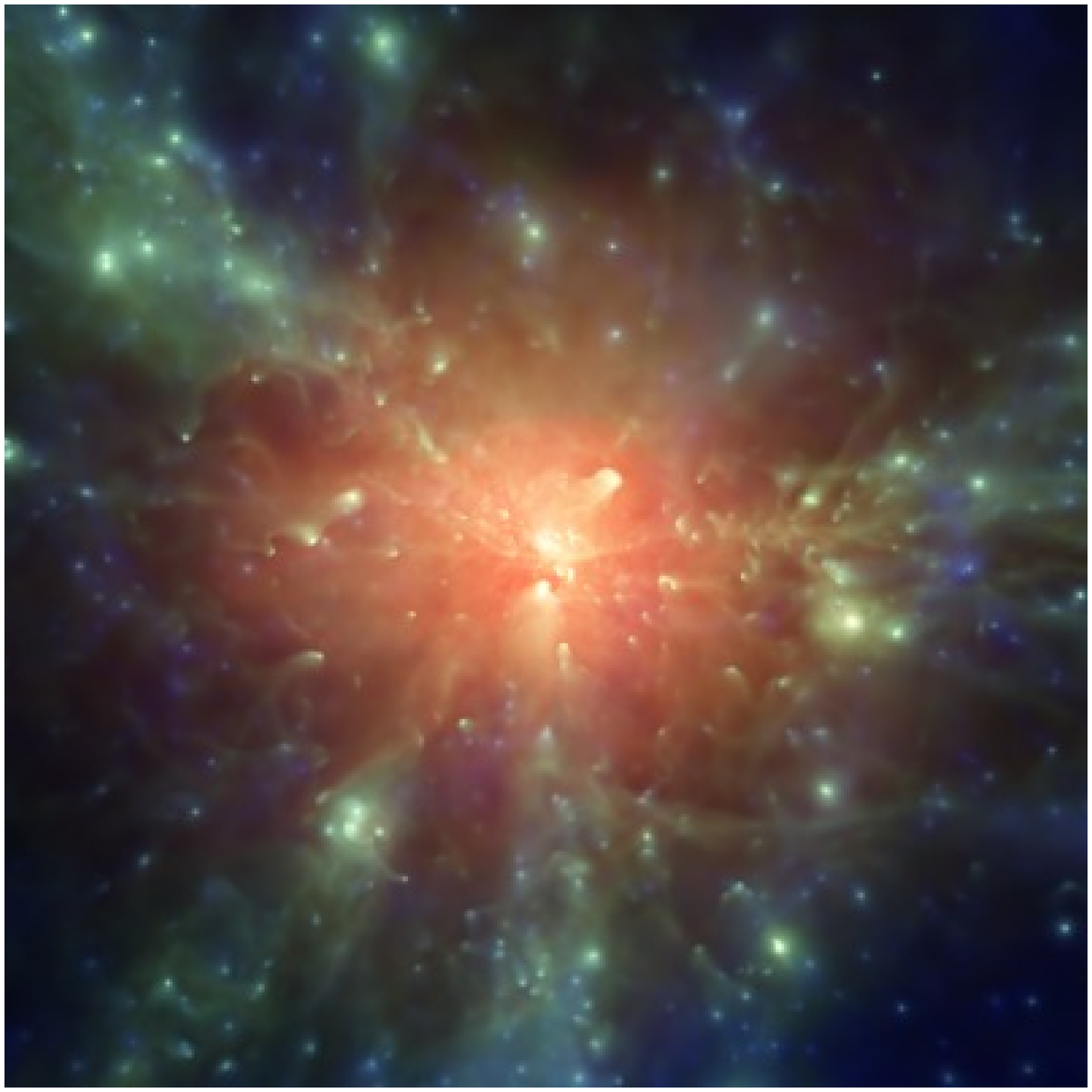}
\includegraphics[width=0.195\textwidth]{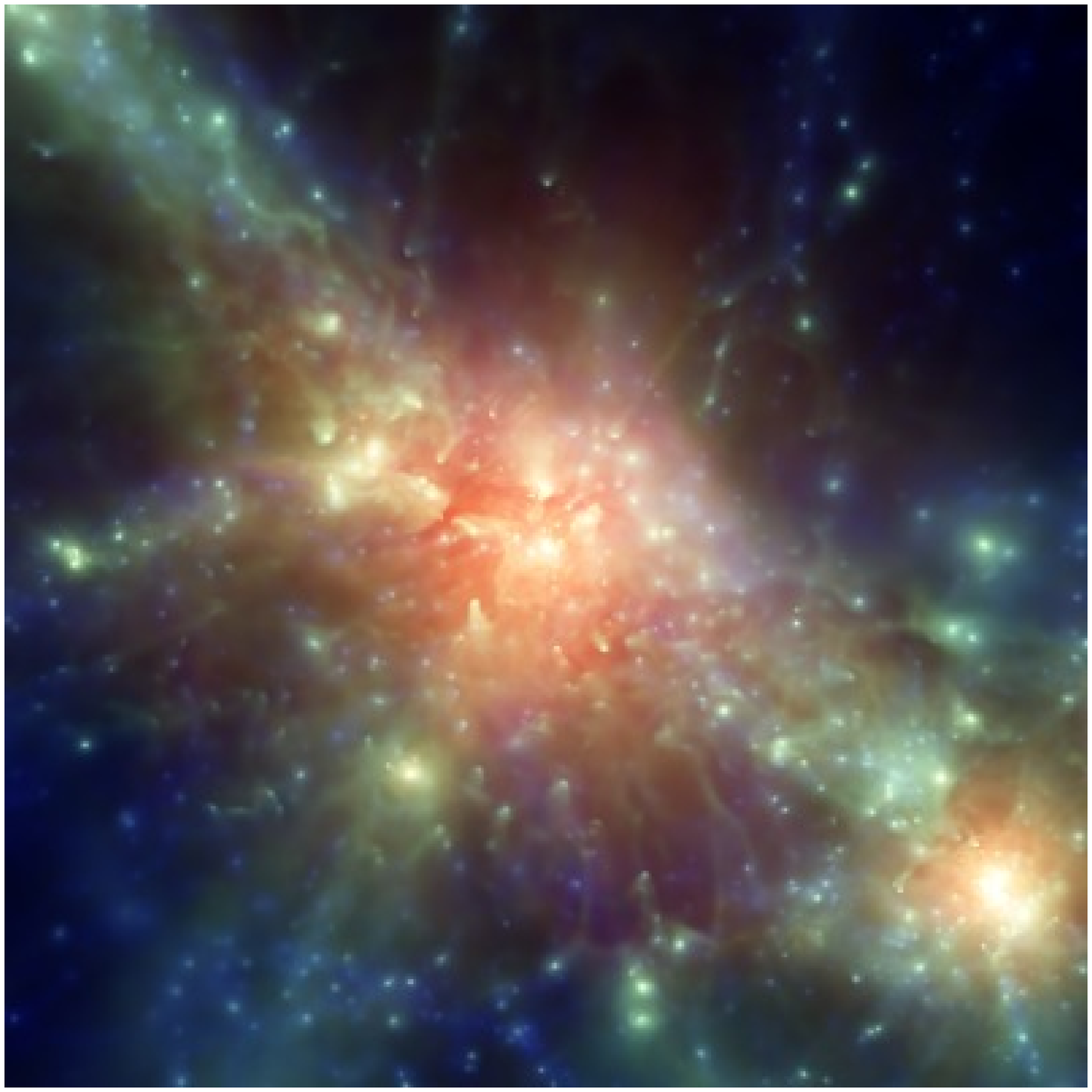} \\
\includegraphics[width=0.195\textwidth]{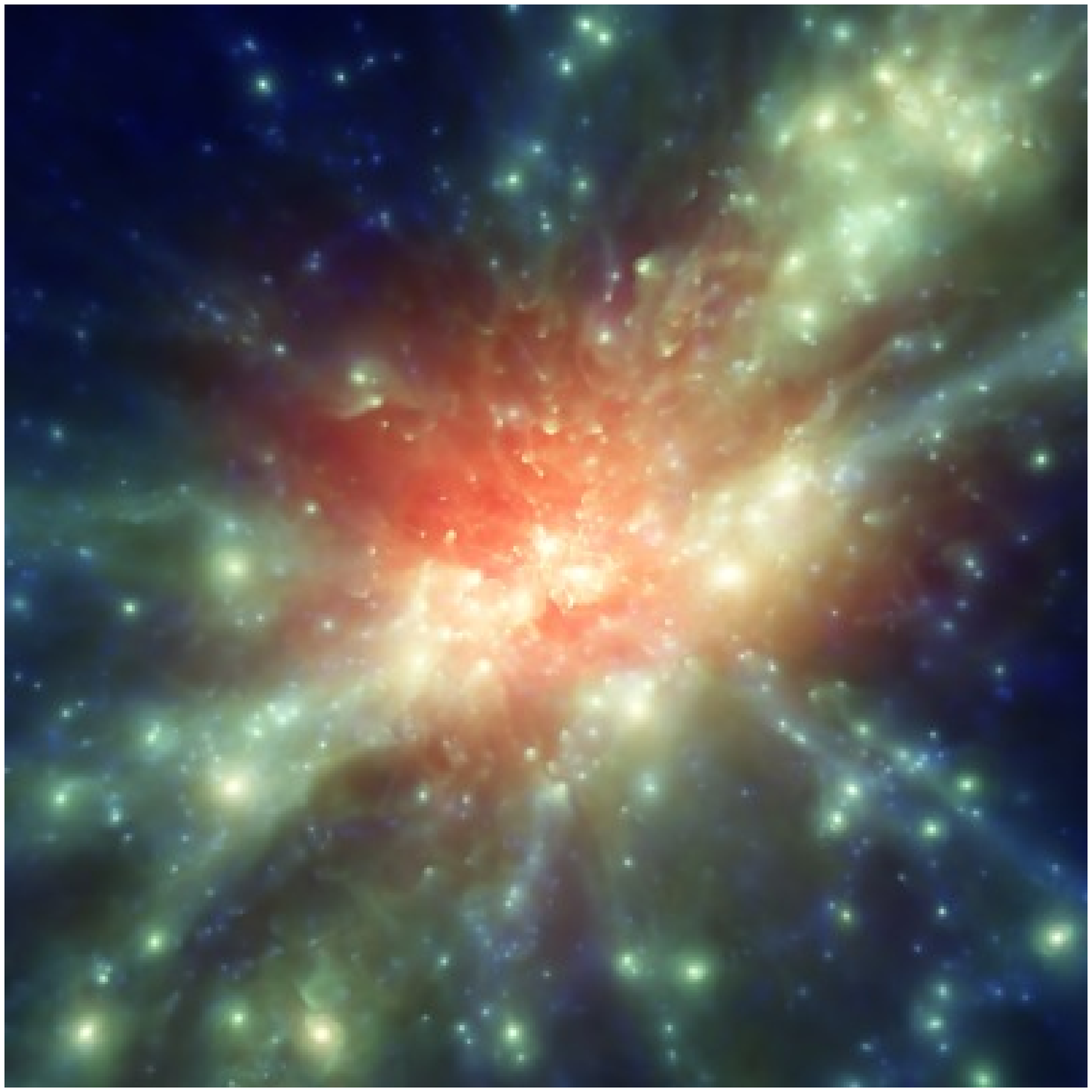}
\includegraphics[width=0.195\textwidth]{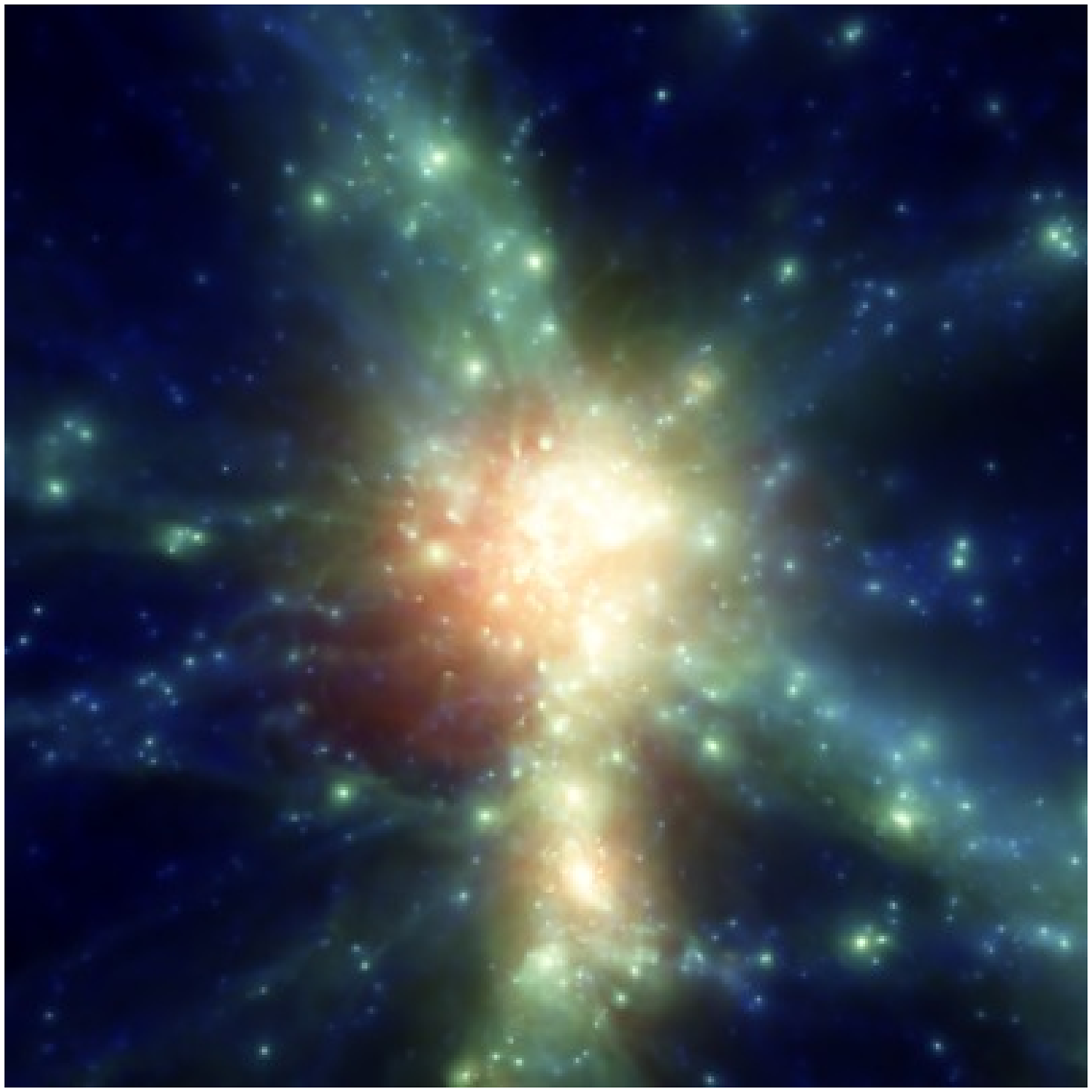} 
\includegraphics[width=0.195\textwidth]{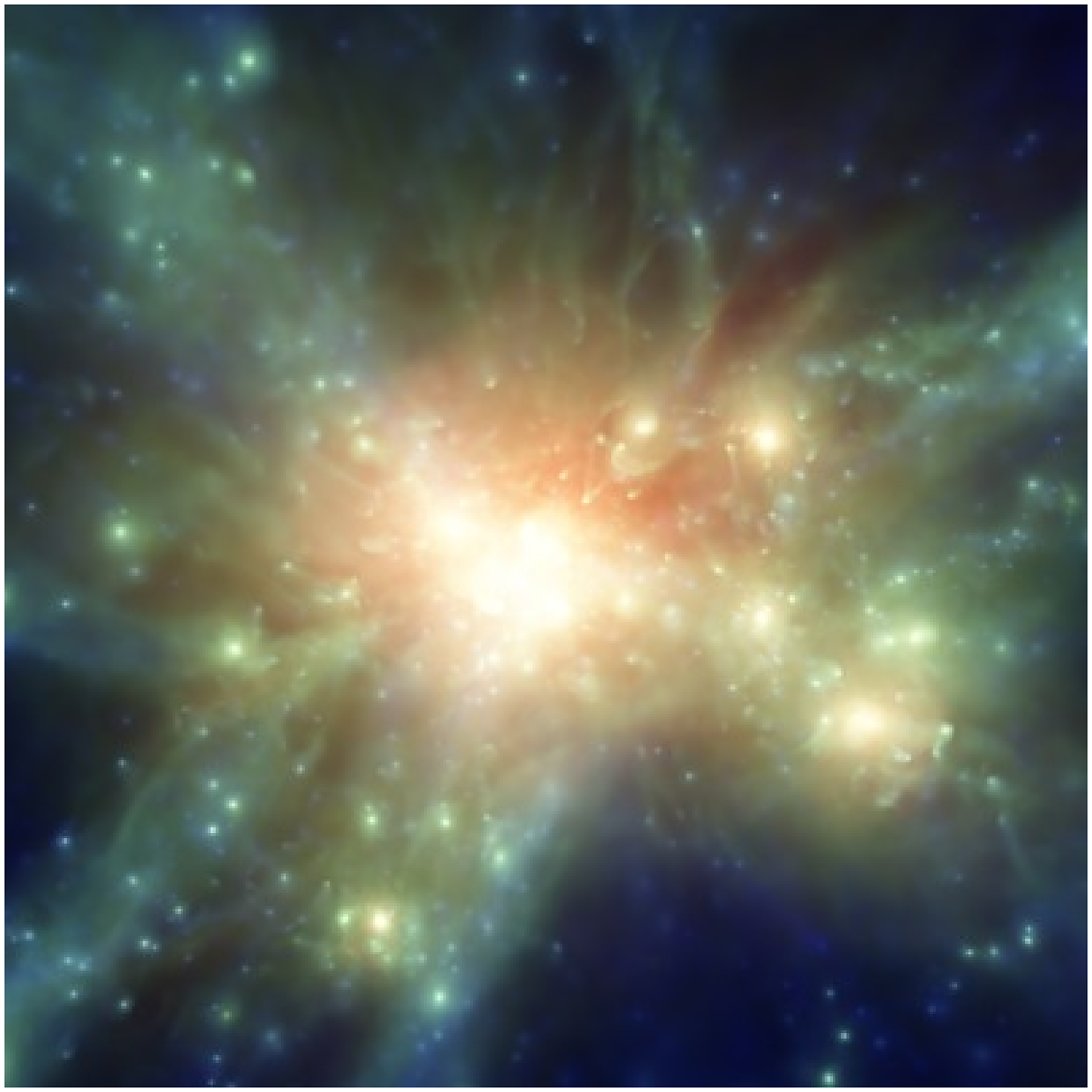}
\includegraphics[width=0.195\textwidth]{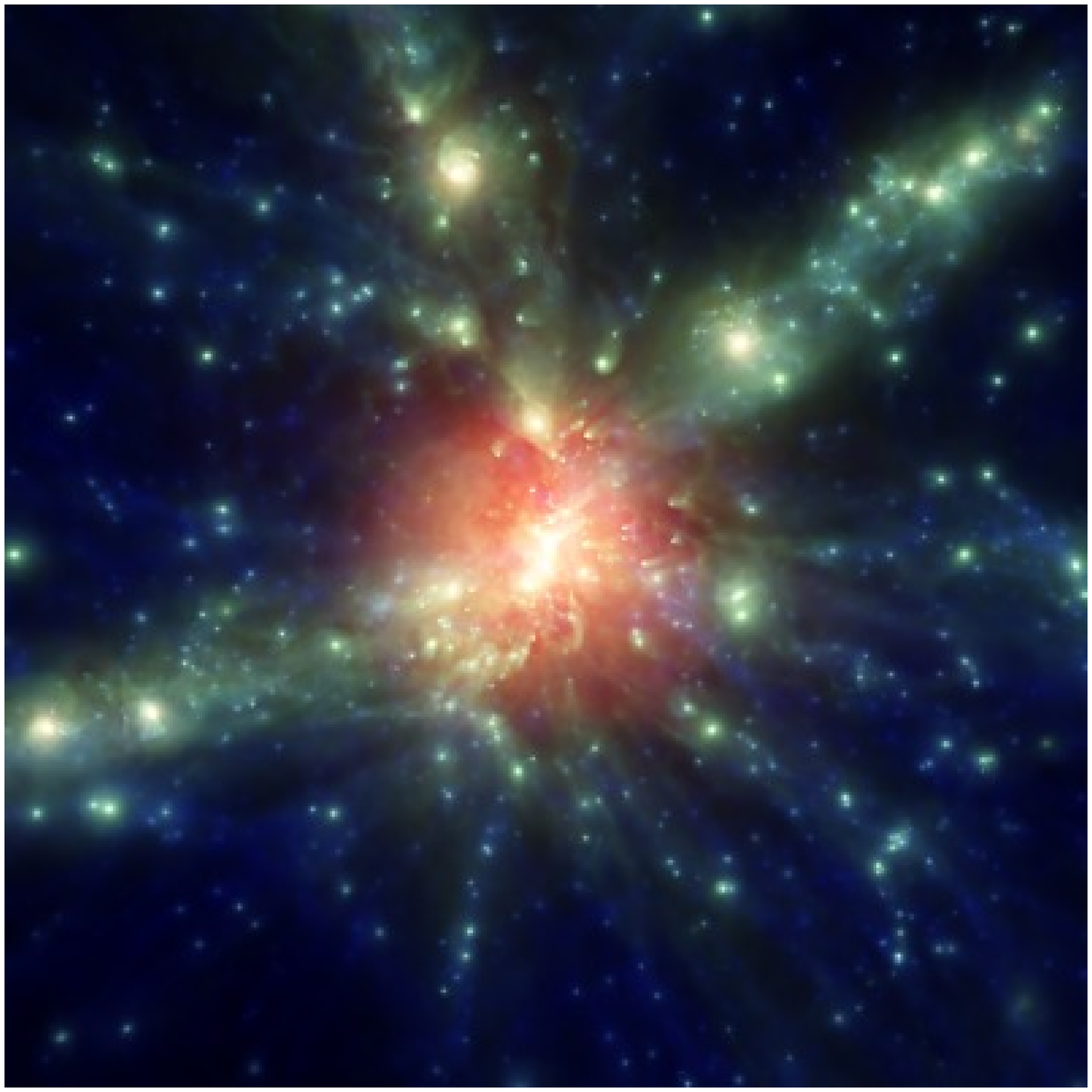} \\
\includegraphics[width=0.195\textwidth]{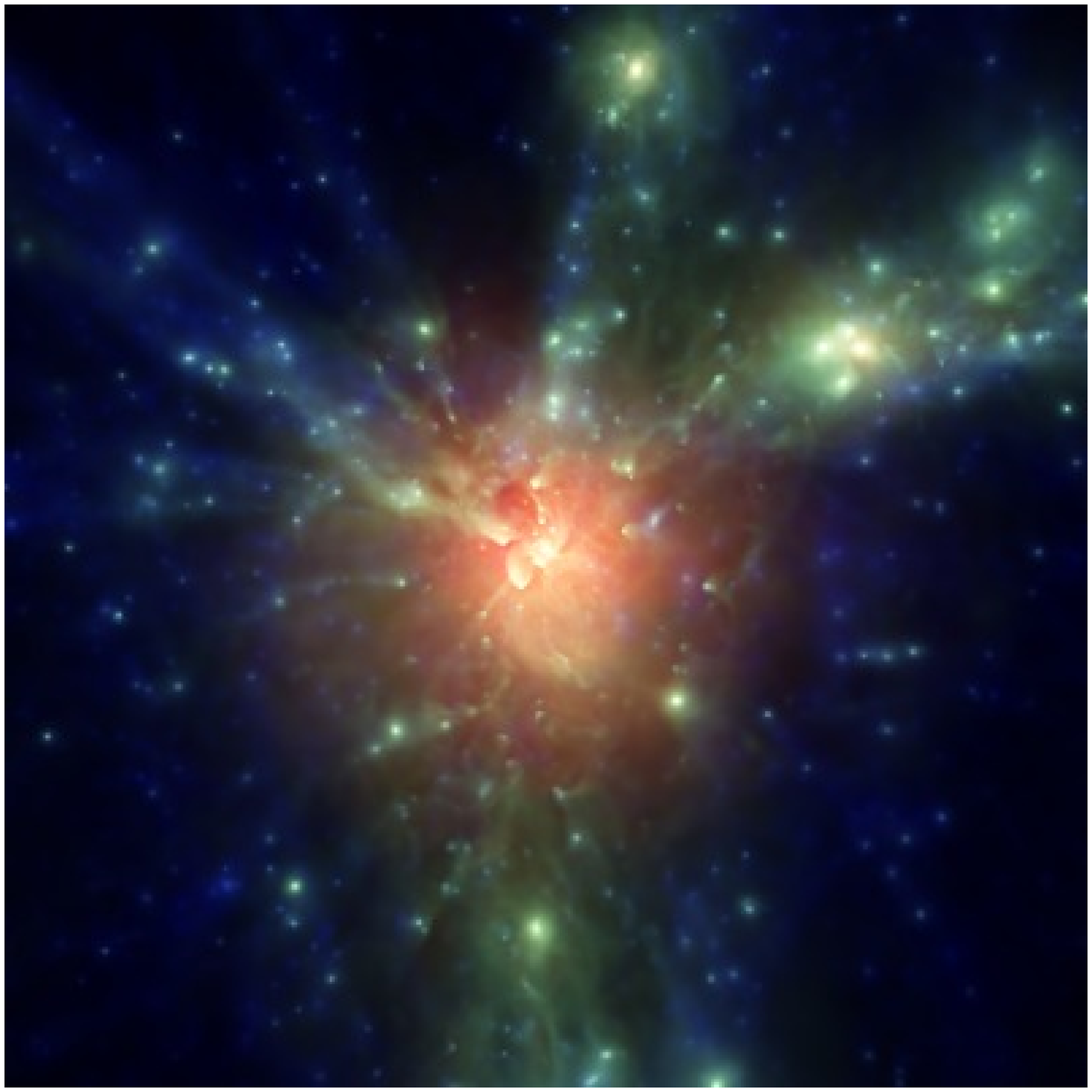}
\includegraphics[width=0.195\textwidth]{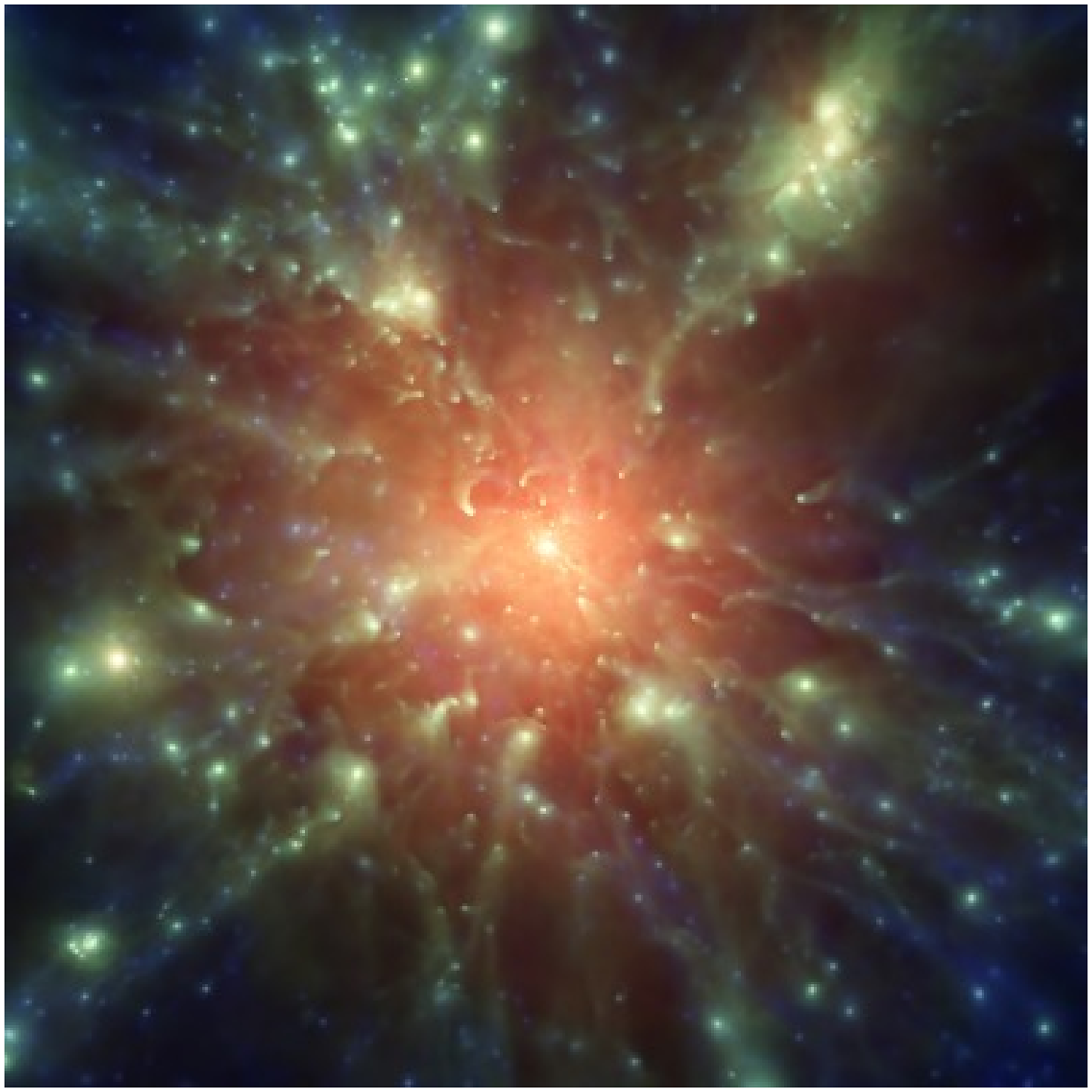}
\includegraphics[width=0.195\textwidth]{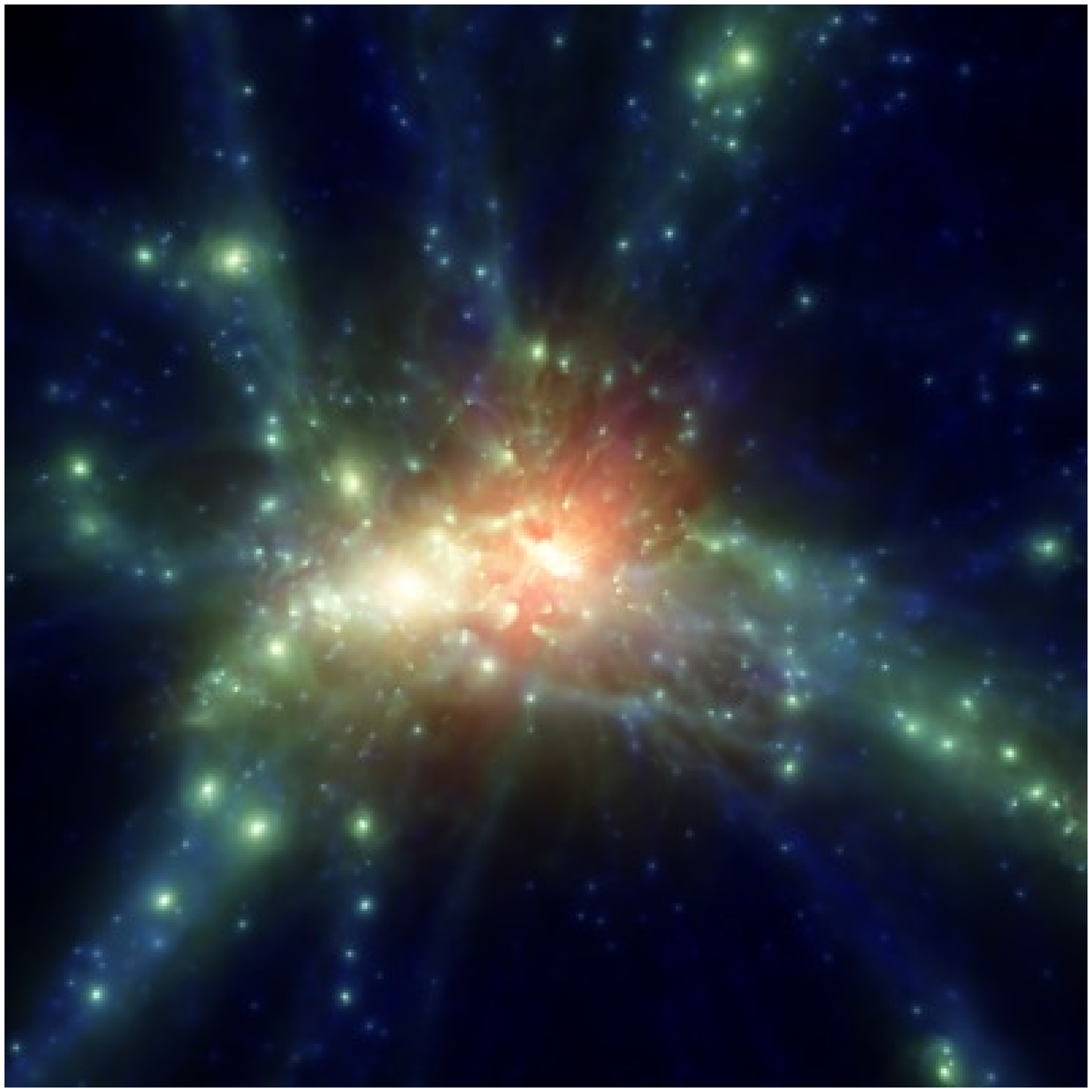} 
\includegraphics[width=0.195\textwidth]{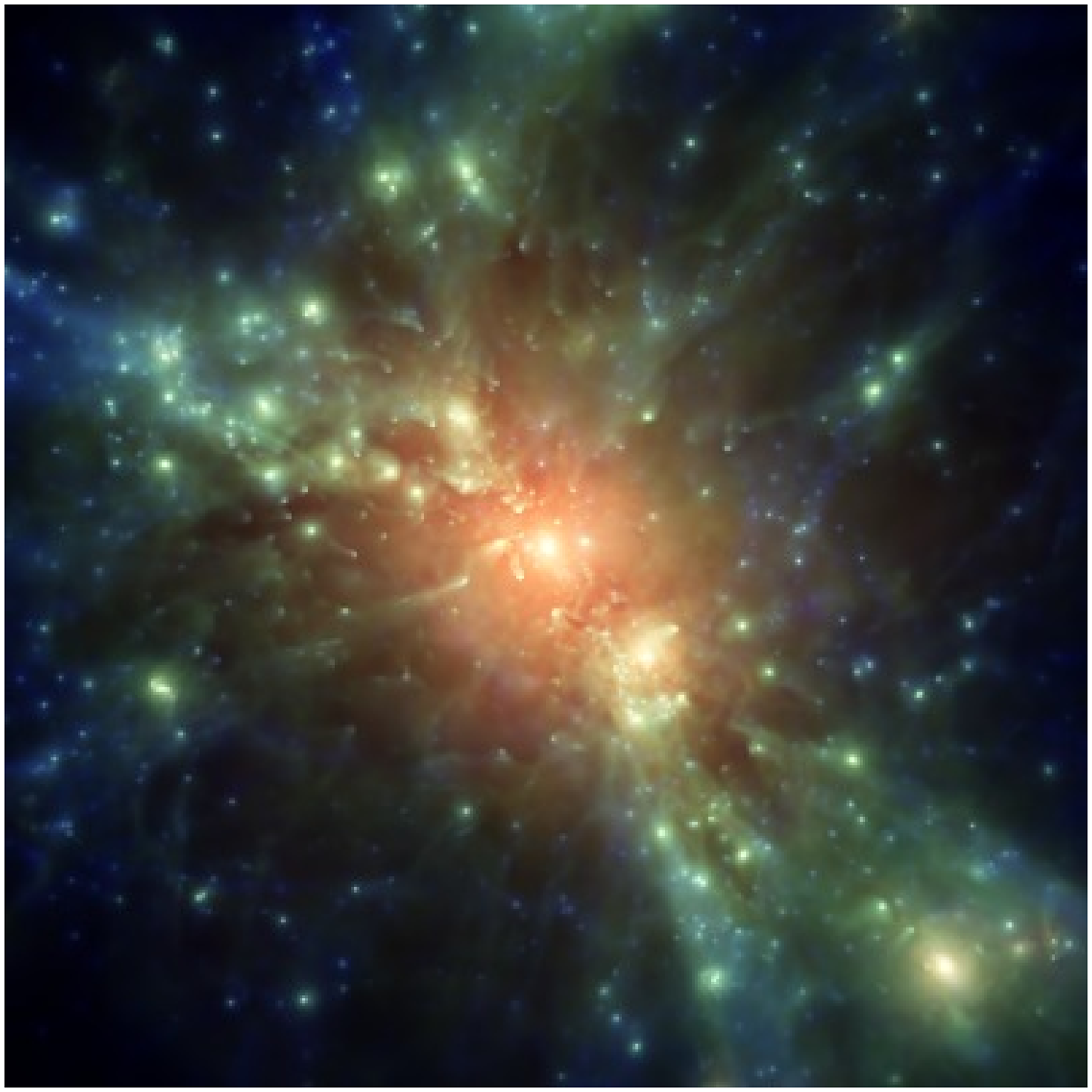} \\
\includegraphics[width=0.195\textwidth]{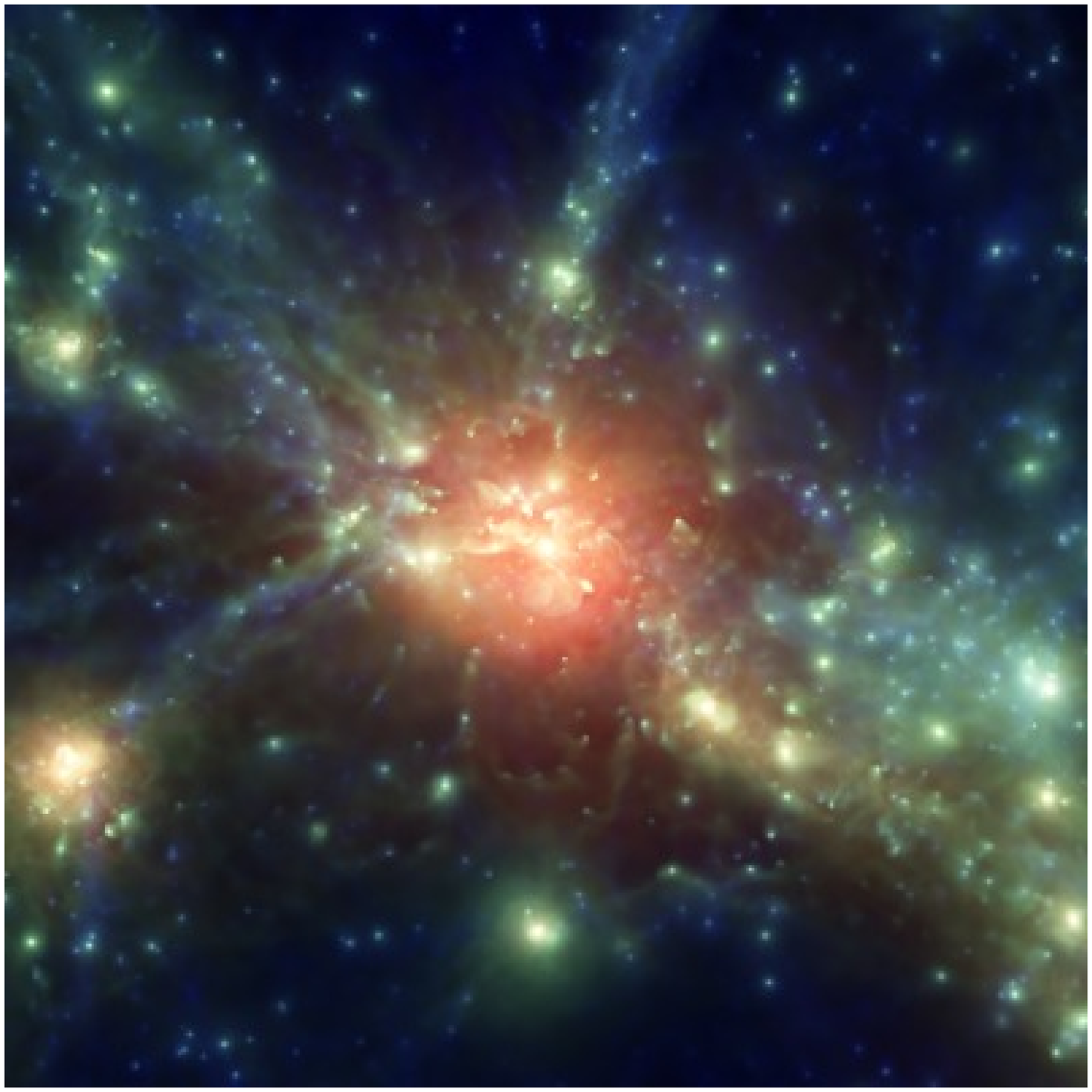}
\includegraphics[width=0.195\textwidth]{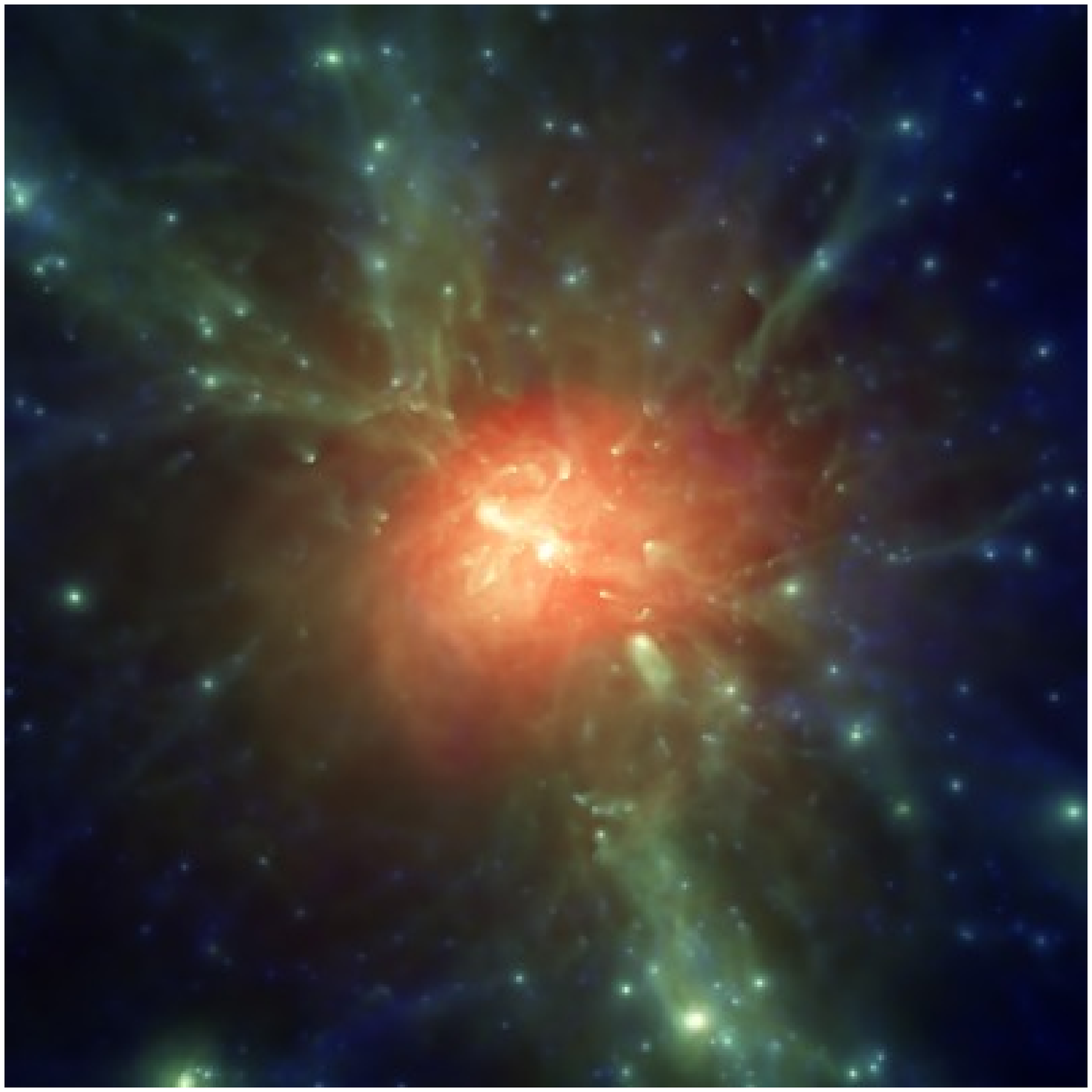}
\includegraphics[width=0.195\textwidth]{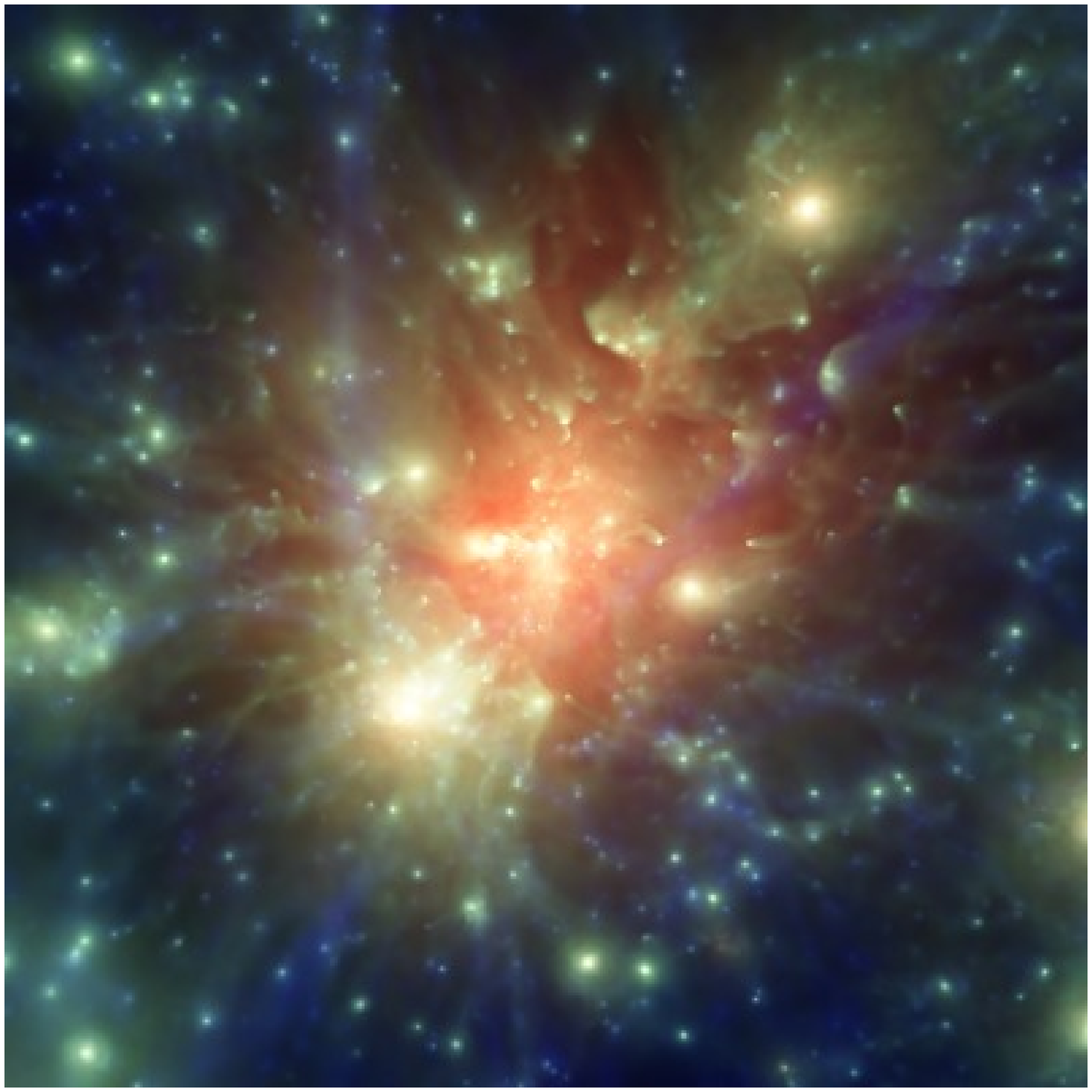}
\includegraphics[width=0.195\textwidth]{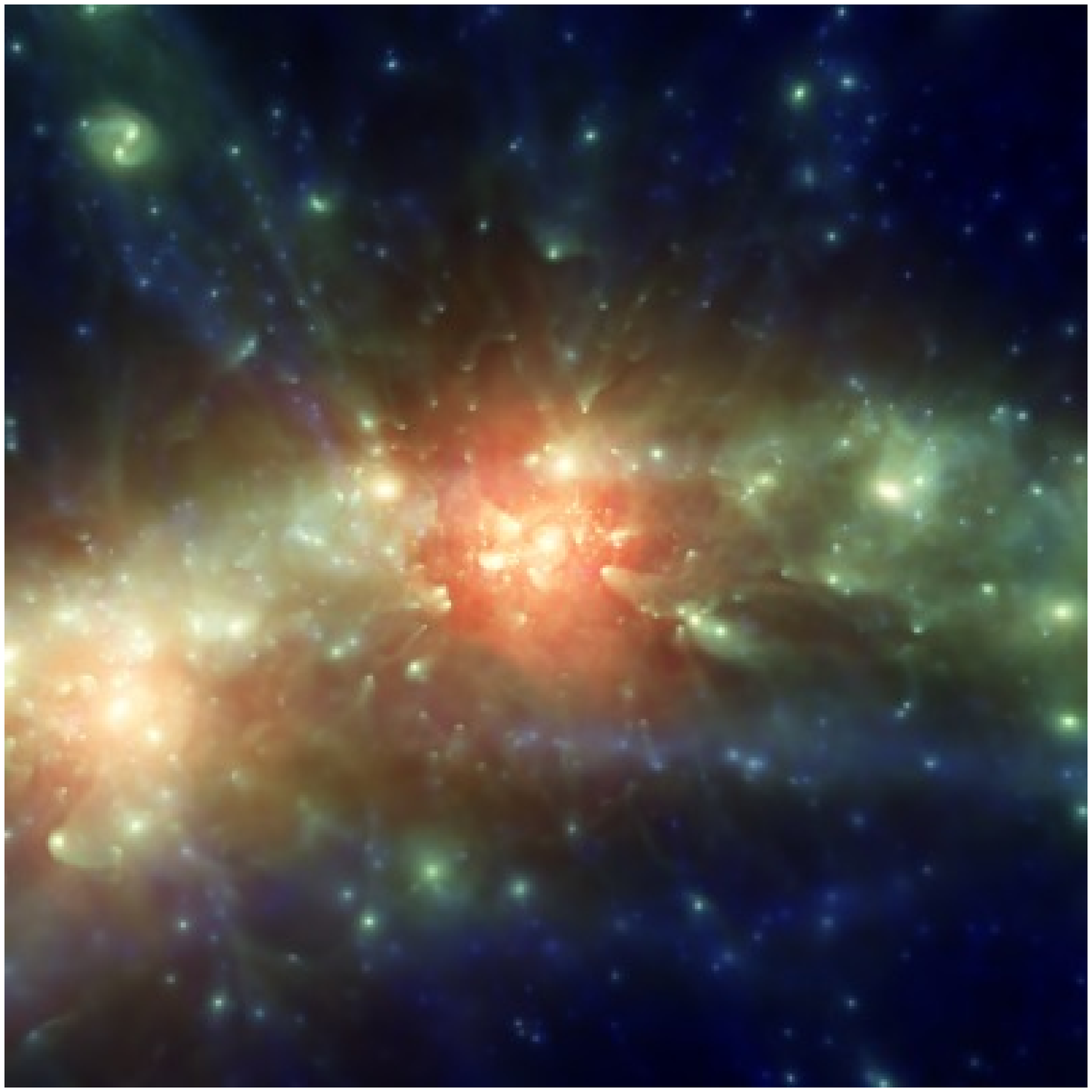} \\
\includegraphics[width=0.195\textwidth]{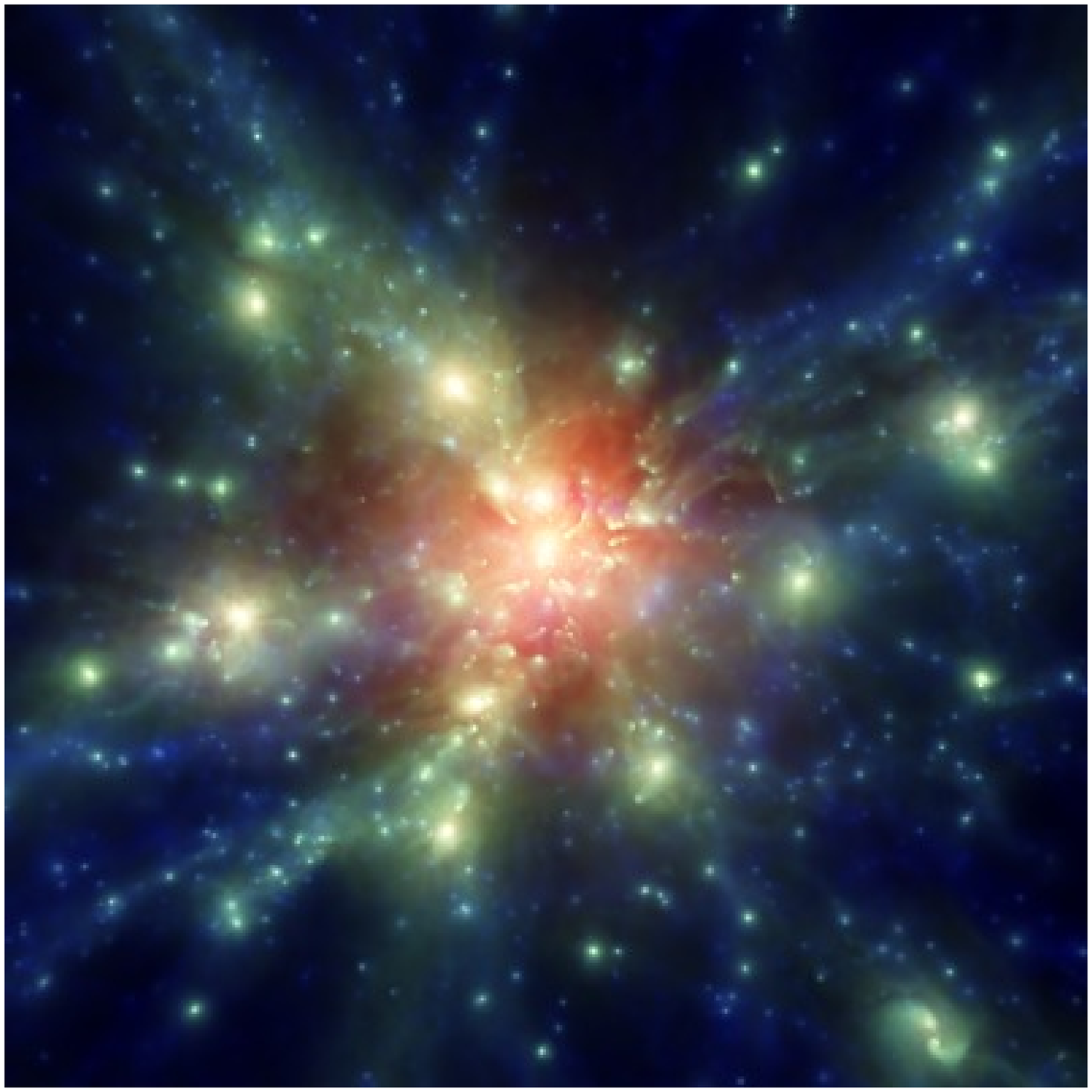}
\includegraphics[width=0.195\textwidth]{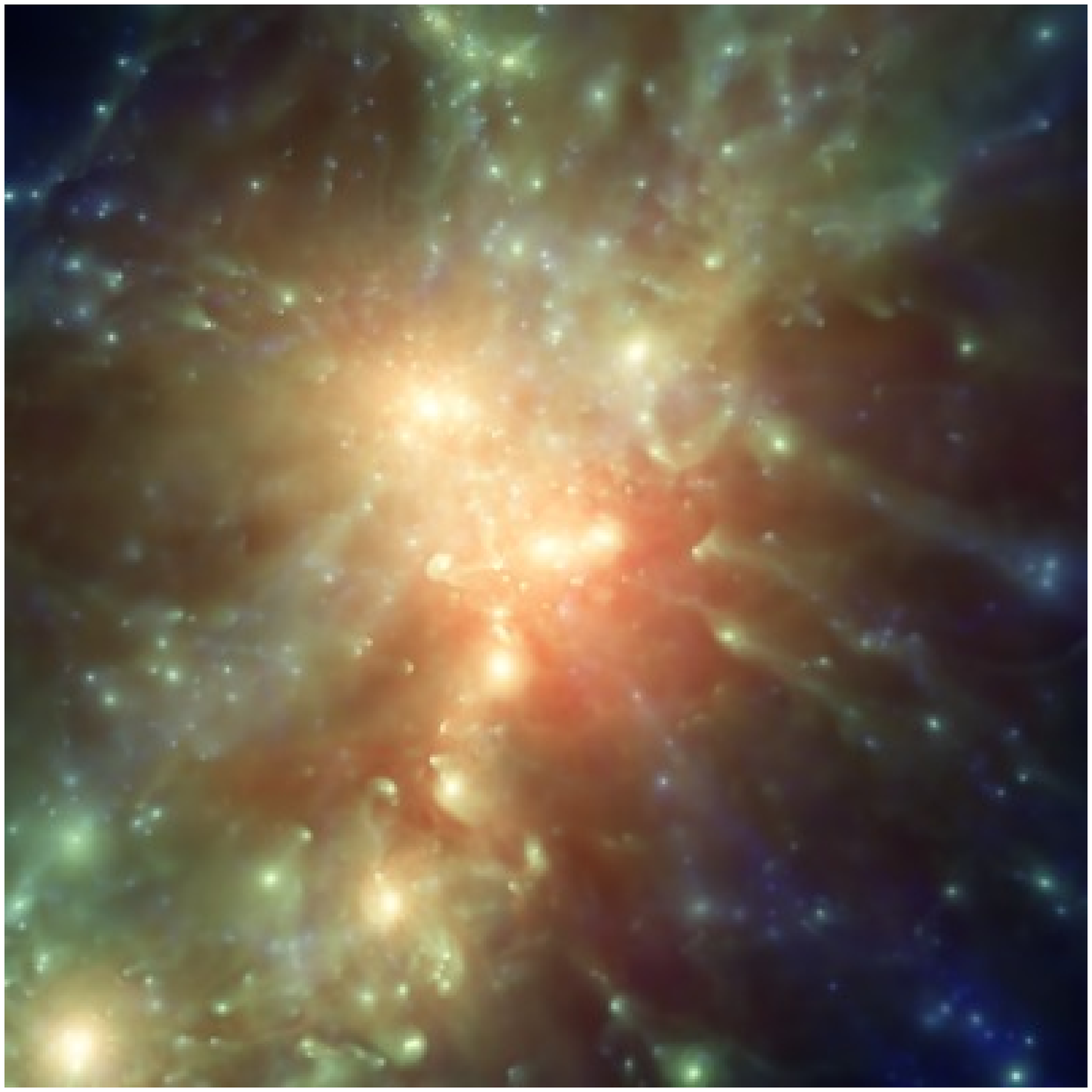}
\includegraphics[width=0.195\textwidth]{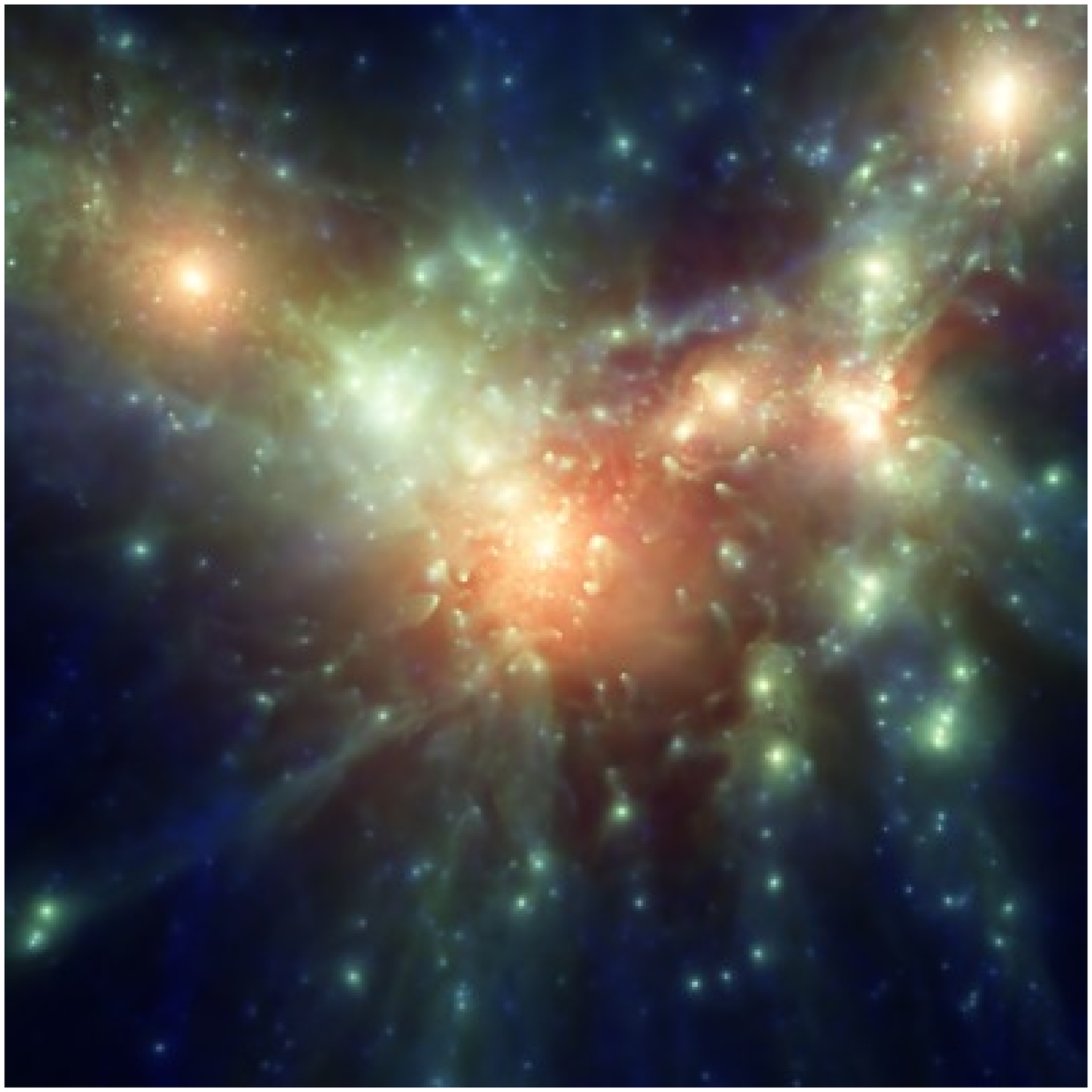}
\includegraphics[width=0.195\textwidth]{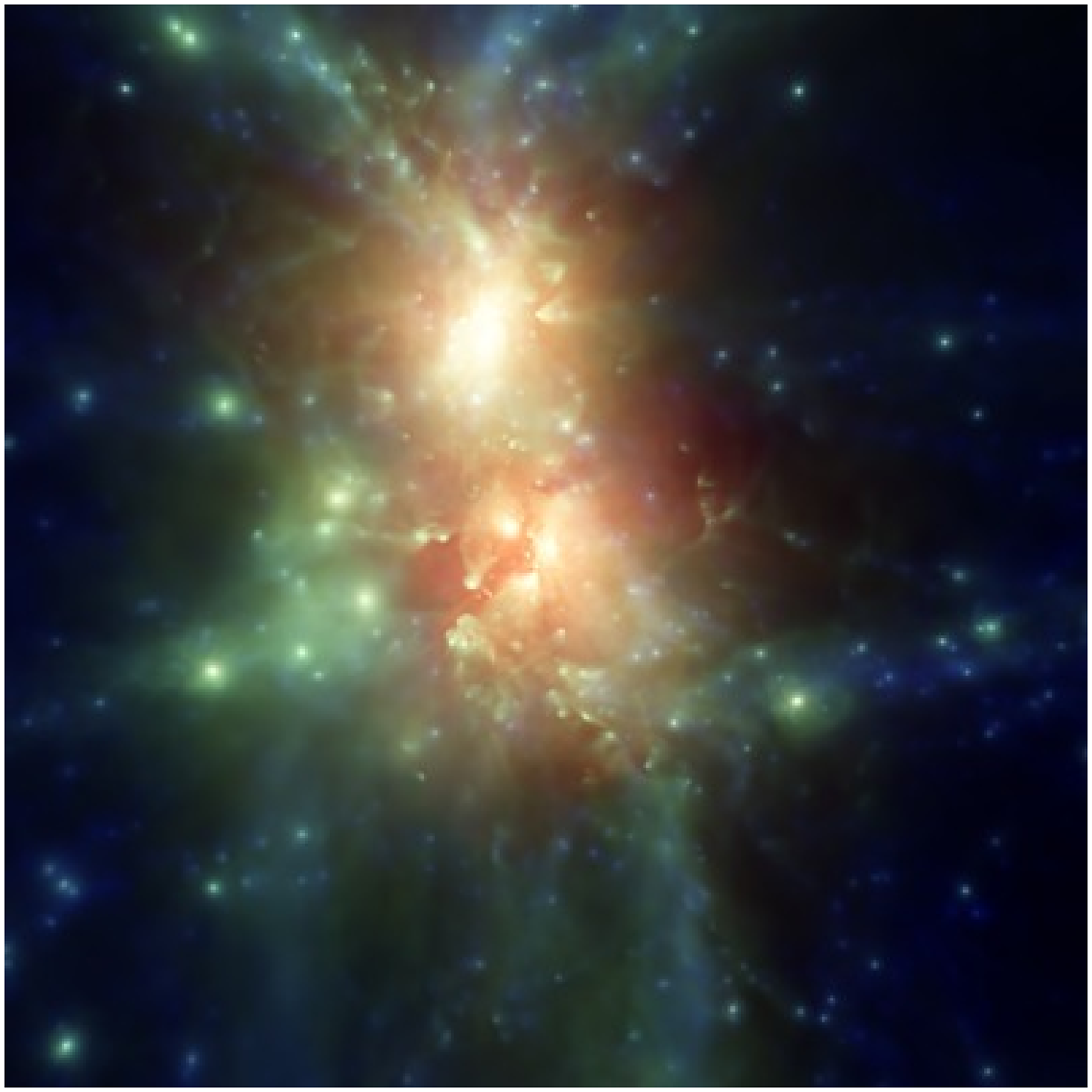}
\end{center}
\caption{Ray-tracing images of a 15 Mpc$h^{-1}$ regions around the 
center of the individual clusters. Color coded is the temperature of the gas.}
\label{fig:clusters}
\end{figure*}

\begin{table*}
\caption{Clusters with mass $10^{15} h^{-1}$\msun
  $>M_\mathrm{vir}>10^{14} h^{-1}$\msun free of low-resolution
  particles.}
\label{tab:Dianoga_all}
\centering
\begin{tabular}{| c c c c c | } 
\hline\hline 
Cluster name  &   $M_\mathrm{vir}$  & $M_{gas}$  & $R_\mathrm{vir}$ & $R_{\rm cleaned}$\\
              &   $M_{\odot} \rm{h}^{-1}$ &$h^{-1}M_{\odot}$  & kpc h$^{-1}$& $R_\mathrm{vir}$\\
d1\_9   &  1.436E+14 &   2.067E+13   & 1117.11 & $>$5 \\
d2\_2    & 1.708E+14 & 2.506E+13 & 1183.26 & $\geq$5\\
d2\_5   & 1.646E+14 & 2.331E+13 & 1168.81 & $\geq$5 \\
d2\_6   & 1.070E+14 & 1.540E+13 & 1013.07 & $\geq$5\\
d3\_4   &  5.249E+14 &   7.466E+13   & 1723.11 & $>$5\\
d3\_23  &  1.219E+14 &   1.709E+13   & 1057.81 & $>$5 \\
d5\_2    & 7.707E+14 & 1.104E+14 & 1955.10 & 4\\
d5\_6    & 1.768E+14 & 2.642E+13 & 1197.06  & $\geq$5 \\
d5\_11   & 2.266E+14 & 3.327E+13 & 1300.91  & 4\\
d5\_25   & 2.259E+14 & 3.301E+13 & 1299.61  & 4\\
d6\_6  &  1.434E+14 &  2.117E+13   & 1116.47   & $>$5 \\
d6\_11   & 2.286E+14  &  3.191E+13   & 1304.73  & $>$5 \\
d6\_18  & 1.310E+14  &  1.971E+13   & 1083.39 & $ 1$ \\
d6\_26   &  5.762E+14 &  8.107E+13   & 1777.52  &  $\geq$5   \\
d7\_11  &  1.520E+14 &  2.232E+13  & 1138.18  & $>$5  \\
d8\_1   & 4.884E+14 & 7.196E+13 & 1682.05 &  3\\
d8\_6    & 4.993E+14 & 6.883E+13 & 1694.58 &  $\geq$5  \\
d8\_8   & 2.112E+14 & 2.993E+13 & 1270.59 & 2 \\
d8\_29  & 1.081E+14 & 1.485E+13 & 1016.29 &  $\geq$5  \\
d10\_2  & 1.838E+14 & 2.532E+13 & 1212.96 &  $\geq$5 \\
d10\_3  & 8.074E+14 & 1.131E+14 & 1984.74 & 3\\
d10\_4    & 2.889E+14 & 4.126E+13 & 1411.13 & 3 \\
d10\_6   & 1.233E+14 & 1.802E+13 & 1061.86 &  $\geq$5  \\
d10\_12  & 1.017E+14 & 1.475E+13 & 996.10  &  $\geq$5 \\
d10\_16  & 1.057E+14 & 1.346E+13 & 1008.81 &  $\geq$5 \\
d11\_3   & 1.119E+14 & 1.562E+13 & 1028.04 &  $\geq$5  \\
d12\_1    & 2.622E+14 & 3.596E+13 & 1366.08 &  $\geq$5\\
d12\_4  & 3.815E+14 & 5.462E+13 & 1548.76 &  $\geq$5\\
d13\_1  & 4.930E+14 & 6.871E+13 & 1687.93  &$\geq$5\\
d13\_2 & 3.808E+14 & 5.518E+13 & 1548.19  &$\geq$5\\
d13\_3 & 4.868E+14 & 7.024E+13 & 1680.78  & 1\\ 
d13\_7  & 2.426E+14 & 3.830E+13 & 1331.19  &$\geq$5\\
d14\_2 & 1.344E+14 & 2.043E+13 & 1092.54 &  $\geq$5\\ 
d14\_3  & 3.065E+14 & 4.416E+13 & 1439.40 &  $\geq$5\\
d14\_5   & 1.754E+14 & 2.519E+13 & 1194.04 &  $\geq$5 \\
d15\_7  & 1.419E+14 & 2.037E+13 & 1112.67 &  $\geq$5  \\
d17\_4 & 1.072E+14 & 1.501E+13 & 1013.47 &  1 \\ 
d18\_1   & 2.982E+14 & 4.633E+13 & 1426.32 &  $\geq$5\\
d19\_5   & 2.573E+14 & 3.267E+13 & 1357.47 &  $\geq$5 \\
d20\_2  & 6.489E+14 & 8.920E+13 & 1849.37  &$\geq$5 \\ 
d20\_4 & 1.135E+14 & 1.615E+13 & 1033.07  &$\geq$5\\ 
d21\_2  & 2.083E+14 & 2.978E+13 & 1264.69 &  4\\
d21\_3 & 3.019E+14 & 4.324E+13 & 1432.09 &  3\\ 
d21\_19 & 1.031E+14 & 1.687E+13 & 1000.69 & $\geq$5  \\
d21\_23 & 1.526E+14 & 2.141E+13 & 1139.66 & $\geq$5 \\ 
d23\_2   & 1.134E+14 & 1.619E+13 & 1032.80 & $\geq$5  \\
d23\_4  & 2.970E+14 & 4.217E+13 & 1424.53 & $\geq$5 \\
d23\_7  & 1.172E+14 & 1.681E+13 & 1044.22 & $\geq$5   \\
d24\_1   & 7.822E+14 & 1.096E+14 & 1964.50 & 3 \\
d24\_22   & 1.149E+14 & 1.485E+13 & 1037.26 & $\geq$5 \\
d24\_363  & 3.341E+14 & 4.041E+12 & 1479.48 & $\geq$5\\

\hline

\multicolumn{5}{l}{\scriptsize Col. 1: Cluster name; Col. 2: Total
  mass of the cluster inside the virial radius}\\ \multicolumn{5}{l}{
  \scriptsize Col. 3: Mass of the gas component inside the virial
  radius; Col 4: Virial radius;} \\ \multicolumn{5}{l}{ \scriptsize
  Col 5: Number of virial radii cleaned by LR particles.}
\end{tabular}
\end{table*}


\begin{figure*}
\begin{center}
\includegraphics[width=0.49\textwidth]{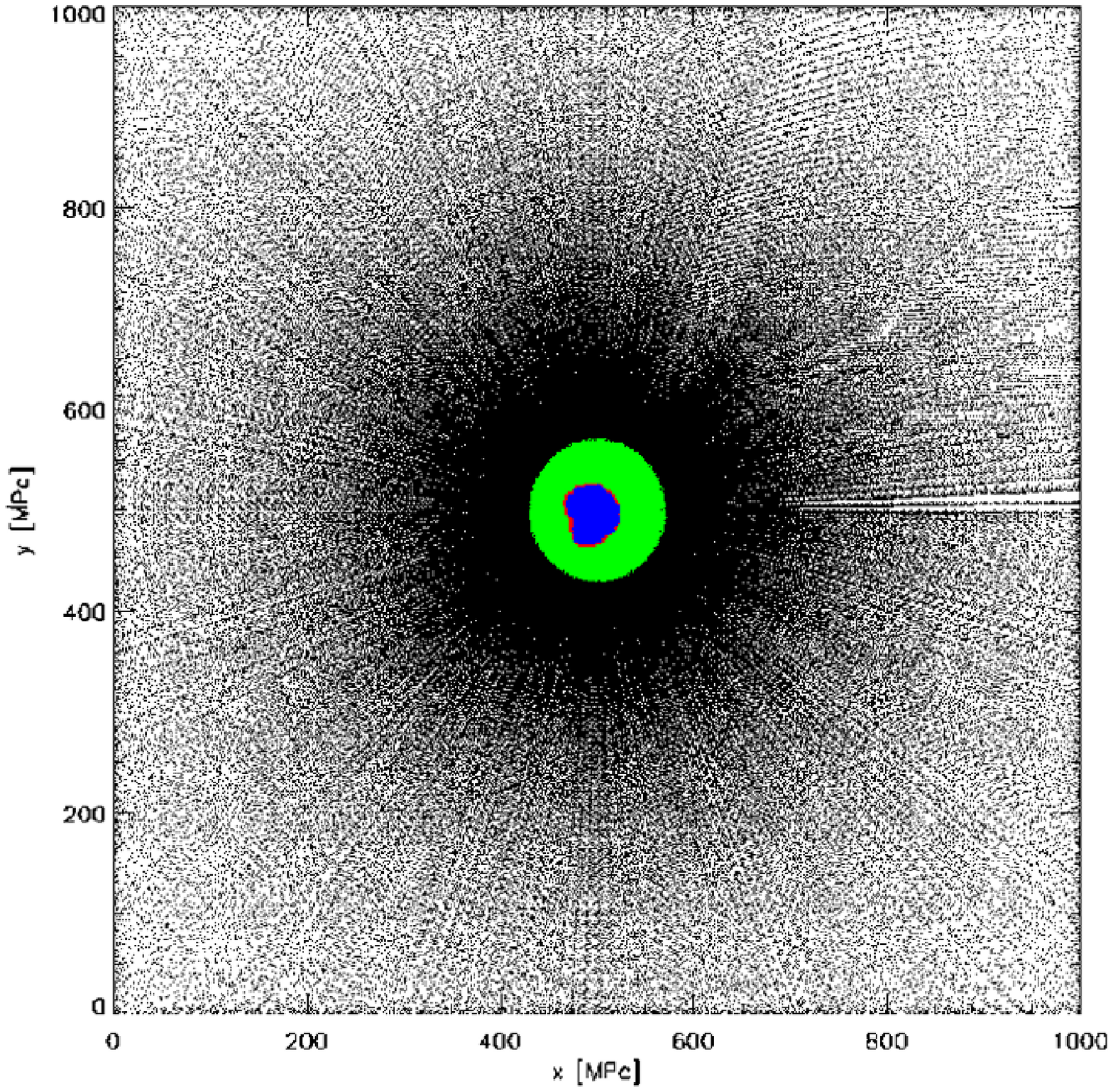}
\includegraphics[width=0.49\textwidth]{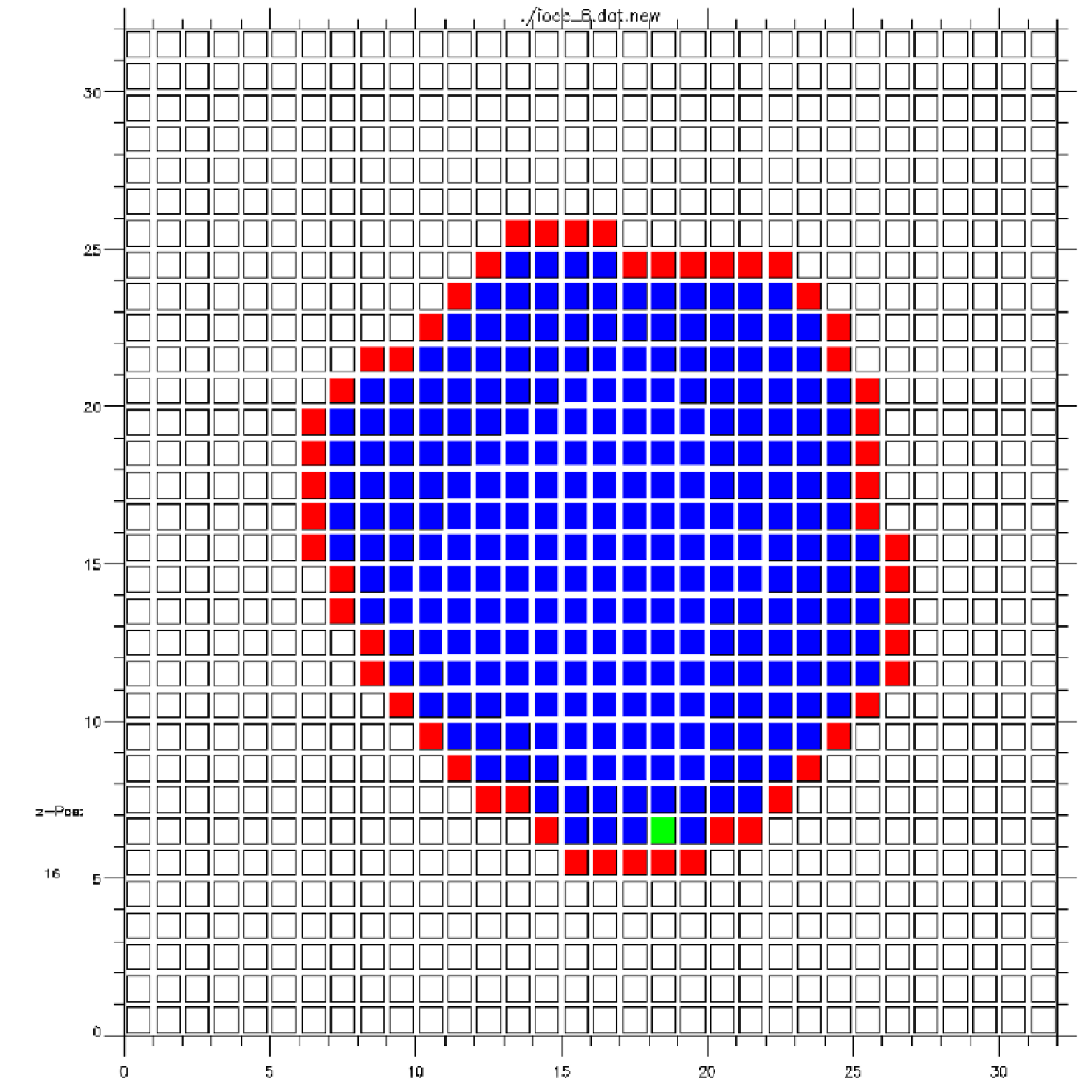}
\end{center}
\caption{Initial condition region for the high resolution
  simulations. Left: Black: DM particles with degraded mass resolution
  outside the HR region, \eg grained version of the original IC region
  used in the parent simulation, with increasing mass toward the outer
  regions. Green: DM particles outside the HR region with the same
  mass resolution than the parent simulation. This represents a
  ``safety region'' where a normal grid is used and particles have the
  same mass that the parent simulation. Red: HR region. Blue: region
  where high resolution DM particles have been splitted into gas and
  DM particles. Right: A slice through the HR initial condition
  region. Blue boxes refer to the position of the particles traced
  back, which where at $z=0$ falling within 5 $R_\mathrm{vir}$ of the
  target cluster. Red boxes are the cells that are included
  automatically to obtain a concave region.  The green box refer to
  cell which was added by hand to avoid holes within the HR region.}
\label{fig:app_gas}
\end{figure*}

\subsection{Adding the baryonic component}
Once the IC for the DM particles have been obtained, the baryonic
component was added. The high resolution dark matter particles are
splitted into one gas and one DM particle. The mass of the initial DM
particle is splitted according to the cosmic baryon fraction,
conserving the center of mass and the momentum of the parent DM
particle. We displaced them by half the mean interparticle
distance. Here we added a further optimization. Taking the ``cleaned
region'' around all clusters of interest within the high resolution
region, we traced back the corresponding Lagrangian region into the
initial conditions. To associate concave volume to the selected
particles, we measured their distance to the center of the high
resolution region and calculated the maximum distance found in each
direction by sampling the sphere using a HealPIX discretization
\citep{2005ApJ...622..759G}.  Only those dark matter particles which
are found within such a volume (including a very small safety buffer)
were splitted into one gas and one DM particles. This typically saves
$\approx 20\%$ of gas particles while still having splitted dark
matter (and, accordingly, gas particles) within the full extent of the
``clean region''.\\ In the left panel of figure \ref{fig:app_gas}, in
the central part, the spatial extent of the whole high resolution
region compared to the extent of baryon-filled region is visible.
\end{appendix}

\end{document}